\documentclass[osajnl,showpacs,10pt,twocolumn]{revtex4-2}

\usepackage{amsmath,amssymb,graphicx}
\usepackage{float}
\usepackage{multirow}
\usepackage{colortbl}
\DeclareMathOperator{\sech}{sech}
\usepackage[section]{placeins}

\begin{document}

\title{Unidirectional flow of composite bright-bright solitons through asymmetric double potential barriers and wells}
\author{$^{1}$ Amaria Javed}
\author{$^{1}$ T. Uthayakumar}
\author{$^{2}$ M. O. D. Alotaibi}
\author{$^{3}$ S. M. Al-Marzoug}
\author{$^{3}$ H. Bahlouli}
\author{$^{1}$ U. Al Khawaja}
\email{u.alkhawaja@uaeu.ac.ae}
\affiliation{$^{1}$ Department of Physics, United Arab Emirates University, P.O. Box 15551, Al-Ain, United Arab Emirates}
\affiliation{$^{2}$ Department of Physics, Kuwait University, P.O. Box 5969 Safat, 13060 Kuwait}
\affiliation{$^{3}$ King Fahd University of Petroleum and Minerals, Dhahran 31261, Saudi Arabia}

\begin{abstract}
We investigate the dynamics of two component bright-bright (BB) solitons through reflectionless double barrier and double well potentials in the framework of a Manakov system governed by the coupled nonlinear Schr\"{o}dinger equations. The objective is to achieve unidirectional flow and unidirectional segregation/splitting, which may be used in the design of optical data processing devices. We observe how the propagation of composite BB soliton is affected by the presence of interaction coupling between the two components passing through the asymmetric potentials. We consider Gaussian and Rosen-Morse double potential barriers in order to achieve the unidirectional flow. Moreover, we observe a novel phenomenon which we name ``\textit{Polarity Reversal}" in the unidirectional flow. In this situation, the polarity of the diode is reversed. To understand the physics underlying these phenomena, we perform a variational calculation where we also achieve unidirectional segregation/splitting using an asymmetric double square potential well. Our comparative study between analytical and numerical analysis lead to an excellent agreement between the two methods.
\end{abstract}

\maketitle

\section{Introduction}\label{intro}

Solitons appear in diverse fields of science and engineering, most importantly in nonlinear physics, data communication systems, biophysics, hydrodynamics, quantum field theory, etc., owing to their invariant shape and velocity, before and after collisions \cite{booka}. In optical systems, they manifest themselves as `optical solitons', realized through a delicate balance between nonlinearity and dispersion/diffraction which play an indispensable role in long haul communications systems, data processing devices, switching devices, routers and computers \cite{books1, books2, books3, books4, Yuri}. In order to realize these functionalities, diverse configurations, namely, photonic circuitry \cite{gates1}, silicon-on-silicon waveguides \cite{gates2}, X-junctions \cite{gates3}, fused-fiber couplers \cite{gates4}, liquid crystals \cite{gates5,gates6}, nonlinear waveguide arrays \cite{gates7, gates8}, Mach-Zehnder interferometer \cite{gates9}, etc., \cite{gates10,gates11,gates12,am1,am2,am4} have been proposed.

Particularly, the evolution and scattering of solitons through various external scattering potentials, namely, walls or barriers \cite{a1,a2,a3,a4,a5,a6,a7}, steps \cite{a8,Goodman,Sakaguchi,Stoychev,Morales,Lee,a9}, wells \cite{Aceves,Yuri2,a10}, and surfaces \cite{a11,a12} necessitates the nonlinear interaction and wave nature of the solitons. Optimal channelization of such potentials allow the possibility to procure soliton based switches, signal processing devices,  routers, and diodes \cite{gates1,gates7,Cuevas,usamaasad}. Diverse studies in nonlinear materials also reported enhancing the nonlinear properties for applications in photonic devices, optoelectronics and optical amplifier \cite{rev2a,rev2b,rev2c}.  Nonlinear control and soliton management, employing complex Ginzburg-Landau equation has been investigated for diverse nonlinear systems to explore their role on the transmission speed, pulse width and period of the solitons\cite{rev2d,rev2e,rev2f,rev2g,rev2h}. Analytical form for an external potential has been used to investigate the possibility of stable solitonic propagation regimes as well as an unstable regime, where the initial soliton profile decays to breathing decaying solitons \cite{rev3a}. Two-dimensional self-trapped optical waves in the presence of Laguerre-Gaussian and harmonic potentials were also investigated considering the closed-form expressions for the soliton solutions and the conservation laws for the norm and the Hamiltonian. The nonlinear modes obtained are found to be stable below a certain threshold norm value \cite{rev3b}. Further, the parameter which allows controllability of the solitonic properties in PT-symmetric Mathieu lattices, namely, shape, dynamics and stability has also been pointed out for efficient light control \cite{rev3c}. Recently, the unidirectional flow of bright solitons through a specific combination of asymmetric potential wells was demonstrated where transmission of solitons were defined by the critical velocity, below which reflection dominates \cite{usamaasad}. Moreover, such special flow of solitons were also found to occur in the solitons scattering through the parity-time symmetric potentials. The physics behind such phenomenon was found to be related to an energy exchange interaction that occurred between the soliton internal modes and its center of mass dynamics during unidirectional scattering \cite{usamayuri}. Furthermore, Reference \cite{usamaandrey} predicted the realization of such unidirectional flow through photonic waveguide arrays with two asymmetric potential wells by appropriate modulation of the coupling coefficients. In this line of research, Ref.~\cite{1} realized the discrete version of this problem and accomplished the discrete solitons based operation of all-optical logic gates, switches, filters and optical diodes through suitable adjustment of a control soliton power in the waveguide arrays with an effective potential. A detailed study describing the interplay between the bound states of the potential well and incident soliton revealed the insignificance of trapping on the unidirectional flow of the solitons. Moreover, it hints to the origin of the unidirectional flow as being related to the excitation of breathing modes in the scattering region \cite{recentpre}.

Recently, Li et. al \cite{Li} investigated the atomic interactions in a two-component Bose-Einstein condensate through the propagation of vector two solitons like matter waves passing across a Gaussian barrier. This study reported the importance of the interspecies interactions which allows the wave packet to propagate like a breather. Also, elaborated the role of the interspecies interaction, relative velocity, barrier and relative phase on the collision dynamics. Furthermore, the splitting of composite solitons scattering through a narrow potential barrier were proposed for designing a two-component soliton interferometer in presence of self-attraction and cross-attraction between the components. This analysis identified the existence of substantial parameter range over which one component undergoes full transmission through the barrier, while the other one undergoes full reflection \cite{Malomed}. Additionally, the diode functionality has also been realized through diverse nonlinear discrete lattices, such as a layered nonlinear, non-mirror-symmetric model \cite{rev1a}, two parallel-coupled discrete nonlinear Schr\"{o}dinger inhomogeneous chains \cite{rev1b}, and nonlinear lattice with asymmetric defects \cite{rev1c}. Further numerical and analytical investigations reported the significance of the scattering of solitons as well as plane waves through the localized nonlinear potentials to account for the best operational regimes of the interferometer \cite{rev1d}. The one-dimensional model for a cavity equipped with saturated gain and Kerr nonlinearity has also been investigated to sustain the state of shuttle motion of the solitons \cite{rev1e}.

In the present work, we are motivated to investigate the scattering of composite solitons which are solutions of the coupled nonlinear Schr\"{o}dinger equations (NLSE) known also as Manakov system \cite{Manakov}, in the presence of external potentials. The combination of bright-bright (BB ) solitons which are exact solutions of the Manakov system are used as an initial pulse and then we observe the effect of interaction coupling on the two components during their time evolution. We consider two types of potentials for numerical simulations. One is an asymmetric Rosen-Morse (RM) double barrier potential and the other one is an asymmetric Gaussian double barrier potential. We obtain unidirectional flow in the presence of interaction and observed an exciting behavior which we coined "\textit{Polarity Reversal}" in unidirectional flow, which is characterized by polarity reversal of the diode. Another main objective of our work is to investigate unidirectional segregation of the composite solitons, i.e. composite solitons splitting into two components such that one component is fully transmitted and the other is fully reflected by an asymmetric potential when propagating from one direction and remain unaffected when transmitted from the opposite direction. We investigate the regimes of unidirectional flow and unidirectional splitting/segregation by calculating transport coefficients in terms of the parameters of the potentials. We also perform an alternative study to enrich our results of unidirectional segregation using a variational calculation \cite{Anderson1,Anderson2} and obtained an excellent agreement between analytical and numerical methods.

The rest of the paper is organized as follows. In Section \ref{model}, we present the setup and theoretical model. In Section \ref{body}, we show how the dynamics of the propagation of composite BB solitons through asymmetric barrier is affected by the presence of strong interaction coupling between the two components. In Section \ref{variational}, we  use variational calculation and numerical method to perform a comparative study of the unidirectional segregation. Finally, in Section \ref{results}, we summarize our main results.\\\\

\section{Theoretical Model}\label{model}
\begin{figure}[!htb]
	\centering
 \textbf{~~~~~~Gaussian and RM asymmetric double potential Barriers}
	\includegraphics[width=0.9\linewidth]{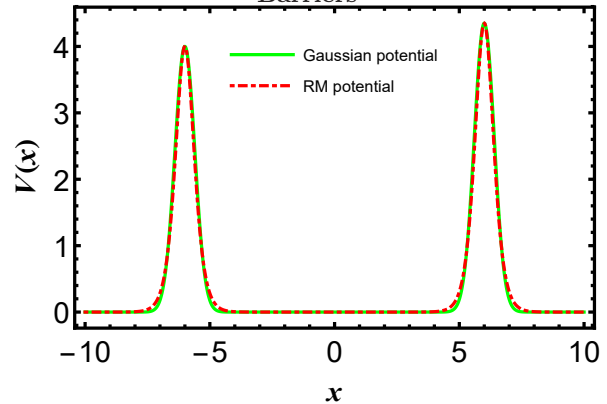}
		\caption{The Gaussian and Rosen-Morse asymmetric double potential barriers plotted for the Eqs. (\ref{Gaussian} $\&$ \ref{rmpotential}), indicated by  solid green and red dot-dashed curves, respectively. The potential barriers are plotted for the parameters: $V_1$ = 4, $V_2$ = 4.35, $\sigma$ = 0.4, $\alpha_1$ = $\alpha_2$ = 2, $x_1$ = -6 and $x_2$ = 6.}
	\label{RM-Gaussianbarriers}
\end{figure}

In the presence of an external scalar potential $V(x)$ the dynamics of bright-bright solitons is governed by the Manakov system of equations,
\begin{eqnarray}
\label{Manakovsystem}
i \psi_{1t} &=& -\dfrac{1}{2} \psi_{1xx} + s [g_1 |\psi_1|^2 + g_{12}|\psi_2|^2] \psi_1 + V(x) \psi_1,\nonumber \\
i \psi_{2t} &=& -\dfrac{1}{2} \psi_{2xx} + s [g_2 |\psi_2|^2 + g_{12}|\psi_1|^2] \psi_2 + V(x) \psi_2. \nonumber \\
\end{eqnarray}
where $\psi_j$, with $j =1, 2$ denotes the wavefunction of the individual components. Repulsive or attractive interactions are accounted for by $s = +1\,\,\text{or}\,\,-1$, respectively. The nonlinear local  interaction strength of the components $\psi_1$ and $\psi_2$ are represented by $g_1$ and $g_2$, respectively. The strength of the interaction coupling the two components is $g_{12}$ and $V(x)$ is the external scalar potential. Since the main focus will be on the effect of the nonlinear interaction $g_{12}$, we set $s=-1, g_1=g_2=1$ and restrict $g_{12}$ to [-1,1]. In addition, we consider also the special case where we set $g_1 \ne g_2$ for $g_{12}$ = 0 and  $g_{12} = -0.5$. With these restrictions on the nonlinearities, the Manakov system describes the composite bright-bright (BB) soliton. For our study, we choose two different types of the potential, $V(x)$, which have similar profiles. One is the asymmetric double Gaussian barrier of the form

\begin{eqnarray}\label{Gaussian}
V(x) = V_1 \exp \big[- \dfrac{(x-x_{1})^2}{2 \sigma^2}\big]\nonumber\\ + V_2\exp \big[- \dfrac{(x-x_{2})^2}{2 \sigma^2}\big],
\end{eqnarray}

\noindent where $V_{1,2}$, $\sigma$, and $x_{1,2}$ determine the height, width, and position of the center of the first and second potential barrier, respectively. The second type of the potential $V(x)$ is of the form

\begin{eqnarray}
\label{rmpotential}
V(x)=V_1\,\sech^2[\alpha_1(x-x_{1})]\nonumber\\
+V_2\,\sech^2[\alpha_2(x-x_{2})],
\end{eqnarray}
\noindent which is a combination of two RM potentials. Here, $V_{1,2}$, $\alpha_{1,2}$, and $x_{1,2}$ determine the height, inverse width, and position of the center of the first and second potential barrier, respectively. Both potentials are asymmetric double barriers and we chose slightly different heights for both potentials to achieve unidirectional behaviour. The profiles of the Gaussian and RM asymmetric double potentials indicated by the solid green and red dot-dashed curves, respectively are displayed in Fig. \ref{RM-Gaussianbarriers}. In our numerical simulation we prepared an initial state far away from the potential region and then let it propagate in real time. We always choose our initial wave functions as the exact solutions of the homogeneous version of Eq.~\eqref{Manakovsystem}, namely
\begin{eqnarray}
\psi_1(x, 0) &=&A~ e^ {i v_1 x} \sech[A(x-x_0)],\nonumber\\
\psi_2(x, 0) &=&A~ e^ {i v_2 x} \sech[A(x-(x_0 + \delta))].
\end{eqnarray}
\noindent where $v_{1,2}$, $x_0$ are the initial center-of-mass velocity and position, respectively, and $\delta$ is a very small shift in the initial position of the component 2, $A$ is an arbitrary real normalization constant that we set to unity. For our investigation, we fixed the separation between the BB soliton components, $\delta$ = 0.001. Now, we present the numerical results of our calculation for the transport coefficients: For the left-to-right-moving soliton with a single potential barrier, we define the reflectance, transmittance and trapping coefficients as follows:
\begin{align}
\label{eq:secII_R,T,L}
\nonumber
R_{i} &= \frac{1}{N} \int_{-\infty}^{-l_{i}} dx |\psi_{i} \left(x,t\right)|^2,  \\ \nonumber
T_{i} &= \frac{1}{N} \int_{l_{i}}^{\infty} dx |\psi_{i} \left(x,t\right)|^2,    \\
L_{i} &= \frac{1}{N} \int_{-l_{i}}^{l_{i}} dx |\psi_{i} \left(x,t\right)|^2,
\end{align}

respectively, where $l_i \approxeq 5/\alpha_i$, $i = 1,2$, from the centre of the barrier and $N$ is the normalization of the total soliton intensity given by $N = \int_{-\infty}^{\infty}(|\psi_1|^2 + |\psi_2|^2)  dx$. For right-to-left-moving soliton, $R$ and $T$ are interchanged but $L$ remains the same. Here, $l_i$ represents the position of measurement of reflectance or transmission, set at a value slightly greater than the position of the boundary of the barrier, which we represented in terms of the inverse width of the barrier ($\alpha_i$). For the considered two potential barriers in series, we choose $-l_i$ to the left of the left barrier and $l_i$ to the right of the right barrier.

\section{Propagation of bright-bright solitons through asymmetric RM potential barriers}\label{body}

\begin{figure}[bt]\centering
\includegraphics[width=8.5cm,clip]{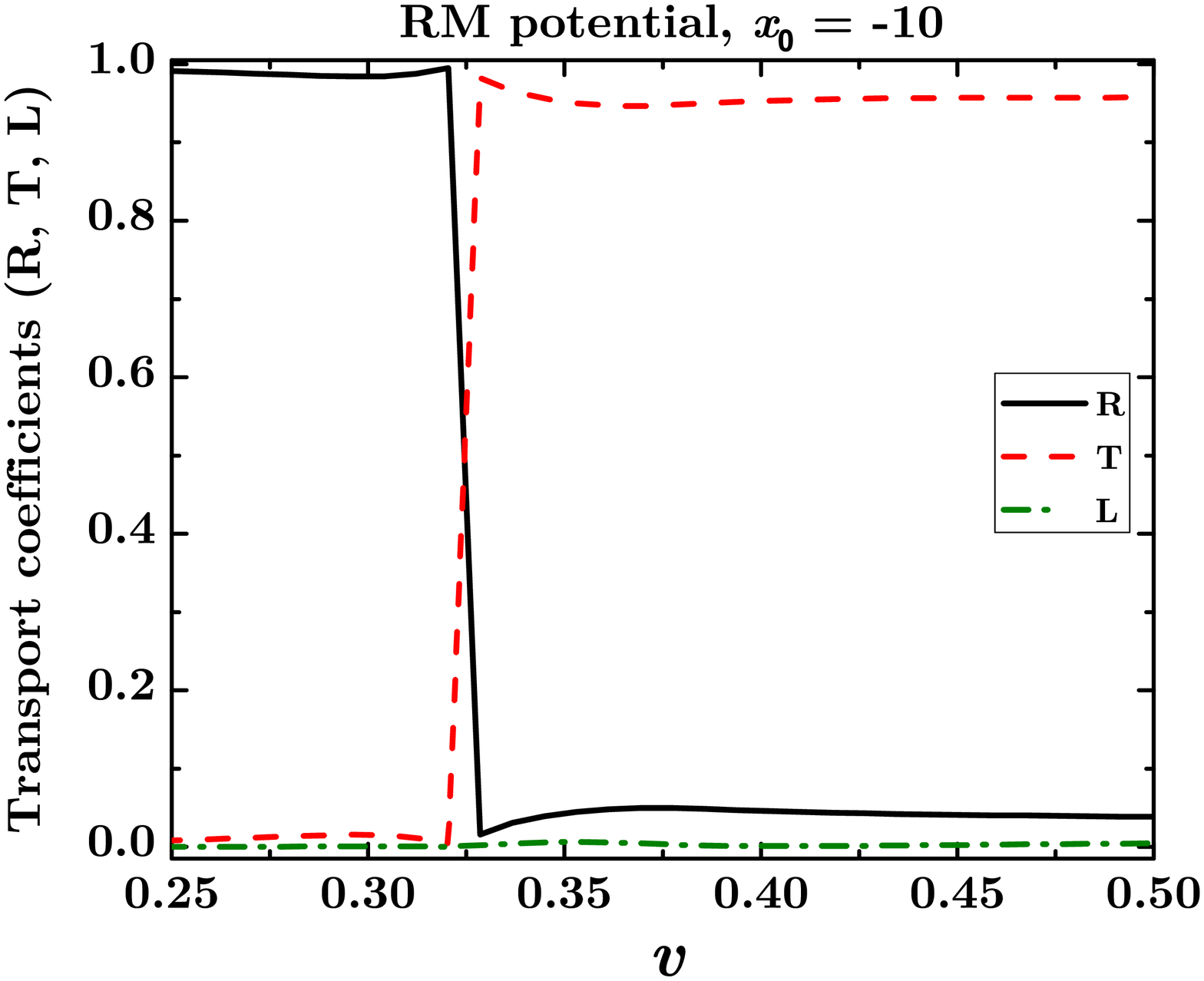}\\
\includegraphics[width=8.5cm,clip]{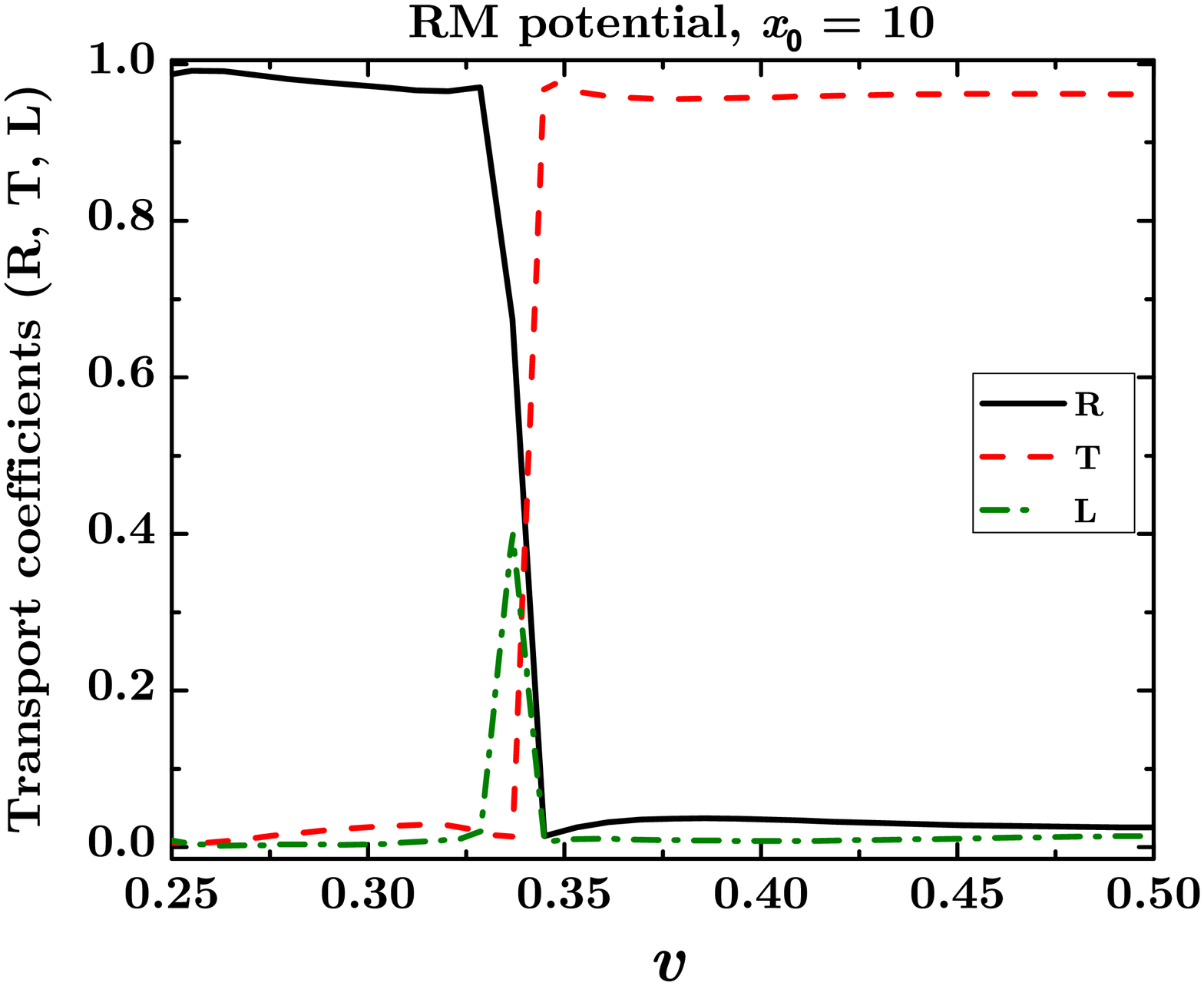}
 \caption{Transport coefficients in terms of velocity for the propagation of the $\psi_1$ component through asymmetric RM potential barriers for $g_1 = g_2 = 1$, and $g_{12} = 0$ from $x_0$ = -10 (upper panel) and $x_0$ = 10 (lower panel). Solid black line represents reflectance, $R$, dashed red line represents transmission, $T$, and dash-dotted green line represents trapping, $L$.}
  \label{fig2}
\end{figure}

The dynamics of the bright-bright solitons described by the two-component Manakov system scattering through an asymmetric RM potential barriers positioned in series with slight difference in the height of the two barriers and finite separation will be presented in this section. As both RM and Gaussian asymmetric potential barriers portray similar behavior throughout the analysis, we explain the entire study using RM potential in this section. In order to avoid redundancy, the results through an asymmetric Gaussian double potential barrier are given in Appendix~\ref{appendix1}. For our numerical investigation, we consider $V_1 = 4$, $V_2 =4.35$, $\sigma=0.4$, $\alpha_1 = \alpha_2 = 2$ and $x_{1} = -6, x_{2} = 6$. The evolution of BB solitons are investigated under the following scenarios: (i) in the absence of mean field coupling, $g_{12} = 0$, (ii) in the presence of an attractive coupling, $g_{12} > 0$, and (iii) in the presence of a repulsive coupling, $g_{12} < 0$.

\subsection{Unidirectional flow for uncoupled components with  $g_{12} = 0$}
The propagation of the BB solitons through an asymmetric RM potential barriers with $g_{12} = 0$ is displayed in Fig.~\ref{fig2} for initial propagation from locations along the axis $x_0 = -10$ and $x_0 = 10$. Fig.~\ref{fig2} displays the transport coefficients namely, reflectance, $R$, transmittance, $T$ and trapping, $L$, in terms of velocity for the left and right moving BB soliton components through an asymmetric RM double barrier potential indicated by solid black, dashed red and dotted green lines, respectively, for the component $\psi_1$. The corresponding curves for $\psi_2$ are identical with those of $\psi_1$. This is obvious because $g_{12}=0$. The transport coefficients obtained reveal the existence of a particular initial velocity value above which there is a sudden drop in reflectance to its minimum value and a sudden rise in the transmittance to its maximum value for both components. Furthermore, this critical velocity required for maximum transmittance appears to be different for right moving and left moving BB solitons. For the right moving BB solitons scattered through RM potential barriers, this critical velocity is found to be $v_c$ = 0.324 whereas that for the left moving BB solitons $v_c$ = 0.339. Such differences in critical velocities for soliton propagation in opposite directions enables the realization of soliton diode using asymmetric RM potential wells \cite{usamaasad}.  While comparing the case of asymmetric potential wells of Ref.~\cite{usamaasad} with the present study, it is observed that the same functionality can be realized through the RM potential barriers. Moreover, the present scheme with RM potential barriers displays larger velocity window of 0.324 $\leq v \leq$ 0.338 for the diode functionality compared to the case of Ref.~\cite{usamaasad} with a smaller velocity window 0.3275$\leq v \leq$ 0.3375.
\begin{figure}[!htb]
	\centering
	 \textbf{RM potential}\par\medskip
\includegraphics[scale=0.45]{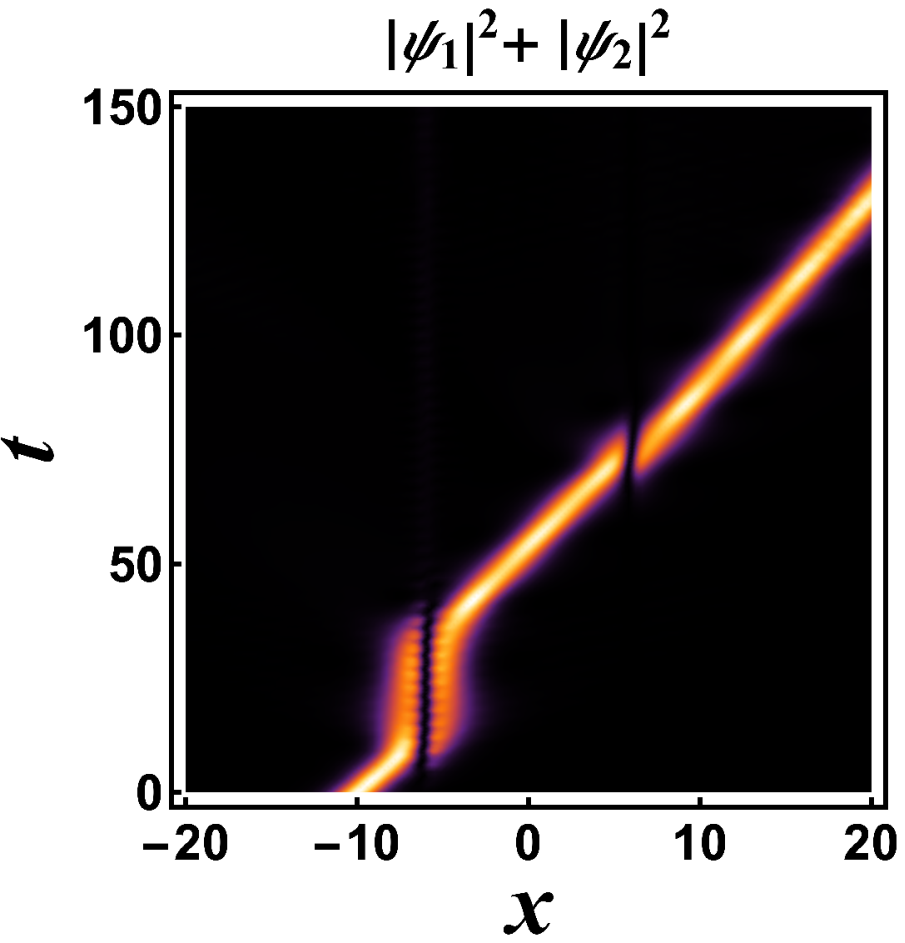}
\includegraphics[scale=0.45]{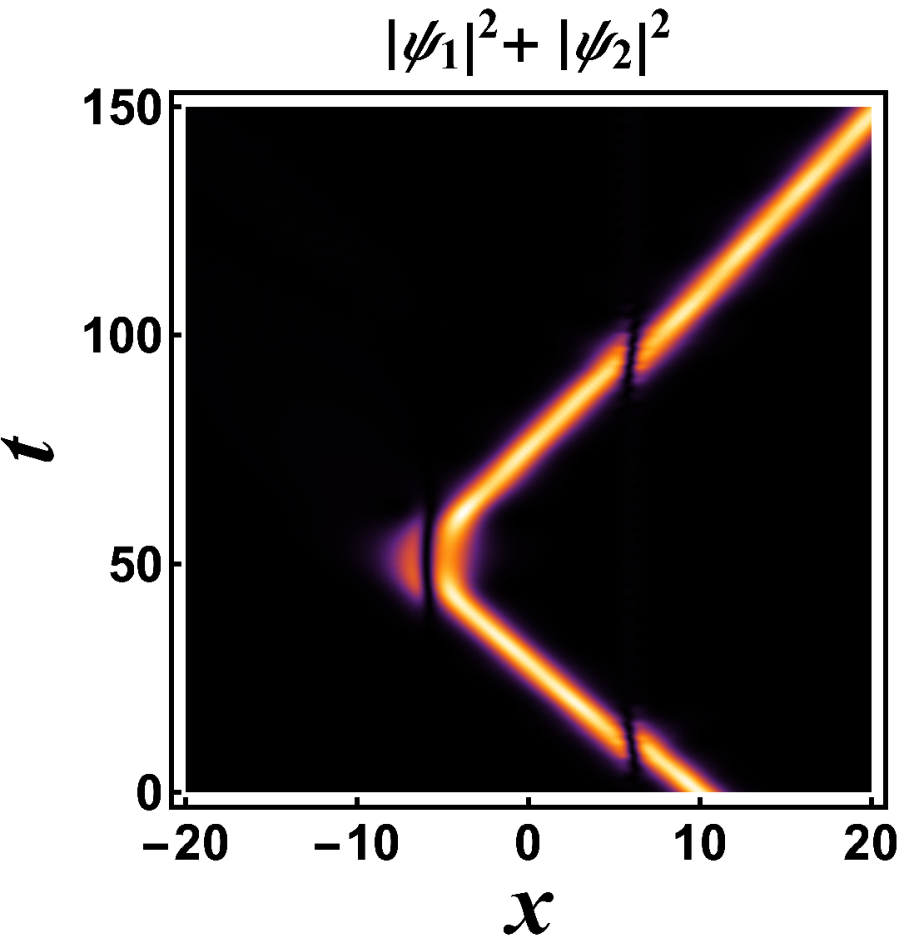}
	\caption{Propagation of composite BB soliton through asymmetric RM potential barriers for $g_1 = g_2 = 1$, and $g_{12} = 0$ at $v$ = 0.324. Both components are identical. Left and right subfigures are results of initial propagation from $x_0$ = -10 and $x_0$ = 10, respectively.}
	\label{fig3}
\end{figure}

\noindent The trapping of solitons within a very narrow velocity window about the critical velocity, as shown by the green dotted curves in Figs.~\ref{fig2} correspond to an unstable trapped state at the center of the potential barrier. The peak at the transition region corresponds to a nonlinear trapped mode by a stationary bound state of the potential. Physically, it corresponds to an unstable equilibrium that separates full transmission from full (quantum) reflection. The transition is so sharp such that the peak is very narrow. Further, trapping of solitons is absent for the right moving solitons from $x$ = -10, but a small velocity range over which soliton trapping is found for the left moving solitons from $x$ = 0. This trapping of left moving soliton appears as a peak near to the critical velocity as indicated by the lower panel of the Fig.~\ref{fig2}.\\
Fig.~\ref{fig3} portray the spatiotemporal evolution of the composite BB solitons for a particular initial velocity. In Fig.~\ref{fig3}, the left subfigure describes the propagation of the BB soliton components $\psi_1$ and $\psi_2$ with a critical velocity $v_c$ = 0.324, incident from $x_0 = -10$ through asymmetric RM double potential barriers. The BB soliton components are first transmitted through the left barrier ($V_1 = 4.0$) positioned at $x_0 = -6$ and then transmitted through the right barrier ($V_2 = 4.35$) positioned at $x_0 = 6$ with an overall transmittance $T\approx0.97$. Considering the right subfigure of Fig.~\ref{fig3} for the left moving BB solitons from $x_0$ = 10, BB solitons are first transmitted through the right barrier ($V_2$) positioned at $x_0 = 6$ and then reflected to the right by the left barrier ($V_1$) positioned at $x_0 = -6$ and finally transmitted through the right barrier to the left barrier with an overall reflectance, $R\approx0.97$. This asymmetrical behavior in the flow of solitons is due to the appreciable velocity reduction of the BB solitons while crossing the first barrier, which plays a decisive role on overall transmission or reflection. For the case of right propagating BB solitons, it suffers a small velocity reduction when it transmits through the first barrier ($V_1$). But this reduced velocity is still sufficient to transmit through the second barrier ($V_2$) and propagates through to the right. On the other hand, the left moving BB solitons suffers an appreciable velocity reduction when it transmits through its first large barrier ($V_2$). This velocity reduction is high enough such that the soliton velocity is less than the critical velocity to overcome the second barrier which makes it impossible to transmit through the second barrier ($V_1$) and results in its reflection towards the right. This unidirectional flow portrays the diode behavior of the composite BB solitons through RM potential barriers similar to the one which we have realized in our previous analysis for a single soliton propagation through an asymmetrical RM potential wells \cite{usamaasad}. Furthermore, we observed a maximum transmission at few lower velocities for $g_{12}$ = 0 for right moving soliton but no soliton trapping as indicated by the appearance of green dot-dashed spike in the lower panel of the Fig.~\ref{fig2}. On the other hand for the left moving soliton there exists trapping as well as maximum transmission for certain lower velocities with  $g_{12}$ $\simeq$ 1.

\subsection{Polarity reversal in unidirectional flow with $g_{12} > 0$}

\begin{figure}[bt]\centering
\textbf{RM potential, $x_0$ = -10}\par\medskip
\includegraphics[scale=0.55]{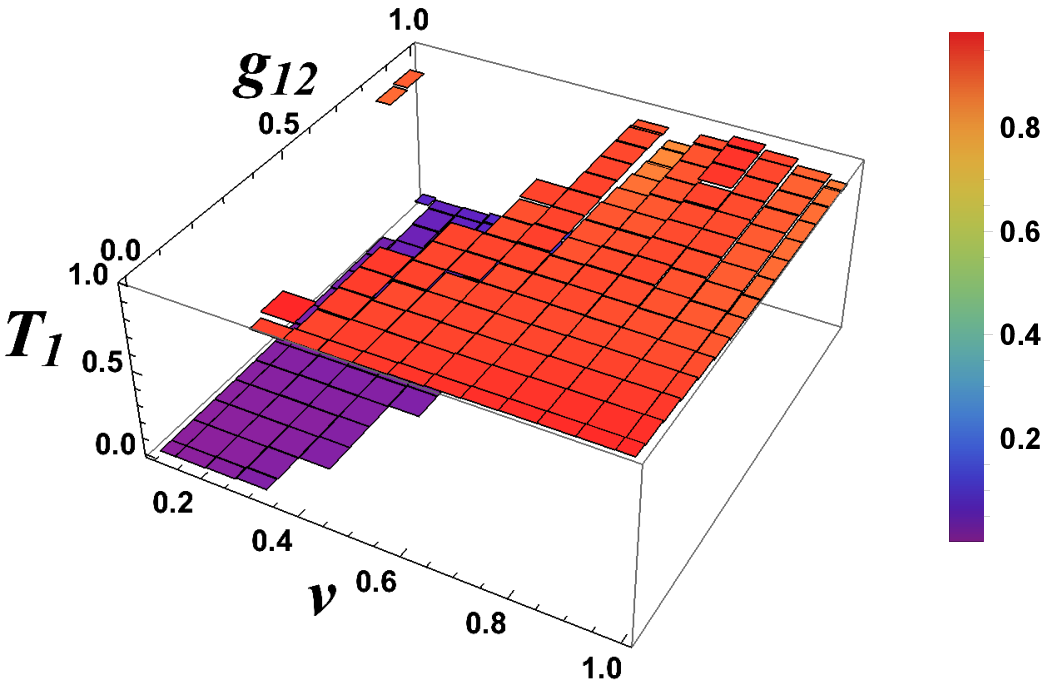}\\
\includegraphics[scale=0.55]{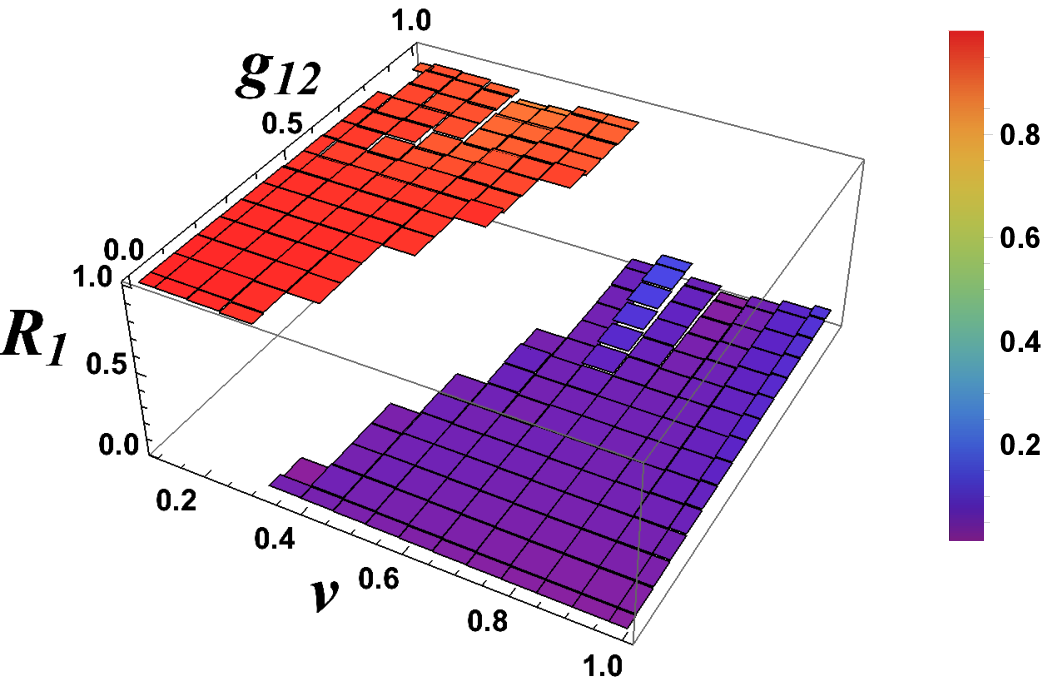}\\
\includegraphics[scale=0.55]{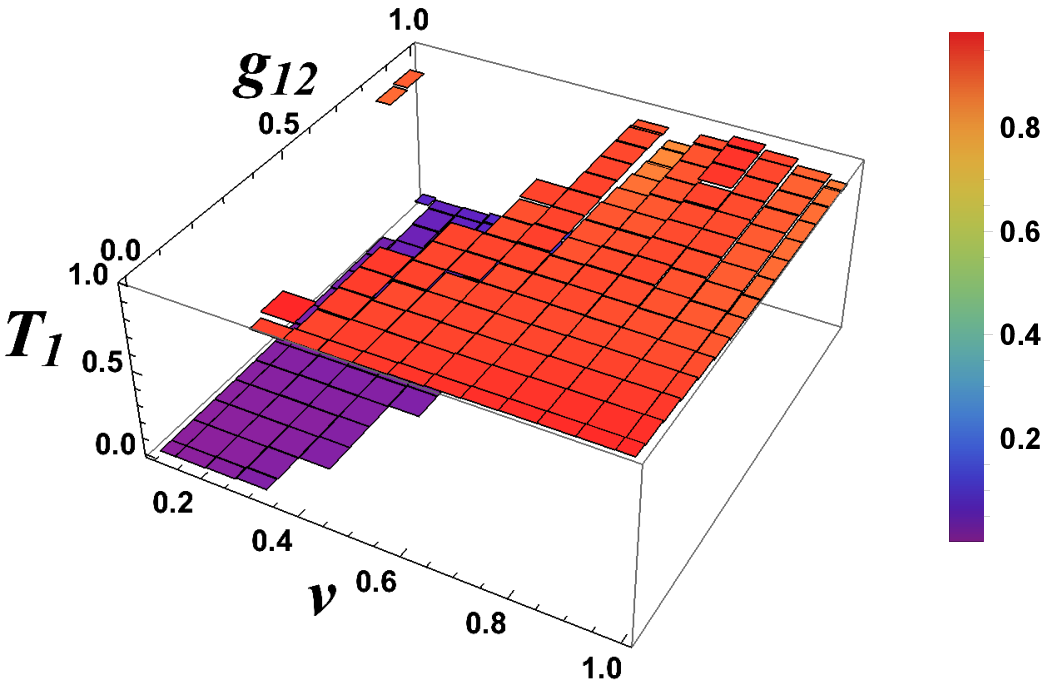}\\
\includegraphics[scale=0.55]{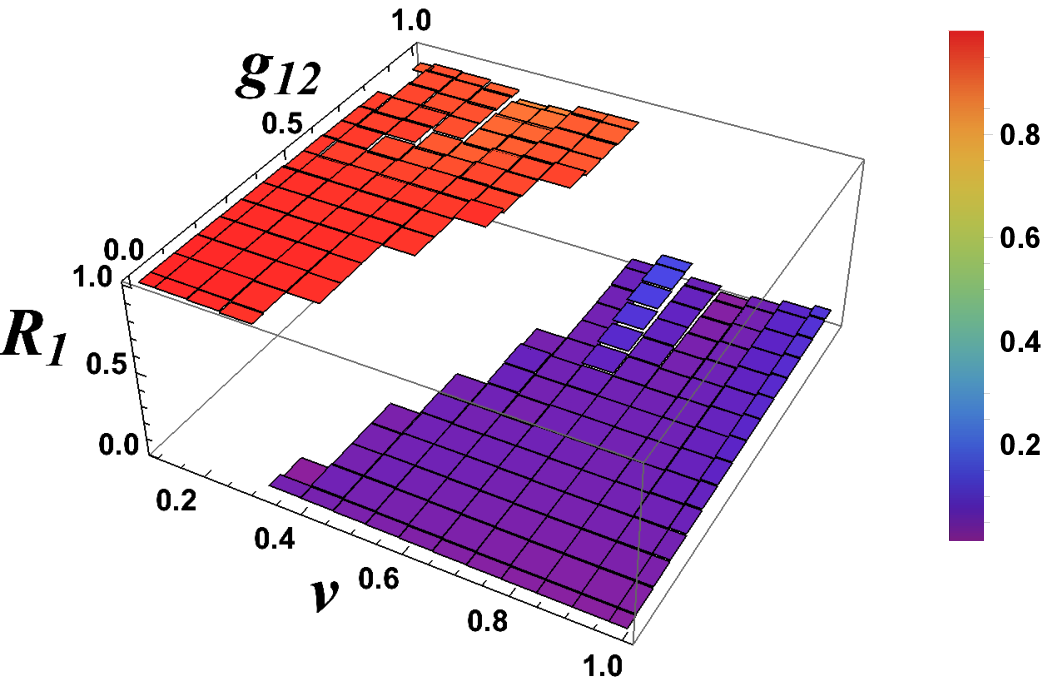}
 \caption{Transmission and reflection coefficients of the component $\psi_1$  for $g_1 = g_2 = 1$, propagating from $x_0$ = -10 (upper two) and $x_0$ = 10 (lower two) through RM potential barriers versus $v$ and $g_{12}$.}
  \label{fig5}
\end{figure}

\begin{figure}[!htb]
	\centering
	 \textbf{RM potential}\par\medskip
\includegraphics[scale=0.45]{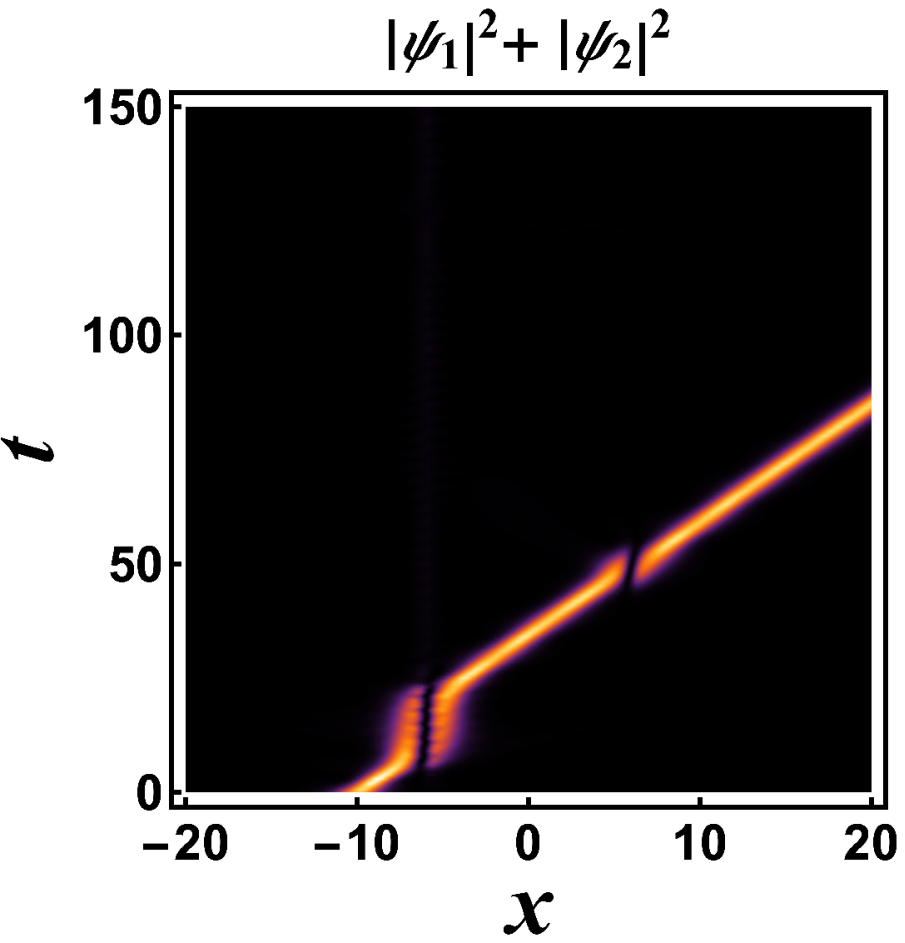}
\includegraphics[scale=0.45]{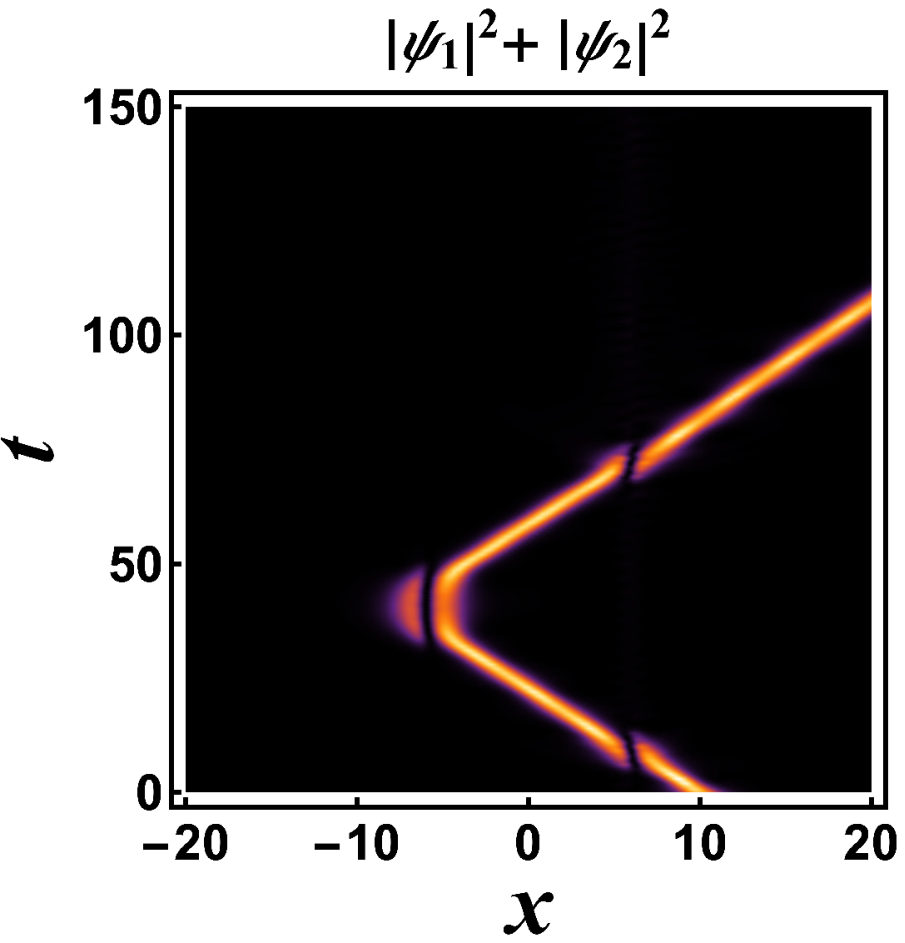}
 \caption{Propagation of composite BB soliton through asymmetric RM potential barriers for $g_1 = g_2 = 1$, and $g_{12} = 0.3$ at $v$ = 0.418. Both components are identical. Left and right subfigures are results of initial propagation from $x_0$ = -10 and $x_0$ = 10, respectively.}
  \label{fig6}
\end{figure}

The role of the attractive mean field coupling ($g_{12} > 0$) on the propagation of BB solitons through the asymmetrical RM potential barriers will be addressed in this section. Since, both components $\psi_1$ and $\psi_2$ exhibit an identical behavior for both directions of propagation, we have described our results with the component $\psi_1$, to avoid redundancy. Fig.~\ref{fig5} describes the transmission and reflection coefficients of the component $\psi_1$ passing through the RM potential barriers from $x_0$= $\pm$10, versus the velocity and positive mean field coupling. In our numerical investigation, we considered varying the incident velocity of the BB solitons from 0.1 to 1 and $g_{12}$ was varied from 0 to 1. The results obtained illustrate that reflection dominates over the transmission for the composite solitons with incident velocities $v$ $\leq$ 0.36 for RM barriers throughout the entire range of $g_{12}$ values. For this range of the incident velocities, reflection is observed to be around $\approx0.98$. A sharp transition from the maximum reflection (R $\approx$ 0.98) to the maximum transmission (T $\approx$ 0.94) is obtained for the BB solitons with an initial incident velocity of $v$ = 0.363 for $g_{12}$ = 0.1. This critical velocity is found to be larger than that for the case with $g_{12}$ = 0, where a minimum critical velocity of 0.324 is required to reach maximum transmission through RM potential barriers. This indicates that the presence of $g_{12}$ coupling introduces a shift in the critical velocity required to reach maximum transmission. For a further increase in $g_{12}$ value, a further shift in the velocity, $v$, is required for maximum transmission to hold.

\begin{figure}[!h]
	\centering
	\textbf{RM potential, $v$ = 0.424}\par\medskip
	\includegraphics[scale=0.45]{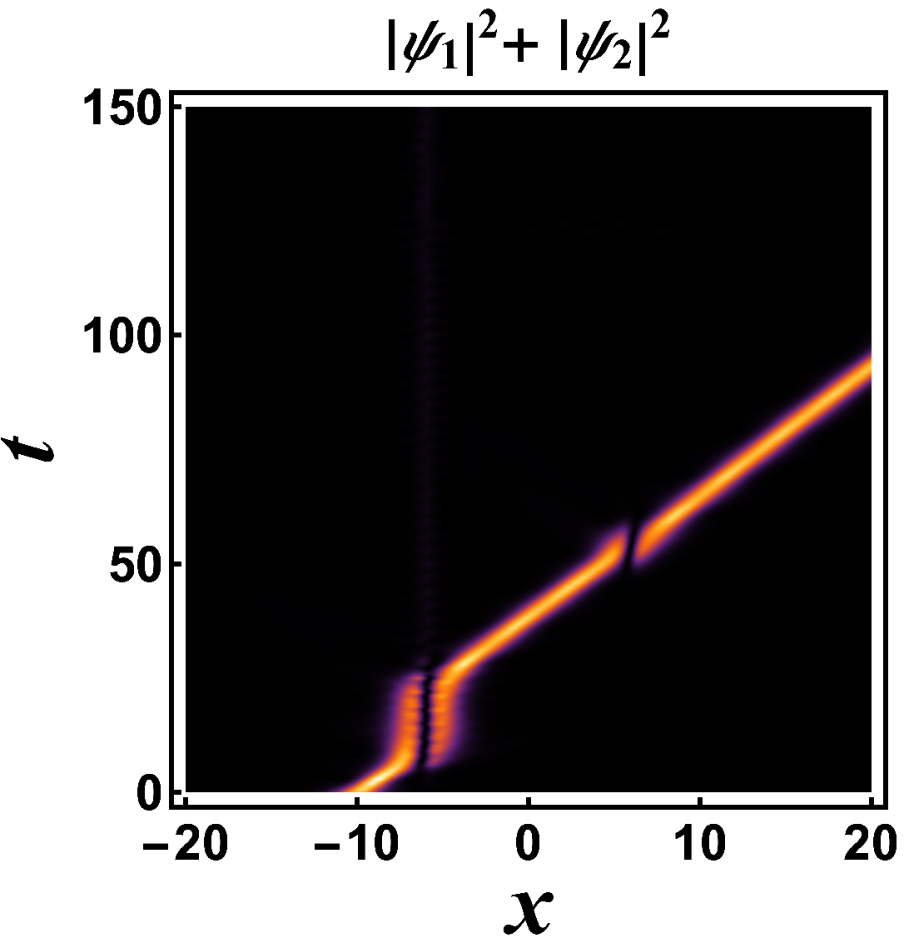}
	\includegraphics[scale=0.45]{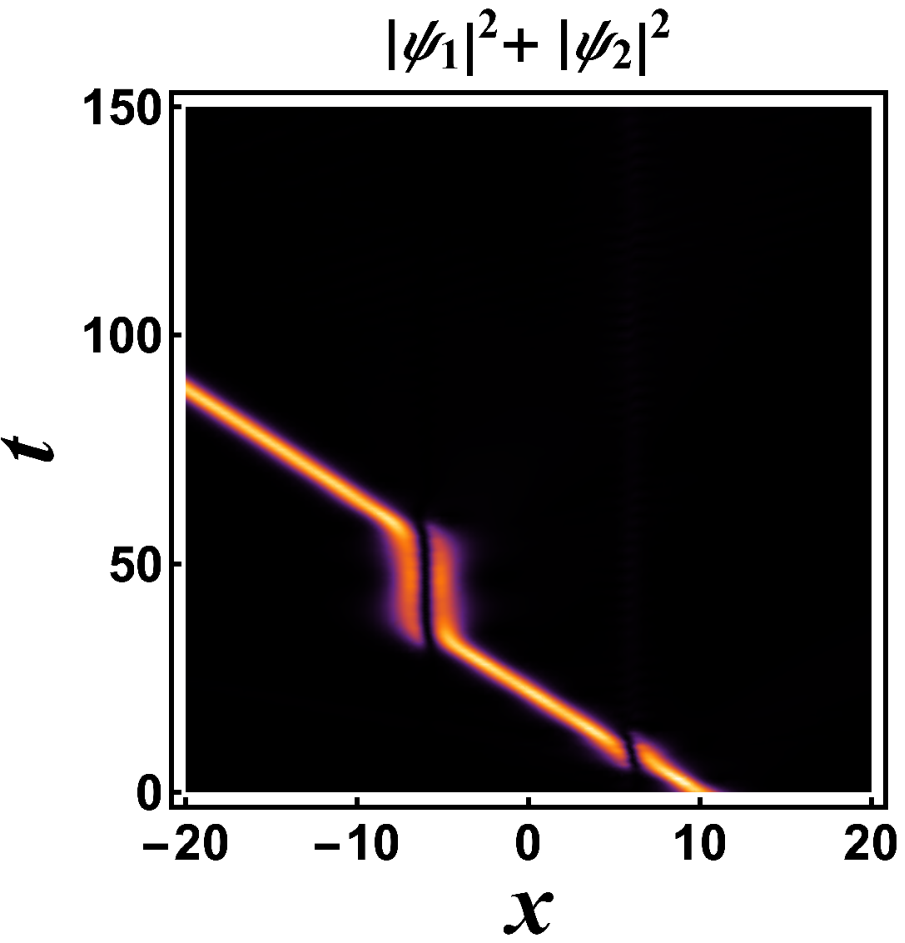}\\
    \includegraphics[scale=0.45]{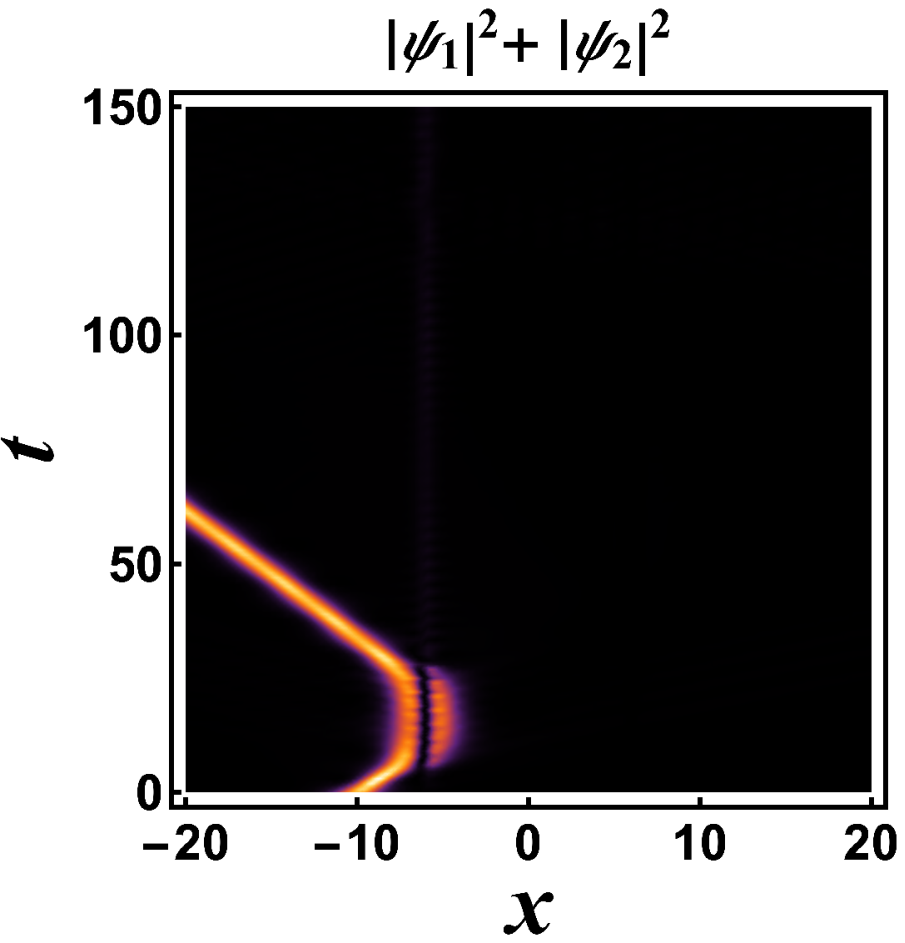}
	\includegraphics[scale=0.45]{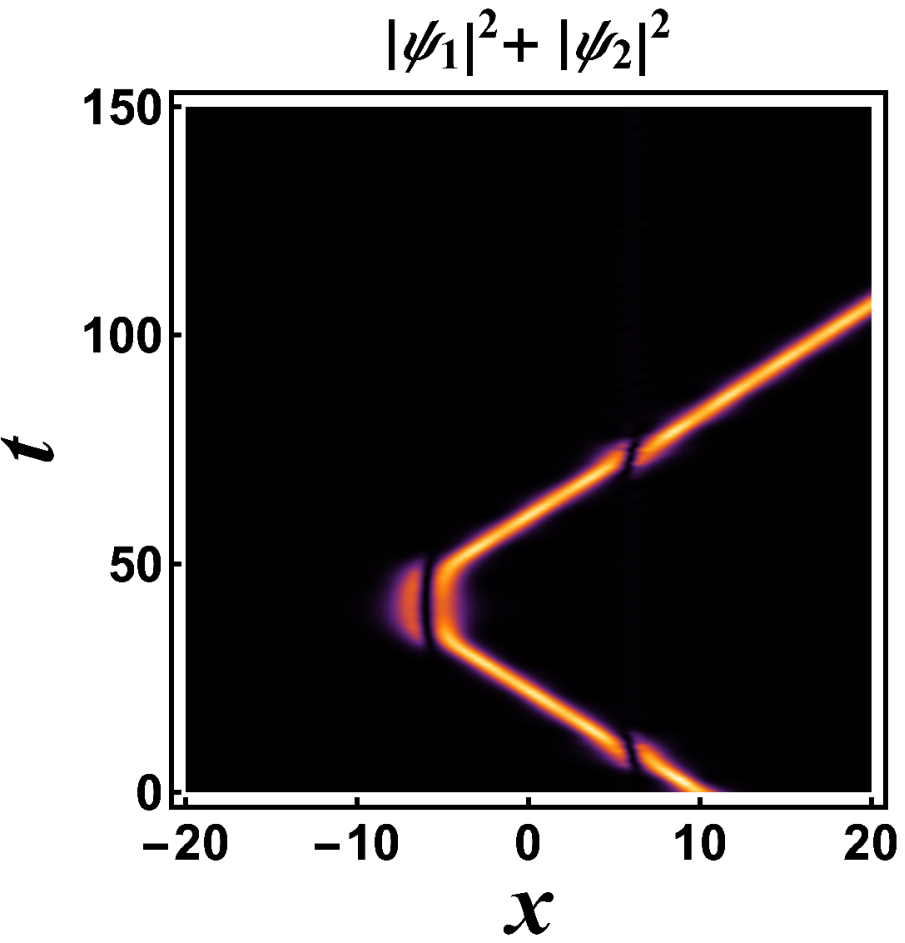}
	\caption{Propagation of composite BB soliton through asymmetric RM potential barriers for $g_1 = g_2 = 1$, and $g_{12} = 0.325$. Upper panel shows full transmission at $v$ = 0.424 while lower panel shows full reflection at $v$ = 0.423, from both left and right directions. There is no unidirectional flow in the range of coupling strength 0.324 $\le$ $g_{12}$ $\le$ 0.329.}
	\label{fig7}
\end{figure}

Next, the propagation dynamics of the BB solitons from $x_0$ = 10 is considered which is shown in the lower two panels of Fig.~\ref{fig5}. Like the previous case, here also reflection is found to dominate for BB solitons with incident velocities $v$ $\leq$ 0.39 (for RM barriers). Thereafter, it shows a sudden sharp transition from higher reflection (R $\approx$ 0.95) to the higher transmission (T $\approx$ 0.95) for incident velocities $v$ = 0.391 for RM potential barriers. Here also, a shift in critical incident velocity is observed with the shift in $g_{12}$ which allows the diode functionality similarly to the case of $g_{12} = 0$.\\
The important characteristic noticed here is a reverse shift in critical velocity (reduction in critical velocity) for $g_{12}$ $\geq$ 0.35, for the left moving solitons. The velocity window for the diode functionality at different $g_{12}$ values are tabulated in Table \ref{RMtable}. For this direction of propagation of BB solitons, the trapping is also found to be negligible.

\begin{figure}[!htb]
	\centering
	 \textbf{RM potential}\par\medskip
\includegraphics[scale=0.45]{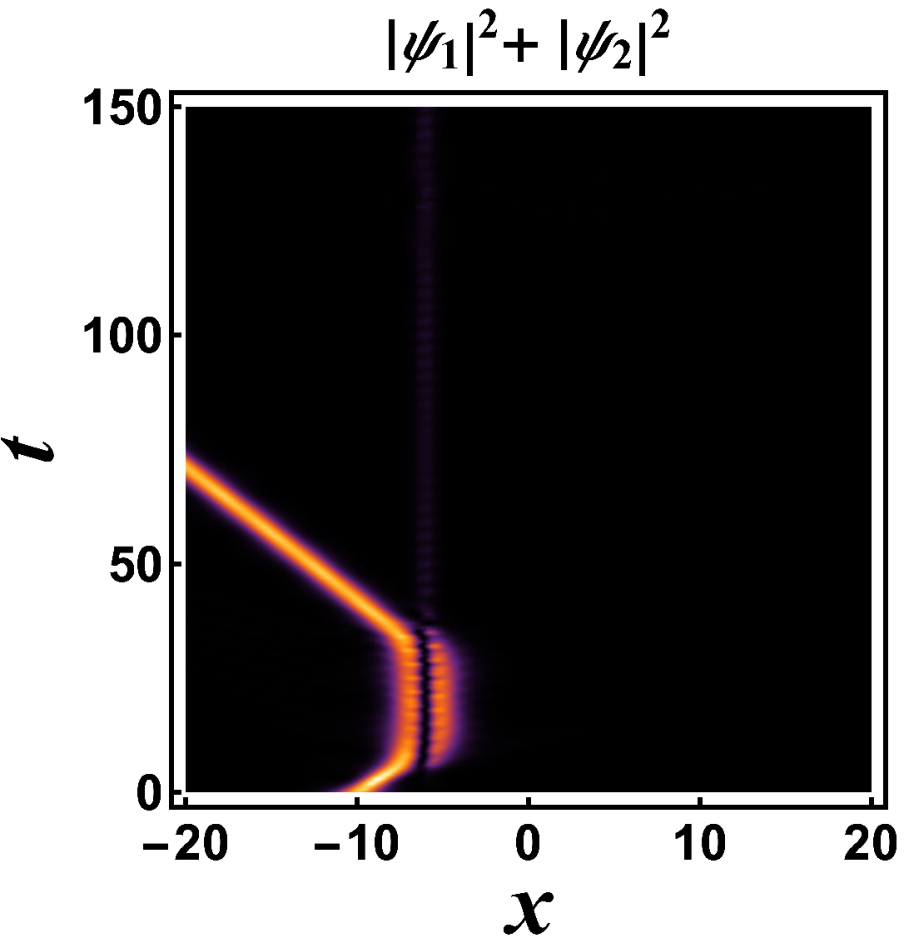}
\includegraphics[scale=0.45]{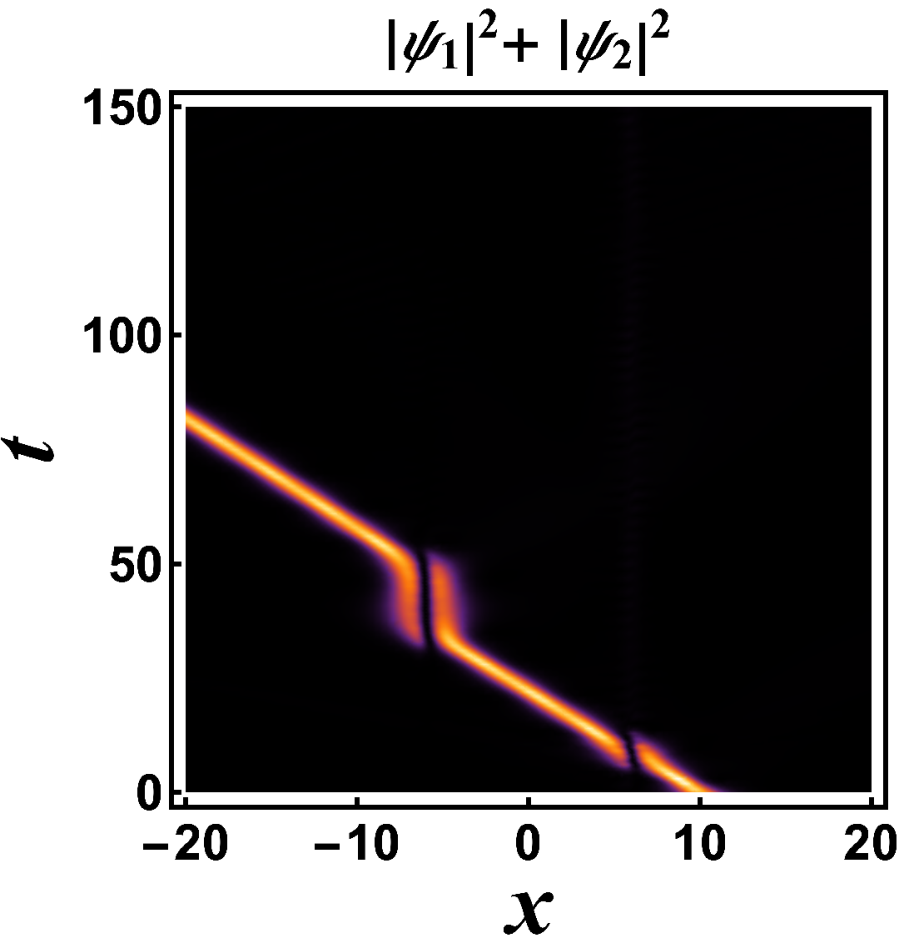}
 \caption{Propagation of composite BB soliton through asymmetric RM potential barriers for $g_1 = g_2 = 1$, and $g_{12} = 0.33$ at $v$ = 0.425. Both components are identical. Left and right subfigures are results of initial propagation from $x_0$ = -10 and $x_0$ = 10, respectively. The polarity reversal phenomenon in unidirectional flow is achieved by comparing with Fig.~\ref{fig6}.}
  \label{fig8}
\end{figure}

\begin{figure}[!h]
	\centering
 \textbf{~~~~~~Polarity Reversal through RM potential}
	\includegraphics[width=1\linewidth]{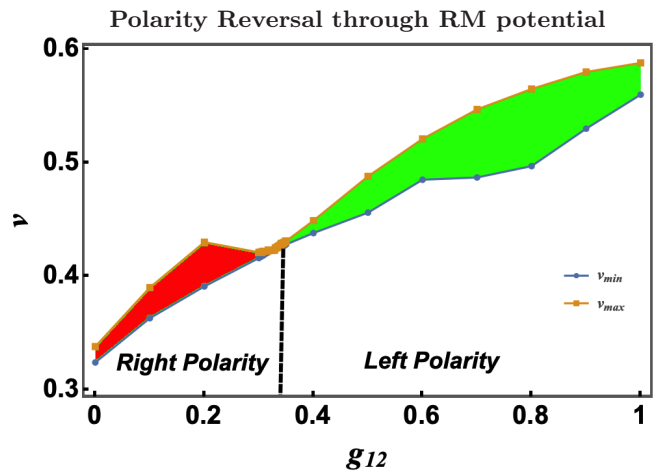}
		\caption{Borders of velocity window for the unidirectional flow ($v_{min}, v_{max}$) vs $g_{12}$ with $g_1 = g_2 = 1$, through the RM potential barriers. Full reflection for $v$ $\leq$ 0.423 and full transmission for $v$ $\geq$ 0.424 is obtained with the range of coupling strength 0.324 $\le$ $g_{12}$ $\le$ 0.329, hence no unidirectional flow is observed at this specific range of $g_{12}$. Away from this point of $g_{12} = 0.329$, we find polarity reversal in unidirectional flow. The shaded region shows the velocity window for the unidirectional flow. The red color shows the right polarity while the green color shows the left polarity of the unidirectional flow. The data used to generate this figure is listed in Table~\ref{RMtable}.}
	\label{fig:rm-reversal}
\end{figure}

Furthermore, we have explored an exciting phenomenon, which we referred to as "\textit{polarity reversal for an optical diode}" and is observed for both kinds of potential barriers considered. In our investigation, we observed that for the right-moving soliton there is a critical velocity over which there is a sudden jump from mostly reflectance to mostly transmittance. A similar behavior exists for the right-moving soliton for another critical velocity. Moreover, it is observed that there exists different critical velocity for the left- and right-moving solitons for full transmission. As a result, there exists an appreciable velocity window or velocity range for which there is almost full transmittance in one direction and nearly zero transmittance in the other direction, i.e., the soliton shows directional propagation for this set of parameters. This behavior (unidirectional flow of solitons) is similar to the diode effect in semiconductor physics. This behavior is observed in forward direction for up to $g_{12} =$ 0.323 and 0.312 for RM asymmetric potential barriers and Gaussian asymmetric potential barriers, respectively (i.e. The right moving solitons, undergoes full transmission and the left moving solitons undergoes full reflection over the obtained velocity window). On the other hand, for $g_{12}$ values above 0.33 and 0.317 for RM asymmetric potential barriers and Gaussian asymmetric potential barriers, respectively, we observed the exact reversal in behavior of diode effect (i.e. The left moving solitons, undergoes full transmission and the right moving solitons undergoes full reflection over the obtained velocity window irrespective of the barrier height. This variation in propagation behavior of right and left moving soliton due the effect of interaction coupling, we refer to as ``polarity reversal of the optical diode or polarity reversal in unidirectional flow''). The "\textit{right polarity}" corresponds to the unidirectional flow towards the right direction in which right moving composite BB solitons transmit through both potential barriers and left moving composite solitons reflect through the potential barrier $V_1$ while "\textit{left polarity}" corresponds to the unidirectional flow towards left direction in which right moving composite BB solitons reflect through the potential barrier $V_1$ and left moving composite solitons transmit through both potential barriers. From the spatiotemporal plots, it is observed that for lower coupling $g_{12}\leq0.323$ (for RM potential), both components exhibit the diode behavior with "\textit{right polarity}" as shown in Fig.~\ref{fig6} obtained for $g_{12}$ = 0.3 which is similar to the one achieved in the case $g_{12}=0$. For $g_{12}$ = 0.324 - 0.329, it is observed that a full transmission exists for BB solitons propagating from both directions with maximum transmission for the critical incident velocities $v$ $\geq$ 0.424 through RM potential barriers. For $v$ < 0.424, it exhibits maximum reflection for both right and left moving composite BB solitons as shown in Fig.~\ref{fig7}. Further, for $g_{12}\geq0.33$ (for RM potential), exactly the reverse phenomena is achieved for $g_{12}\leq0.3$, i.e., the right moving BB solitons passing through a smaller barrier ($V_1$) towards the larger barrier ($V_2$) is getting reflected while that for left moving BB solitons passing through larger barrier ($V_2$) towards the smaller barrier ($V_1$) is getting transmitted. Hence, the polarity of unidirectional flow is reversed from right to left polarity which can be seen by comparing Fig.~\ref{fig8} with Fig.~\ref{fig6} for RM potential. This phenomena is purely due to the increase in $g_{12}$ above certain critical value 0.329 (for RM potential), which is demonstrated by Fig.~\ref{fig:rm-reversal}. Moreover, for an attractive interaction ($g_{12}>0$), the results do not display any segregation or splitting of the BB soliton components and both components remain intact throughout the propagation.

\begin{table*}[!htb]
	\textbf{RM Potential}
	\begin{center}
		\begin{tabular}{ |c|c|c| }
			\hline
			\multicolumn{3}{|c|}{\textbf{Unidirectional flow with composite BB solitons}} \\
			\hline
			\textbf{Interaction strength} & \multicolumn{2}{|c|}{\textbf{Velocity window }} \\
			\textbf{$g_{12}$} & \multicolumn{2}{|c|}{\textbf{ $ v_{min} \le v \le v_{max}$}}  \\
			\hline\hline
			0 & 0.324 $\le$ $v$ $\le$ 0.338& \cellcolor{red}\\\cline{1-2}
			0.1  & 0.363 $\le$ $v$ $\le$ 0.39&\cellcolor{red}\\\cline{1-2}
			0.2 &  0.391 $\le$ $v$ $\le$ 0.430& \cellcolor{red}\\\cline{1-2}
			0.3  & 0.416 $\le$ $v$ $\le$ 0.421&\cellcolor{red}\\\cline{1-2}
			0.301-0.303 & 0.417 $\le$ $v$ $\le$ 0.421&\cellcolor{red}\\\cline{1-2}
			0.304-0.306 & 0.418 $\le$ $v$ $\le$ 0.421&\cellcolor{red}\\\cline{1-2}
			0.307 & 0.418 $\le$ $v$ $\le$ 0.422&\cellcolor{red}\\\cline{1-2}
			0.308-0.310 & 0.419 $\le$ $v$ $\le$ 0.422&\cellcolor{red}\\\cline{1-2}
			0.311-0.313 & 0.420$\le$ $v$ $\le$ 0.422&\cellcolor{red}\\\cline{1-2}
			0.314-0.316 & 0.421$\le$ $v$ $\le$ 0.422&\cellcolor{red}\\\cline{1-2}
			0.317-0.320 & 0.421$\le$ $v$ $\le$ 0.423&\cellcolor{red}\\\cline{1-2}
			0.321-0.323 & 0.423&\multirow{-12}{*}{\rotatebox{90}{\textbf{\textit{Right polarity}\,}}}\cellcolor{red}\\
			\hline\hline
			\textbf{0.324-0.329} &\multicolumn{2}{|c|}{\textbf{ no unidirectional flow}}\\
			\hline\hline
			0.33-0.333 & 0.425&\cellcolor{green}\\\cline{1-2}
			0.334-0.336 & 0.426&\cellcolor{green}\\\cline{1-2}
			0.337-0.339 & 0.426$\le$ $v$ $\le$ 0.427&\cellcolor{green}\\\cline{1-2}
			0.34 & 0.426$\le$ $v$ $\le$ 0.428&\cellcolor{green}\\\cline{1-2}
			0.341 & 0.427$\le$ $v$ $\le$ 0.429&\cellcolor{green}\\\cline{1-2}
			0.342 & 0.427$\le$ $v$ $\le$ 0.428&\cellcolor{green}\\\cline{1-2}
			0.343-0.345 & 0.427$\le$ $v$ $\le$ 0.429&\cellcolor{green}\\\cline{1-2}
			0.346-0.347 & 0.427$\le$ $v$ $\le$ 0.43&\cellcolor{green}\\\cline{1-2}
			0.348 & 0.428$\le$ $v$ $\le$ 0.43&\cellcolor{green}\\\cline{1-2}
			0.349-0.35 & 0.428 $\le$ $v$ $\le$ 0.431&\cellcolor{green}\\\cline{1-2}
			0.4 &  0.438 $\le$ $v$ $\le$ 0.449&\cellcolor{green} \\\cline{1-2}
			0.5 & 0.456 $\le$ $v$ $\le$ 0.488&\cellcolor{green}\\\cline{1-2}
			0.6 &  0.485 $\le$ $v$ $\le$ 0.521&\cellcolor{green} \\\cline{1-2}
			0.7 & 0.487 $\le$ $v$ $\le$ 0.547&\cellcolor{green}\\\cline{1-2}
			0.8 &  0.497 $\le$ $v$ $\le$ 0.565&\cellcolor{green} \\\cline{1-2}
			0.9 &  0.53 $\le$ $v$ $\le$ 0.580&\cellcolor{green} \\\cline{1-2}
			1  & 0.56 $\le$ $v$ $\le$ 0.588&\multirow{-17}{*}{\rotatebox{90}{\textbf{\textit{Left polarity}\,}}}\cellcolor{green}\\
			\hline
			
		\end{tabular}
		\caption {The velocity window for unidirectional flow of composite BB solitons with different coupling strengths. For the RM potential barriers with the range of coupling strength 0.324 $\le$ $g_{12}$ $\le$ 0.329 with $g_1 = g_2 = 1$, we find full reflection for $v$ $\leq$ 0.424 and full transmission for $v$ $\geq$ 0.425, hence no unidirectional flow is observed at this specific range of $g_{12}$. Away from this point, we find polarity reversal in unidirectional flow.}
		\label{RMtable}
	\end{center}
\end{table*}


\subsection{Unidirectional segregation with $g_{12} < 0$}

\begin{figure}[bt]\centering
\textbf{RM potential, $x_0$ = -10}\par\medskip
\includegraphics[scale=0.55]{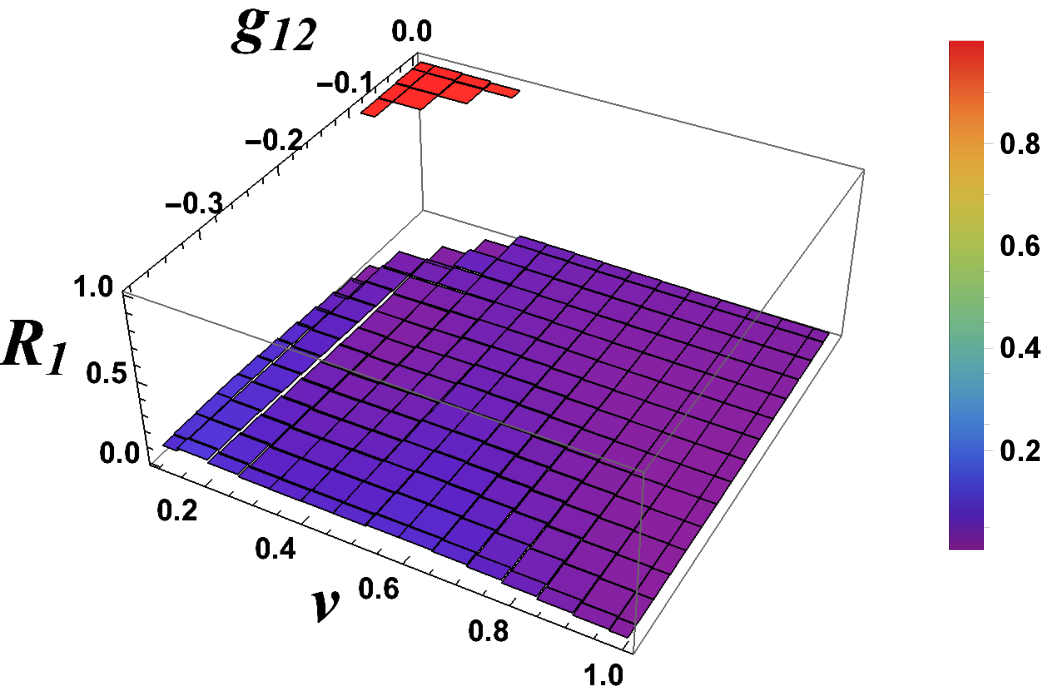}\\
\includegraphics[scale=0.55]{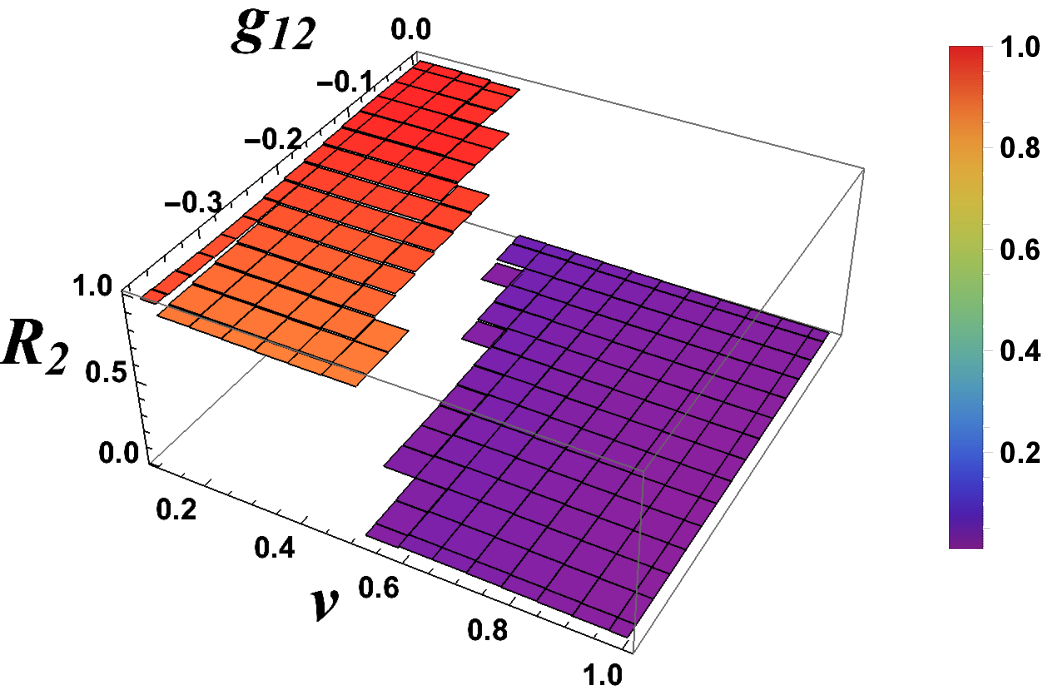}\\
\includegraphics[scale=0.55]{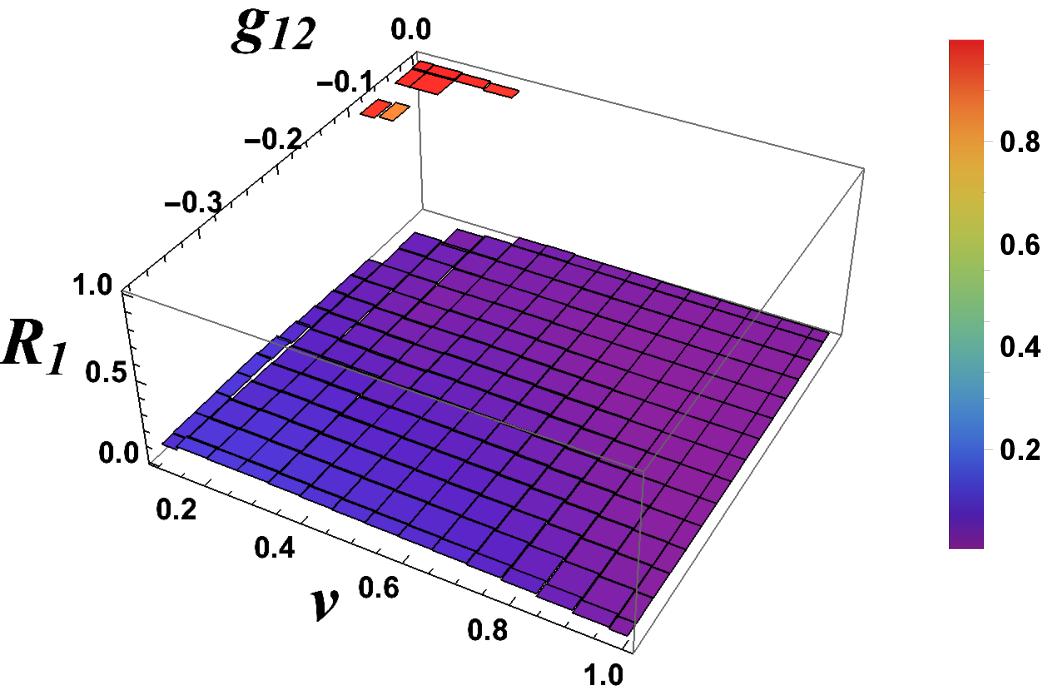}\\
\includegraphics[scale=0.55]{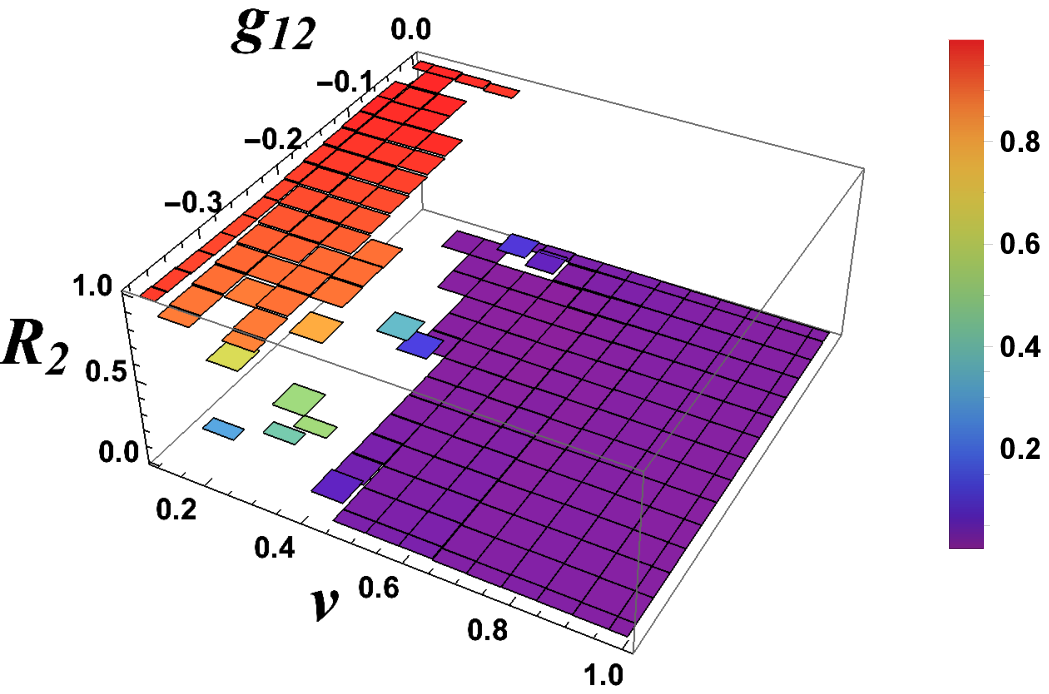}
 \caption{Reflection coefficients of the components $\psi_1$ and $\psi_2$ propagating through RM potential barriers from $x_0$ = -10 (upper two) and $x_0$ = 10 (lower two) versus $v$ and $g_{12}$. Other parameters are $g_1 =1$ and $g_2 = 1$.}
  \label{fig9}
\end{figure}

The influence of repulsive mean field coupling on the propagation of the BB solitons through an asymmetric RM potential barriers will be examined in this section. We find segregation or splitting of composite BB solitons scattering through asymmetric double potential barriers in the presence of a repulsive coupling ($g_{12} < 0$). We also found  unidirectional segregation, i.e., the composite BB soliton components undergo segregation while passing through the potential barriers for propagation in one particular direction and remains intact for  incident propagation from the opposite direction. The reflection coefficients of the components of the BB solitons versus velocity and $g_{12}$ for the initial propagation from $x_0$ = $\pm$ 10  through RM potential barriers are shown in Fig.~\ref{fig9}. We consider varying the incident velocity of the BB solitons from 0.1 to 1 and $g_{12}$ varying from 0 to -0.38 for our numerical investigation. In Fig.~\ref{fig9}, the top surface describes the reflection coefficient of the component $\psi_1$ and the bottom one describes the reflection coefficient of the component $\psi_2$. When the incident velocity is low ($v$ = 0.1) with $g_{12}$ up to  -0.16, reflection is found to be predominant for both the components. In the range $-0.36\leq g_{12}\leq-0.21$, at $v$ = 0.1, the reflection of the component $\psi_1$ is $\approx$ 0.05 and that for the component $\psi_2$ is $\approx$ 0.96. This demonstrates the nearly full separation of the components, i.e., the component $\psi_1$ undergoes maximum transmission and the component $\psi_2$ undergoes maximum reflection. For further decrease in $g_{12}$ from -0.36 to -0.5, we start to have trapping and  $T_1$ gradually reduces and reaches a minimum value of 0.6 ($R_1$ = 0.4) for $g_{12}$ = -0.5, while that of $T_2$ shows slight increase and reaches a transmission $\approx$ 0.17 ($R_2$ = 0.83). For lower $g_{12}$ values down to -0.35, the $R_1$ is found to be low but thereafter increases to the maximum of  $\approx$ 0.4 for further reduction in $g_{12}$ values. The $R_2$ shows a sharp transition from maximum to minimum at velocity $v$ = 0.25 for $-0.35<g_{12}\leq-0.1$. In case $g_{12}$ > -0.35 this transition region shifts to lower velocities around 0.2. In case of $R_2$, this transition to the minimum reflection is gradual and complete drop in reflection occurs at the velocity $v$ = 0.45. With this increase in incident velocity $v$ $\geq$ 0.45, the reflection for both the components are observed to be minimum for the entire range of the $g_{12}$ values. The obtained results illustrate that repulsion dominates for lower velocities and higher $g_{12}$ values. The obtained results demonstrate almost similar dynamics as the one obtained for the propagation from $x_0$ = -10. But a greater reflection window for the component $\psi_2$ over the mid $g_{12}$ values for velocity $\approx$ 0.3 is noticed. Also, a shift in the incident velocity $v$ $\geq$ 0.55, required for the minimum reflection is observed.

\begin{figure}[!htb]
	\centering
	 \textbf{Unidirectional segregation through RM potential }\par\medskip
	\includegraphics[scale=0.45]{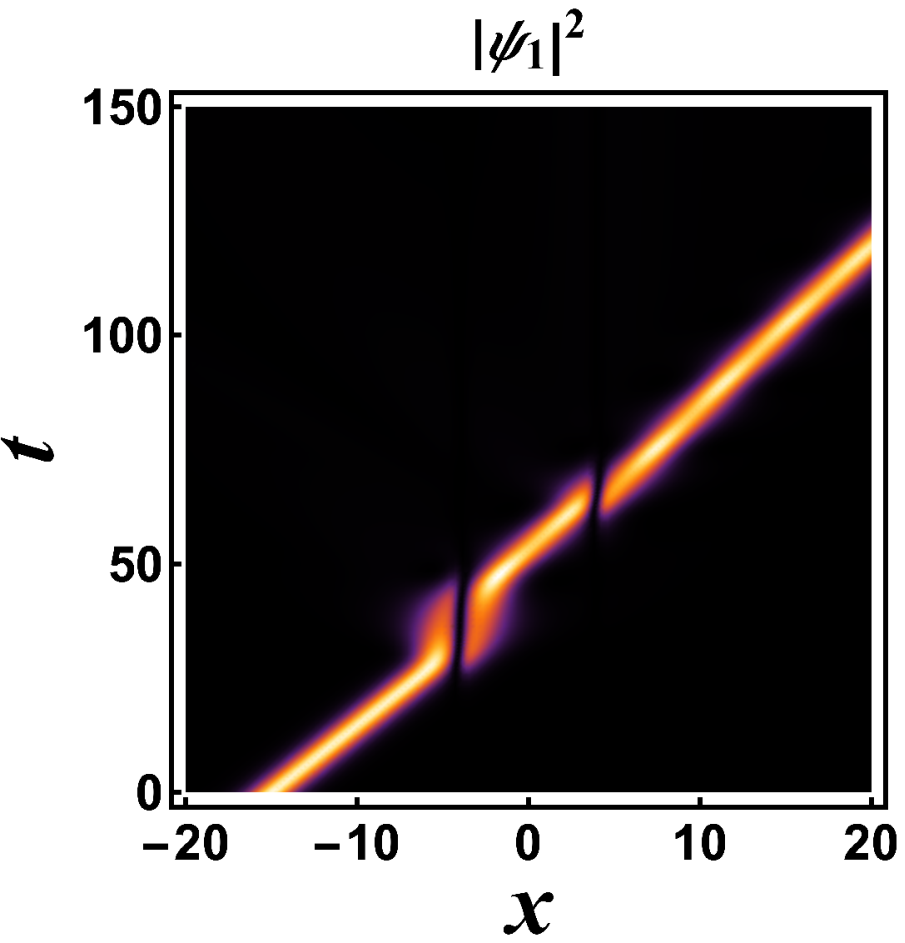}
\includegraphics[scale=0.45]{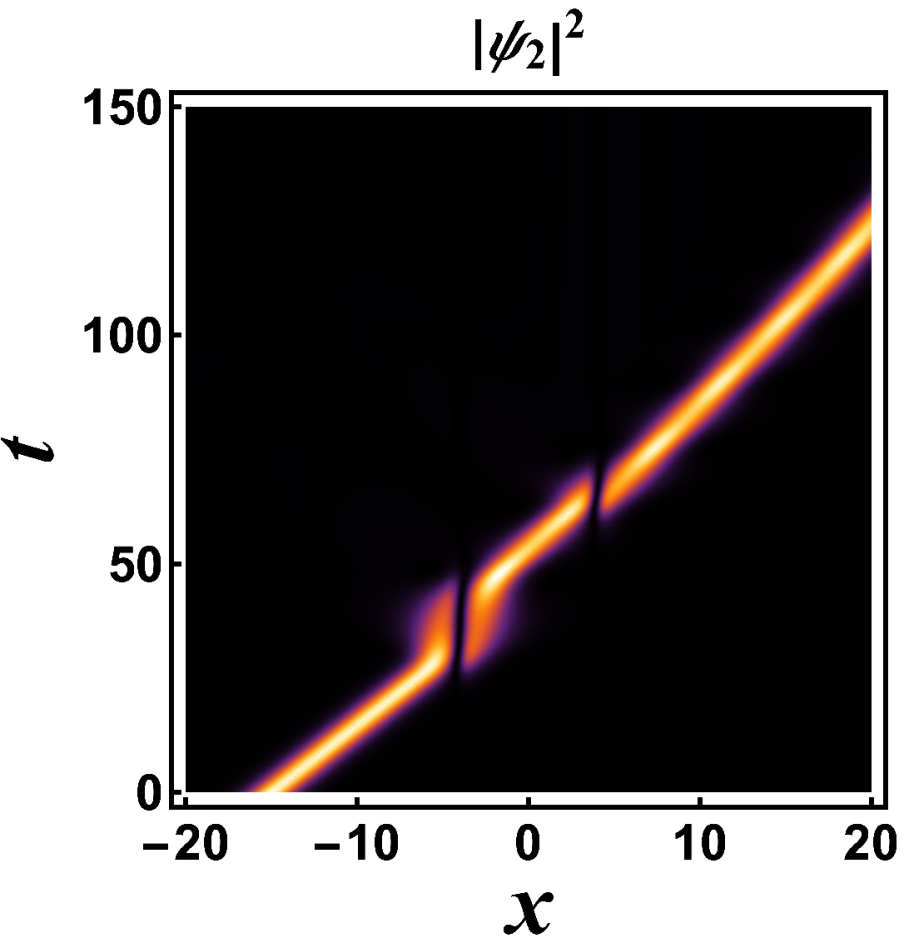}
\includegraphics[scale=0.45]{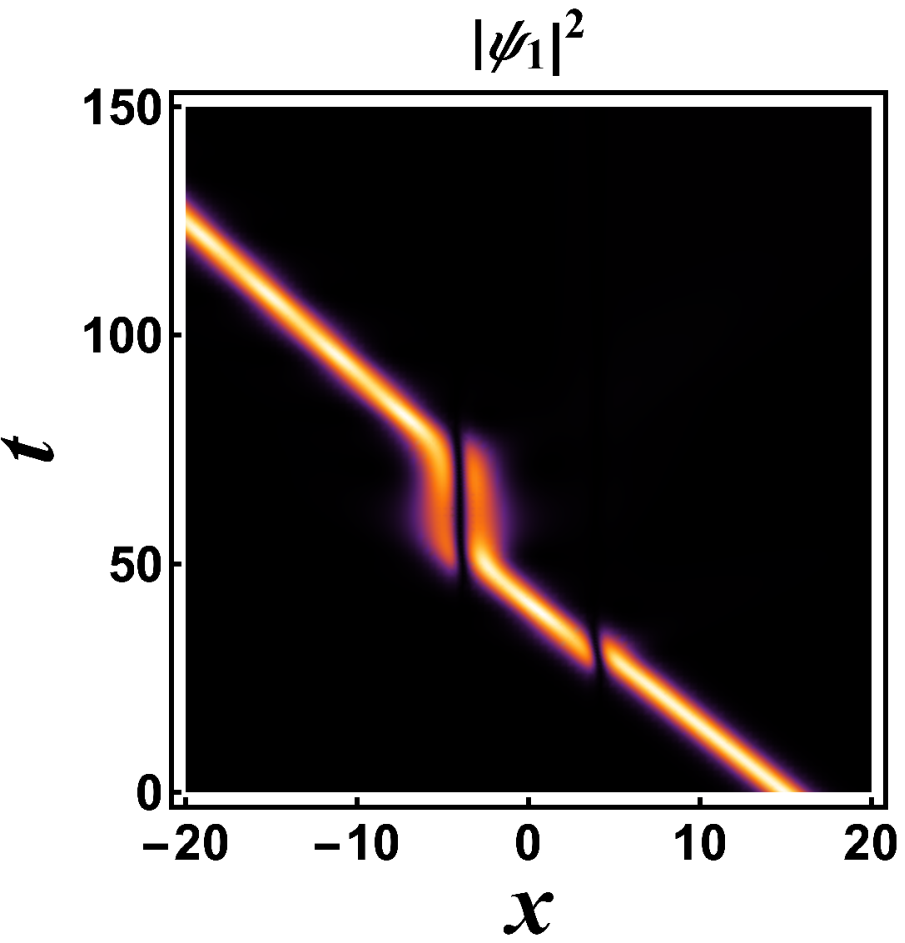}
\includegraphics[scale=0.45]{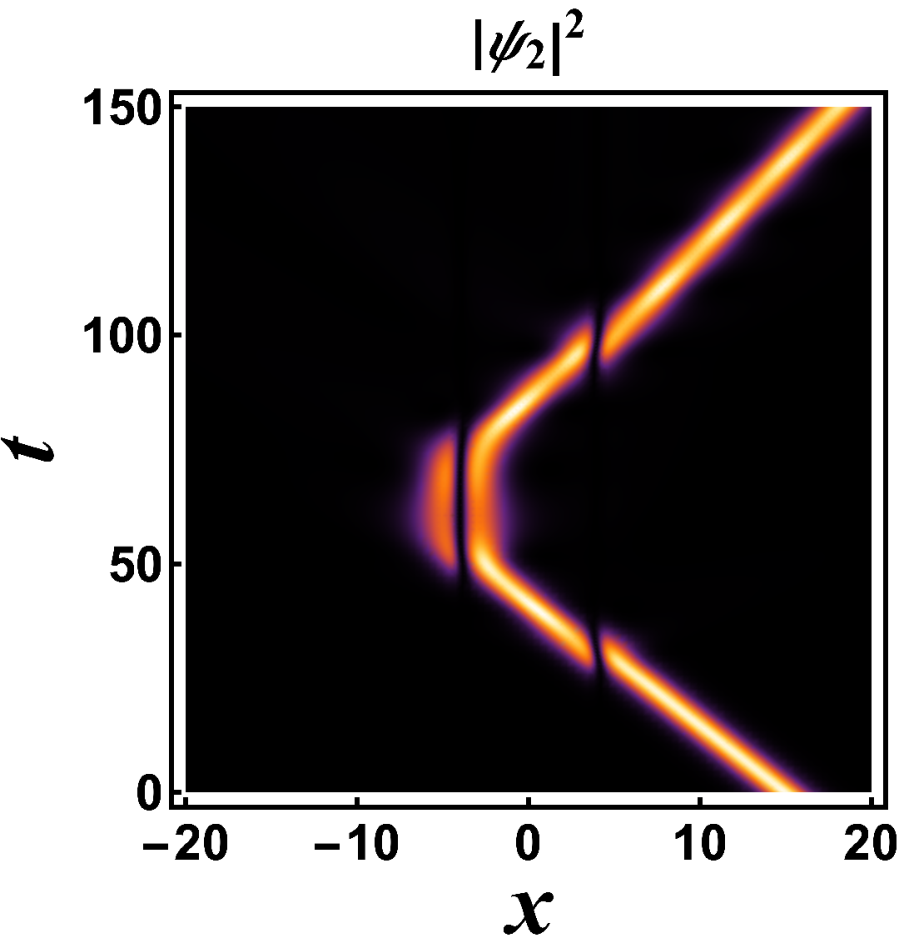}
\includegraphics[scale=0.45]{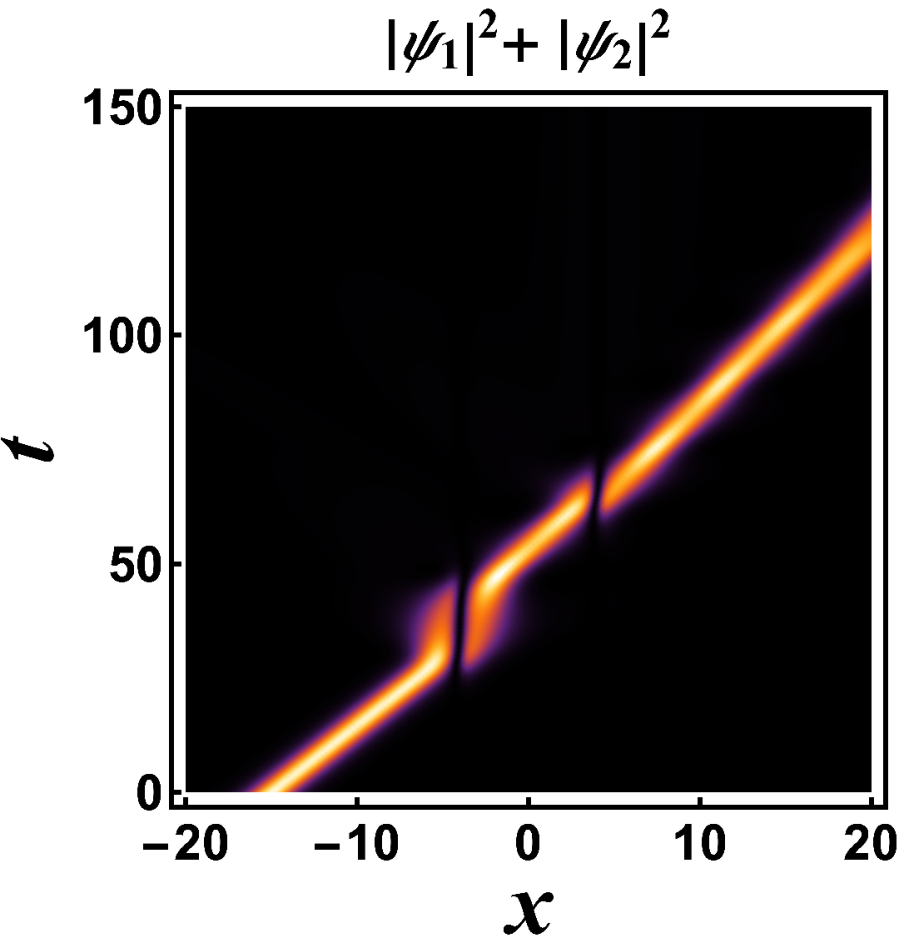}
\includegraphics[scale=0.45]{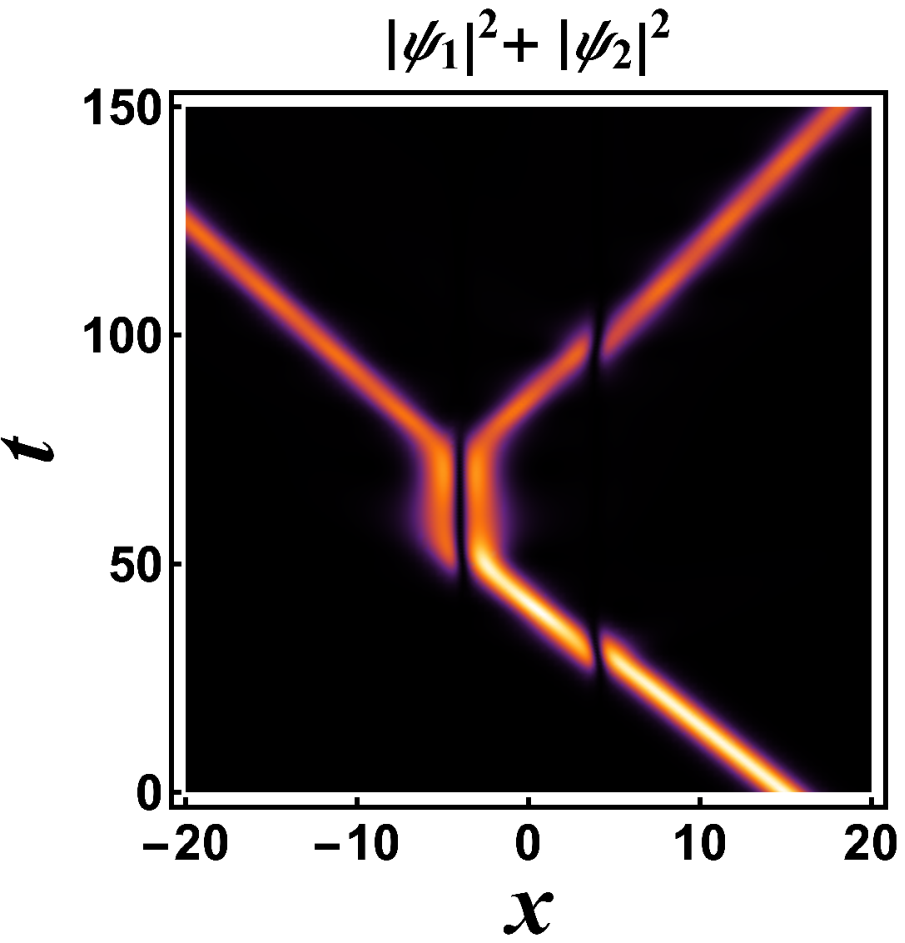}
	\caption{Propagation of components through asymmetric RM potential barriers for $g_1 = g_2 = 1$, and $g_{12} = -0.004$ at $v$= 0.34. The parameter of initial position $x_0$= $\pm$ 15 is used. Unidirectional segregation of composite BB soliton is achieved.}
	\label{fig11}
\end{figure}
Furthermore, we report here a phenomenon of unidirectional segregation as is demonstrated in Fig.~\ref{fig11}. This unidirectional segregation is obtained for BB soliton components passing through an asymmetrical RM potential barriers with barriers positioned at $x_1$ = -4 and $x_2$ = 4 with incident velocity $v$ = 0.34 in the range $-0.0045  \le  g_{12}  \le  -0.0033$. For the right moving BB solitons incident from the initial position at $x_0$ = -15, both components undergo transmission through both barriers and reach the right side with $T_1 = T_2 = 0.98$. On the other hand, for a left moving BB solitons, the component $\psi_1$ passes through both barriers and shows complete transmission $T_1 = 0.98$ and reaches the left side. Whereas the component $\psi_2$ passes through the first barrier and gets reflected by second barrier, then it again transmits through the first barrier and reaches the right side with a reflectance $R_1 = 0.98$. This transition is found to be very sensitive to the parameters of the designed reflectionless potential and the transmitted components of the right moving BB solitons are found to remain intact in a certain time interval, thereafter one of the components is out of phase with respect to the other component with a small phase shift but both the components continue along the same trajectory. This sensitive behavior paves the way to perform a variational calculation in order to achieve unidirectional segregation and to understand the physics underlying this phenomenon which we will discuss in the section \ref{variational}.

\subsection{Special Case:  $g_{1} \ne g_2$}
In this section, we investigate the influence of the nonlinear local interaction strength of each component on the BB solitons dyanamics (i) in absence of interaction coupling ($g_{12}$ = 0) and (ii) in presence of the repulsive interaction coupling ($g_{12} < 0$), respectively. For our study, we consider varying the nonlinear local interaction strength of the component $\psi_1$, by varying $g_1$ from 1.05 to 1.5 and the velocity is varied from 0.1 to 1.

\begin{figure}[bt]\centering
	\textbf{From $x_0$ = -10 }\par\medskip
	\includegraphics[scale=0.55]{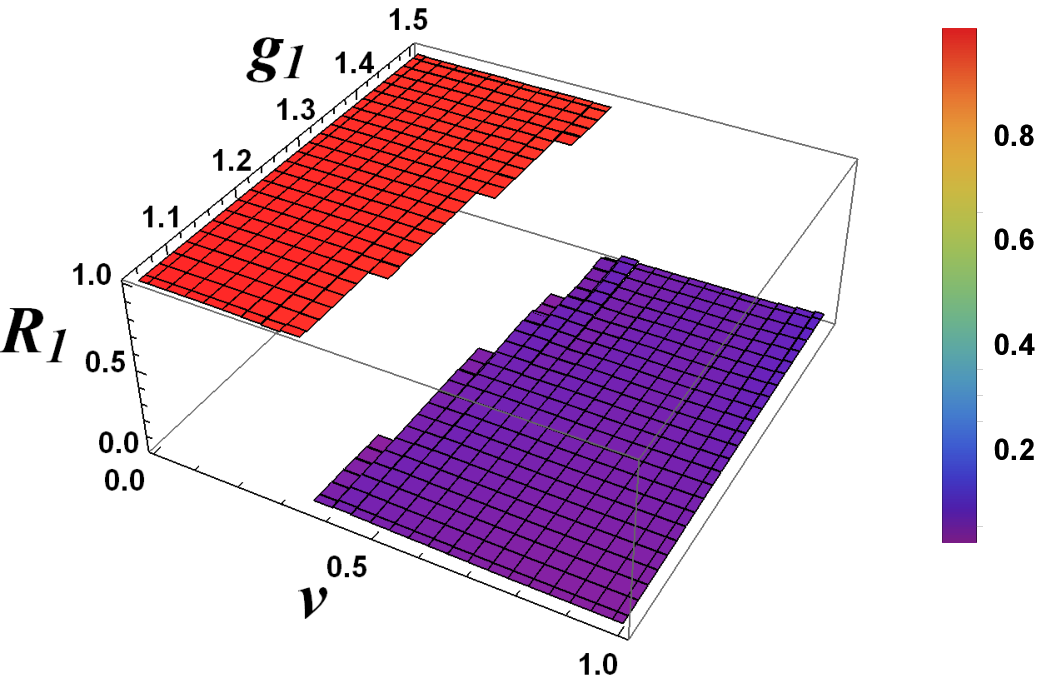}
	\includegraphics[scale=0.55]{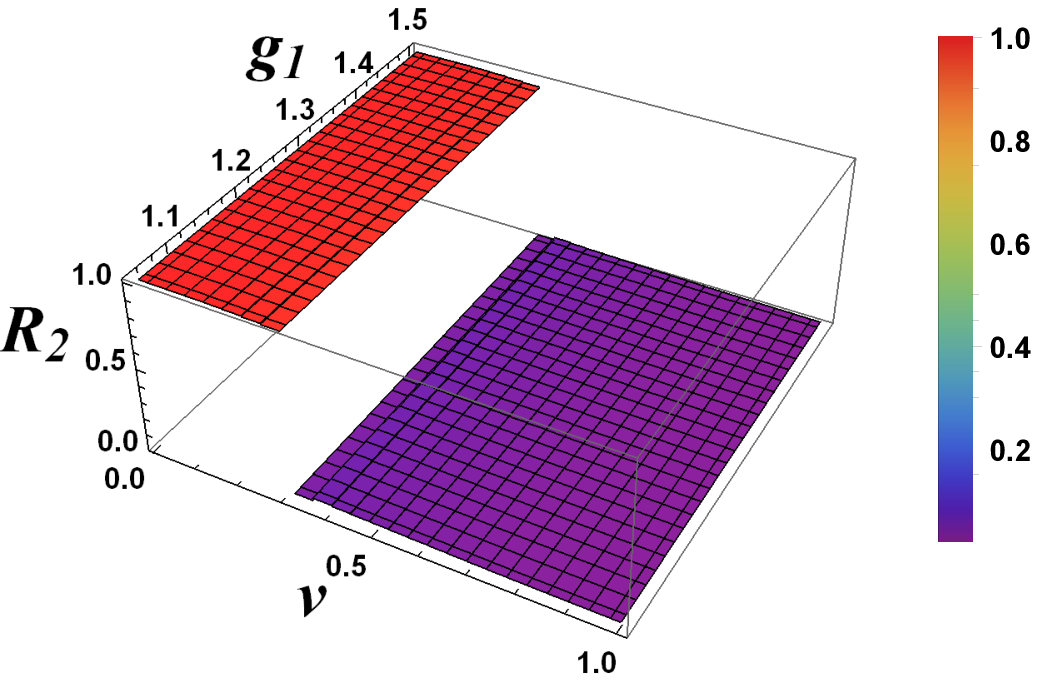}
	\includegraphics[scale=0.55]{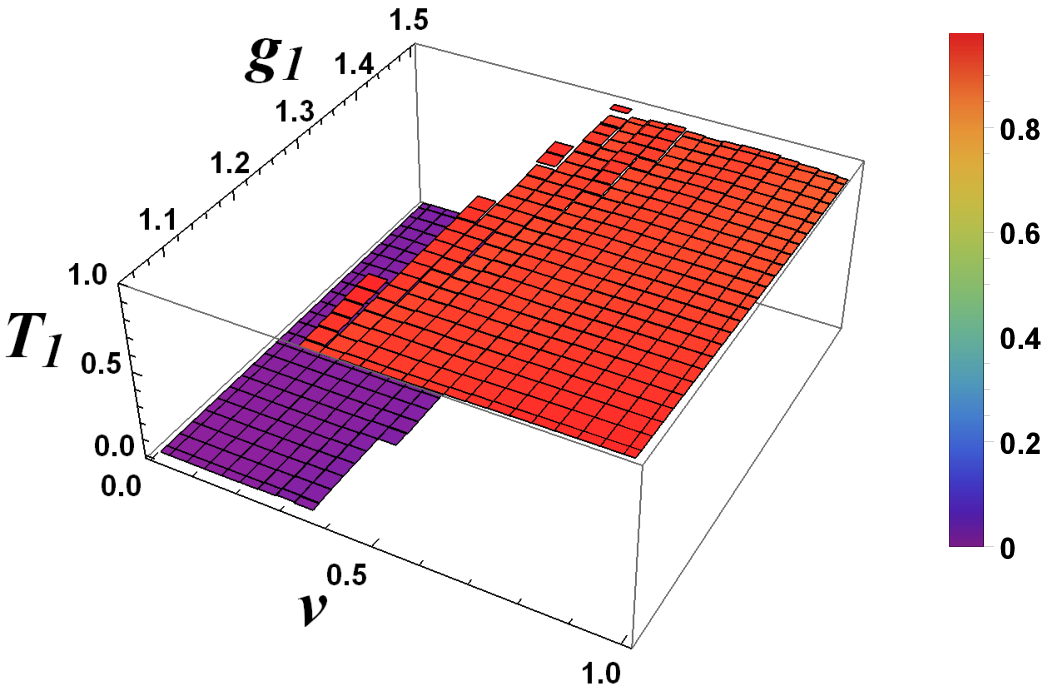}
	\includegraphics[scale=0.55]{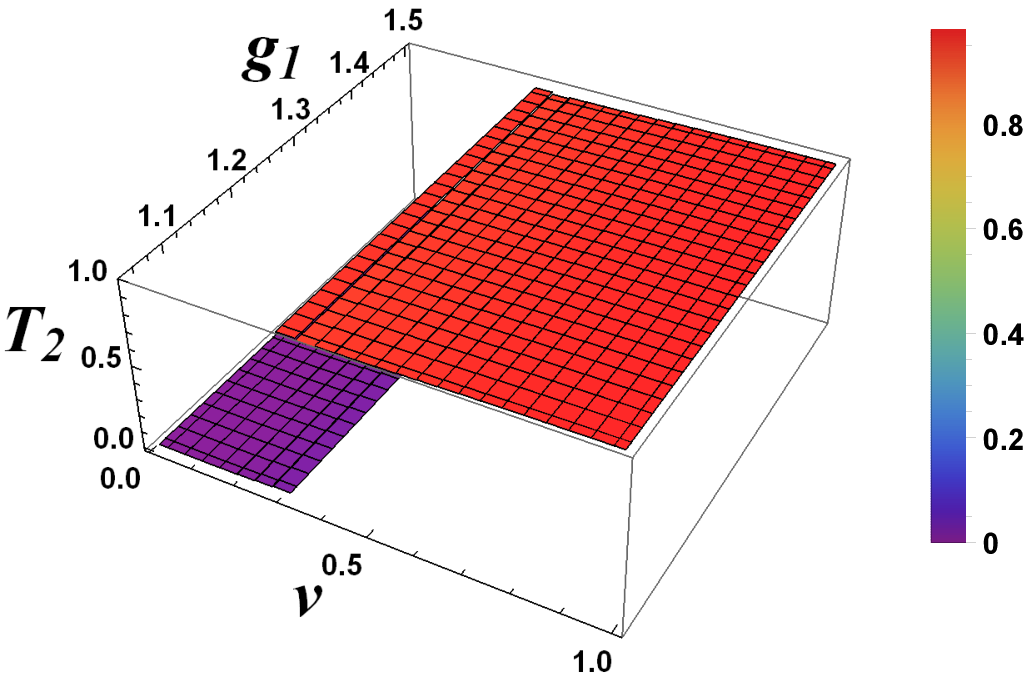}
	\caption{Reflection and transmission coefficients of the components $\psi_1$ and $\psi_2$ propagating through RM potential barriers from $x_0$ = -10 versus $v$ and $g_{1}$. The parameters used are $g_{12} = 0$ and $g_2$ = 1.}
	\label{figR1}
\end{figure}

\textbf{(a) For $g_{12} = 0 $}\\\\
The results obtained for the propagation of BB solitons through the asymmetric RM potential barriers in the absence of interaction coupling ($g_{12}$ = 0), with $g_1$ varying form 1.05 to 1.5 is illustrated in the Fig. \ref{figR1} for initial propagation from $x_0$ = $\pm$10. In Fig. \ref{figR1}, the top two plots display the results of reflectance of the components $\psi_1$ and $\psi_2$, respectively of the BB solitons propagating from the initial position $x_0$ = -10. The lower two plots provide the results of the transmittance of the components $\psi_1$ and $\psi_2$, respectively of the BB solitons propagating from the initial position $x_0$ = -10. For $g_1$ = 1.05, the value of $R_1$ is found to be maximum up to the velocity $v$ = 0.346, thereafter it makes a sharp transition from maximum reflectance to the minimum. On the other hand the $R_2$ maintains its maximum value up to the velocity $v$ = 0.307 and thereafter reaches sharp minimum. We infer that the increase in nonlinear local interaction strength introduces a shift in the velocity required for maximum reflectance. The subsequent shift in the velocity shift for maximum reflectance $R_1$ is observed for $g_1$ values 1.14, 1.31 and 1.42, respectively. A similar behavior for shift in velocity to achieve maximum $T_1$ is also observed, as displayed in the third plot of Fig. \ref{figR1}.

\begin{figure}[bt]\centering
		 \textbf{From $x_0$ = 10 }\par\medskip
	\includegraphics[scale=0.55]{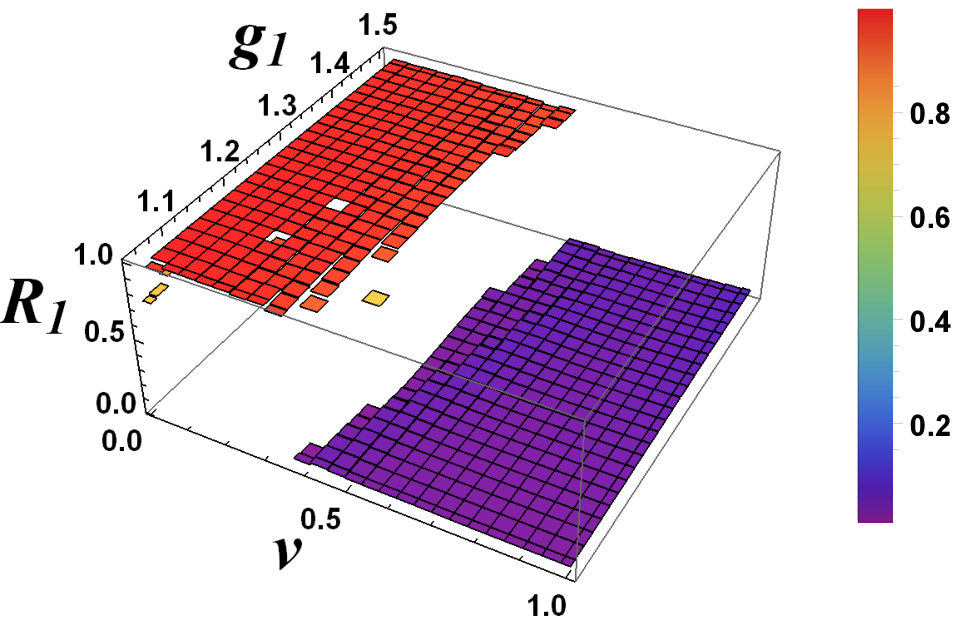}
	\includegraphics[scale=0.55]{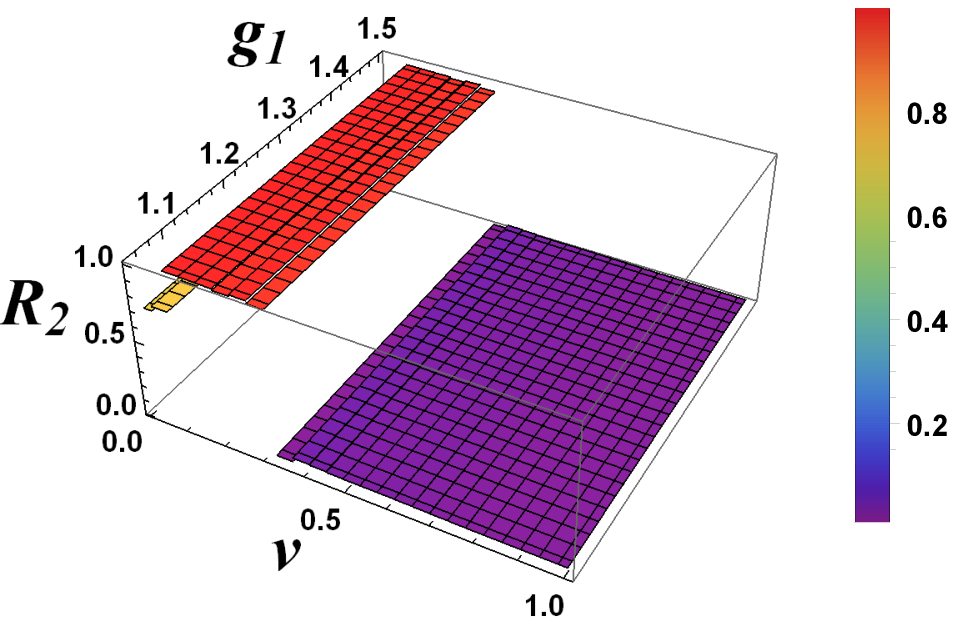}
	\includegraphics[scale=0.55]{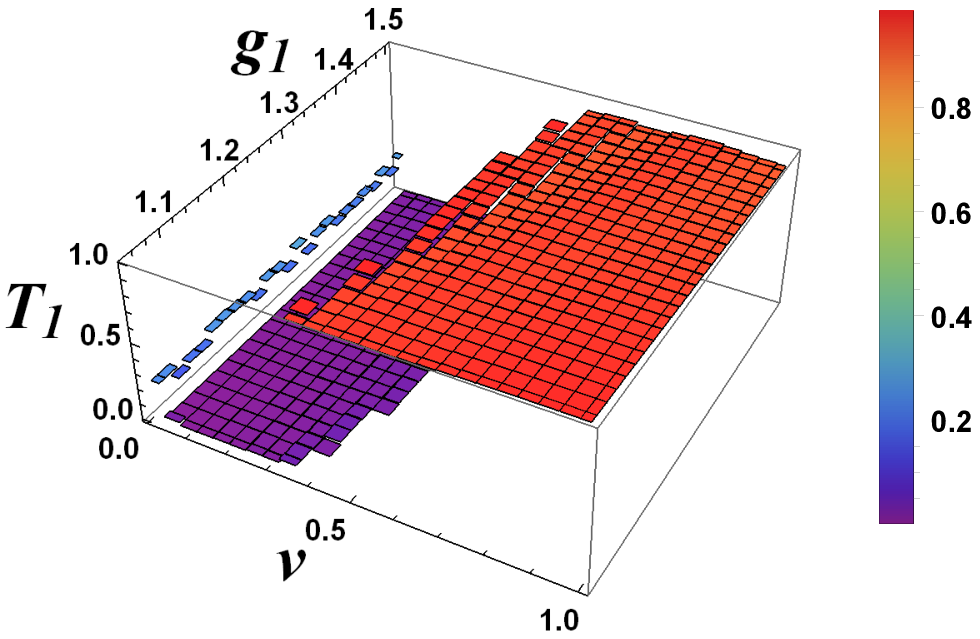}
	\includegraphics[scale=0.55]{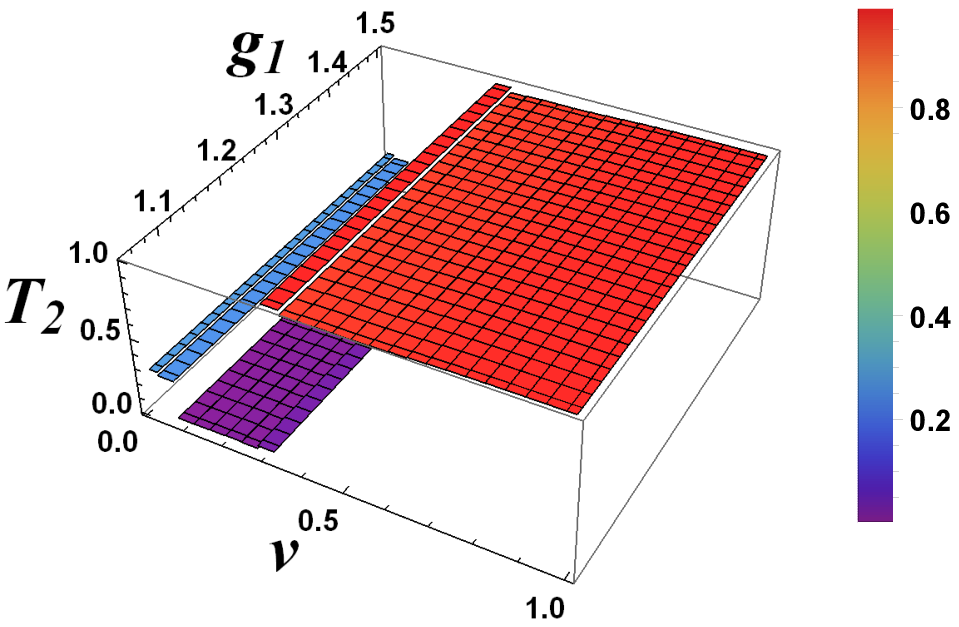}
	\caption{Reflection and transmission coefficients of the components $\psi_1$ and $\psi_2$ propagating through RM potential barriers from $x_0$ = 10 versus $v$ and $g_{1}$. The parameters used are $g_{12} = 0$ and $g_2$ = 1.}
	\label{figR2}
\end{figure}

Next, considering the propagation of the BB solitons from $x_0$ = 10, for extremely low velocities, there exist nonlinear modes with energy trapping around $\approx0.2\,\%$ throughout the variation of $g_1$, as shown by Fig. \ref{figR2}. Moreover, the present case also displays that the existence of the shift in the velocity throughout which the maximum reflectance of $R_1$ is maintained. The reflectance is found to be maximum up to the velocity $v$ equals to 0.32, thereafter it undergoes a sharp transition from maximum reflection to minimum reflection at $g_1$ = 1.05. Thereafter, it displays the right shift in the velocity for maximum reflectance as obtained for the propagation from $x_0$ = -10, but at different $g_1$ values.

\begin{figure}[!htb]
	\centering
	 \textbf{RM potential}\par\medskip
\includegraphics[scale=0.45]{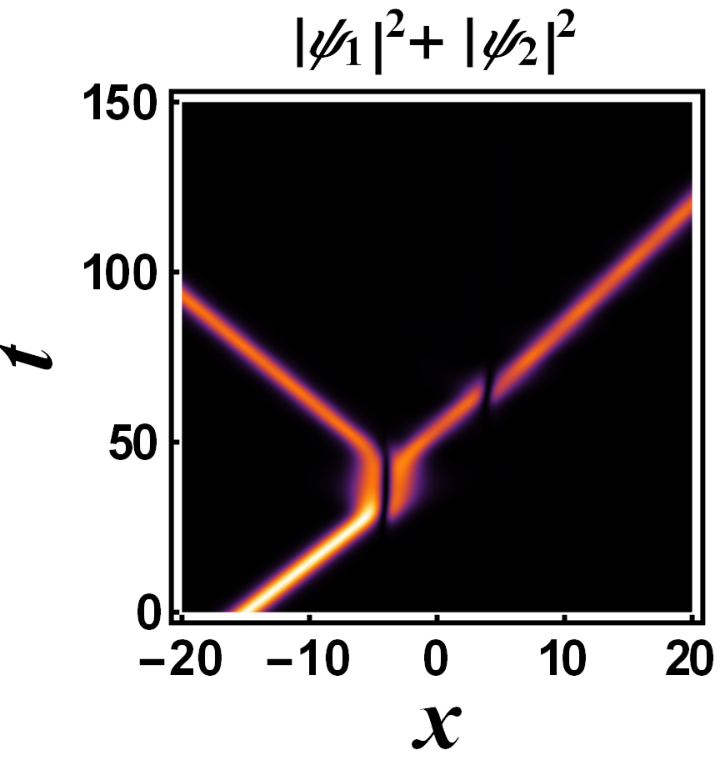}
\includegraphics[scale=0.45]{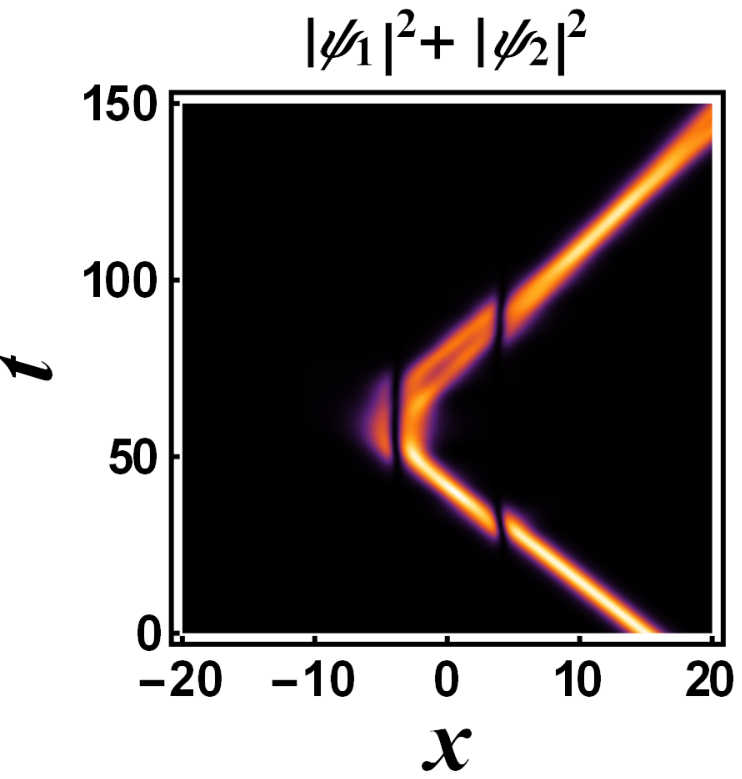}
		\caption{Propagation of components through asymmetric RM potential barriers at $v$= 0.34. The parameter of initial position $x_0$= $\pm$ 15 is used. Unidirectional segregation of composite BB soliton is achieved. Other parameters used are $g_{12} = 0$, $g_1$ = 1.05,  and $g_2$ = 1.}
	\label{figR3}
\end{figure}

Furthermore, for the present case, we also have identified the regimes of the unidirectional segregation as in our previous section. Here, for the $g_1$ values from 1.03 to 1.1, the system exhibits the unidirectional segregation for the BB soliton components propagating from initial positions, $x_0\,=\pm15$ with barrier positions, $\pm4$. The unidirectional flow obtained for these parameters at $v$ = 0.34 is illustrated in Fig. \ref{figR3}. In this case, for the right propagating BB solitons, the component $\psi_1$ undergo complete reflection in first barrier meanwhile the component $\psi_2$ undergoes full transmission crossing both barriers. On the other hand, the left moving solitons undergo full reflection at the first barrier.

\begin{figure}[bt]\centering
	\textbf{From $x_0$ = -10 }\par\medskip
	\includegraphics[scale=0.55]{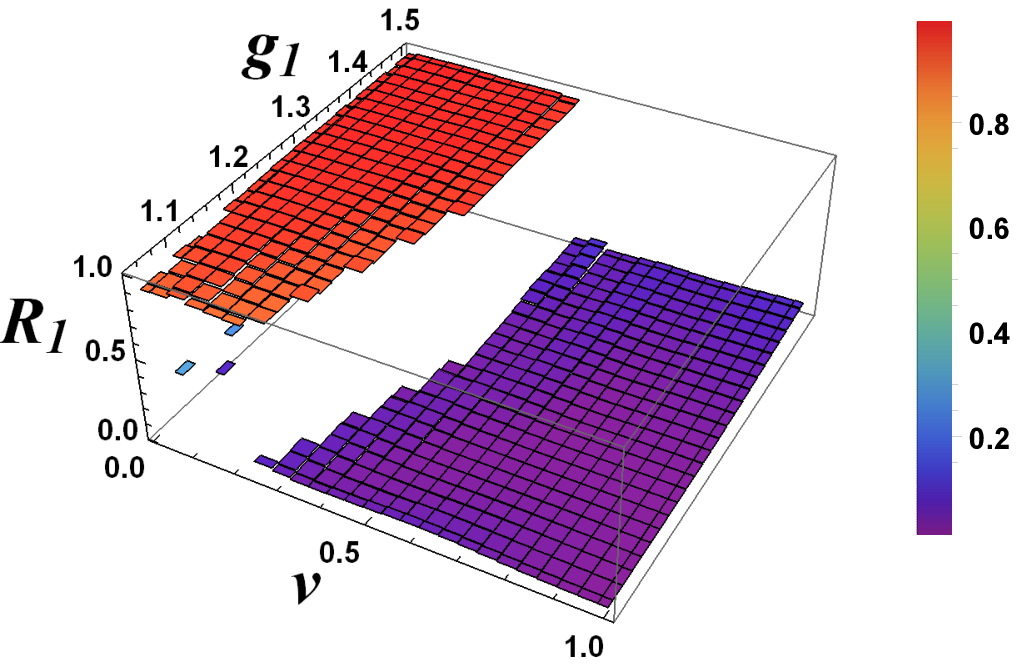}
	\includegraphics[scale=0.55]{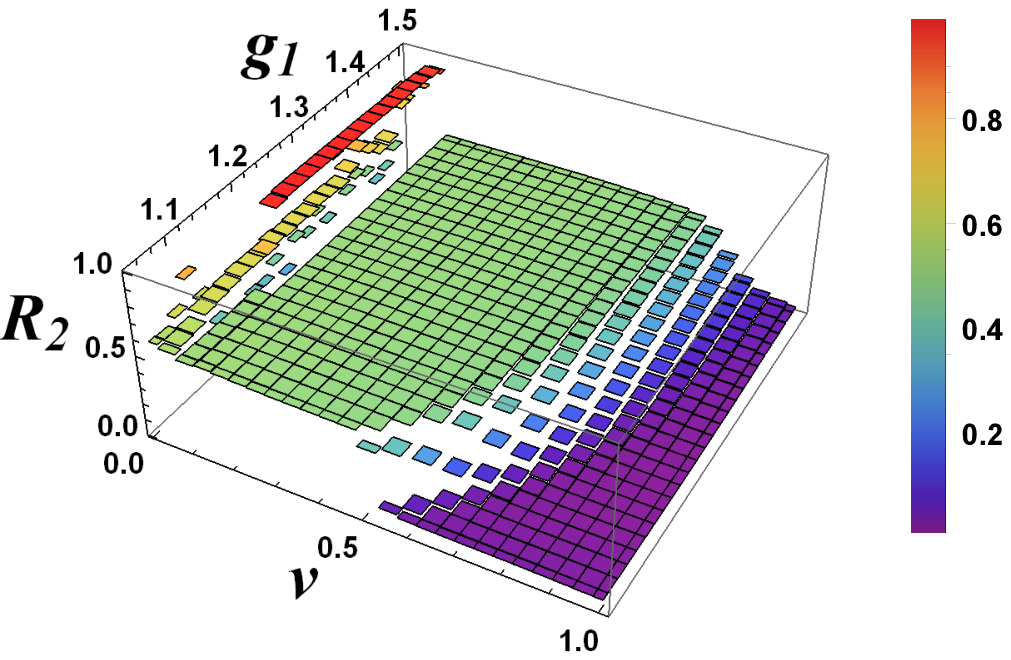}
	\includegraphics[scale=0.55]{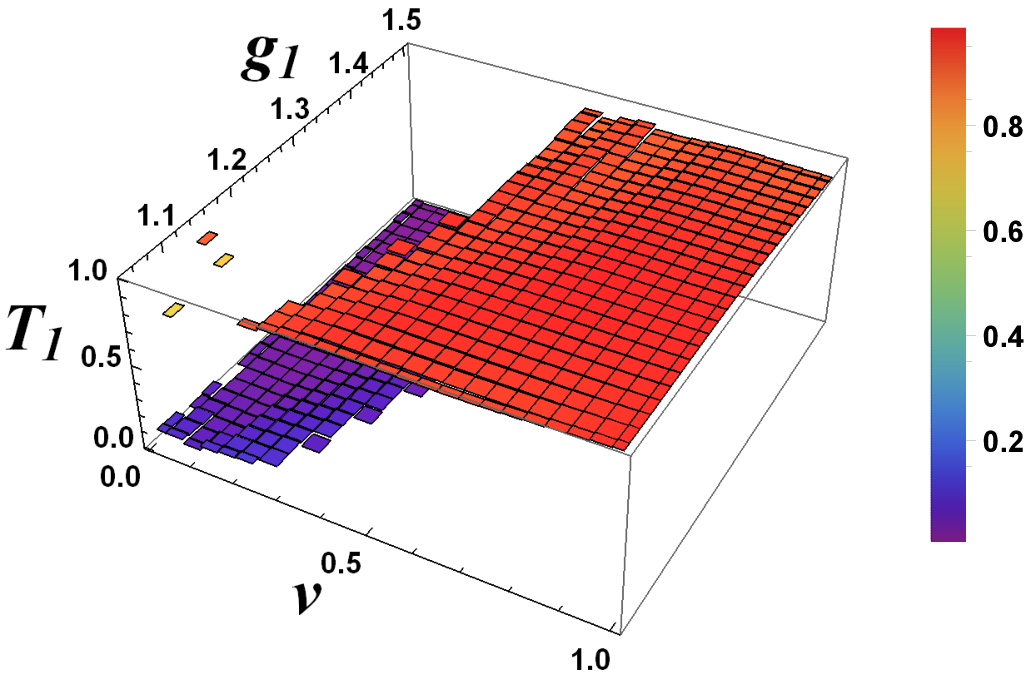}
	\includegraphics[scale=0.55]{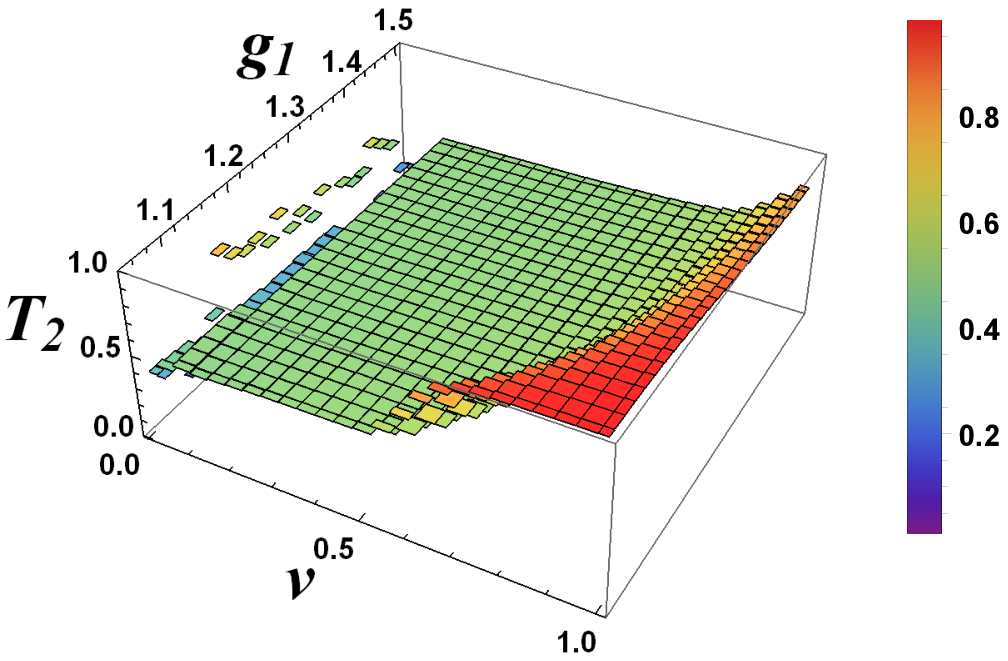}
	\caption{Reflection and transmission coefficients of the components $\psi_1$ and $\psi_2$ propagating through RM potential barriers from $x_0$ = -10 versus $v$ and $g_{1}$. The parameters used are $g_{12} = -0.5$ and $g_2$ = 1.}
	\label{figR1b}
\end{figure}

\textbf{(b) For $g_{12} = -0.5 $}\\\\
Additionally, in order to understand the influence of the nonlinear local interaction coupling in presence of repulsive interaction coupling, here, we consider the case $g_{12}$ = -0.5. The dynamics of the BB solitons through the RM potential barriers is investigated for $g_1$ varying from 1.05 to 1.5. The results obtained for the BB solitons propagating from $x_0$ = -10 is provided by the Fig. \ref{figR1b}. For lower velocities and low $g_1$ values ($g_1$ < 1.16), there exist few nonlinear trapping modes. But for higher $g_1$ values above 1.16, the nonlinear trapping modes are absent. Here also the shift in velocity for maximum reflectance of $R_1$ is found to increase with increasing $g_1$. After certain $g_1$ value above 1.3, the velocity required for maximum reflectance becomes constant. Next considering $R_2$, for lower velocities and lower $g_1$ values, more trapping modes are observed. For $g_1$ values above 1.2, the trapping is found to reduce but with around $90\%$ reflectance. As the velocity increases, reflectance decreases and reaches up to $50\%$, with around $50\%$ trapping until the velocity equals to 0.4. During this regime, the reflectance is constant for the all $g_1$ values. For further increase in the velocity, the reflection as well as trapping found to reduce gradually and reaches minimum. From the transmittance plots of Fig. \ref{figR1b}, it is illustrated that the dynamics of transmittance of the components are complementary to that of the reflectance of the components.

Next, the propagation of BB solitons from $x_0$ = 10 is considered in the Fig. \ref{figR1b}. It is observed that the existence of the nonlinear trapping modes are high for the left moving solitons propagating from $x_0$ = 10, where it encounters the large barrier first. But for higher $g_1$ values, trapping is found to be minimum. Here also the results demonstrate the shift in velocity for maximum reflectance of $R_1$, with increase in the $g_1$ values. Further increase in the velocity above 0.3, the reflectance drops from maximum to minimum sharply. On the other hand, $R_2$ portrays, almost similar dynamical behavior as that of $R_1$, but the drop in reflectance is shifted to the lower velocity, $v\,=\,0.2$. Here, also the transmittance of the components show more or less opposite behavior as that of the reflectance of the respective components. From the results, it is observed that $g_{12}$ increases the number of nonlinear modes and reduces the transmission of the component around $50\%$.

Furthermore, from the density plots shown by Fig. \ref{figR3b}, $g_{12}$ is found to suppress the unidirectional segregation obtained with same numerical values considered for the case $g_{12}$ = 0 and results in bidirectional segregation. For the BB solitons propagating from $x_0 = -15$, the component $\psi_1$ undergoes full reflection while the component $\psi_2$ undergoes full transmission. Similarly for the BB solitons propagating from $x_0 = 15$, we also found that the component $\psi_1$ shows full reflection and the component $\psi_2$ shows full transmission. Moreover, from our results, we found that even an extremely small value of $g_{12}$ = -0.004 is sufficient to suppress the unidirectional segregation.

\begin{figure}[bt]\centering
	\textbf{From $x_0$ = 10 }\par\medskip
	\includegraphics[scale=0.55]{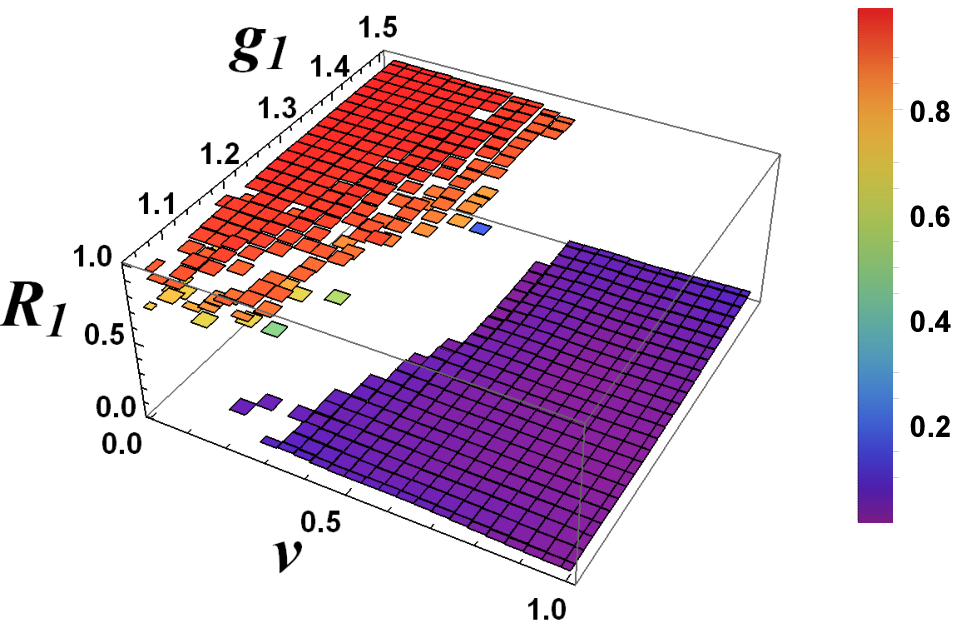}
	\includegraphics[scale=0.55]{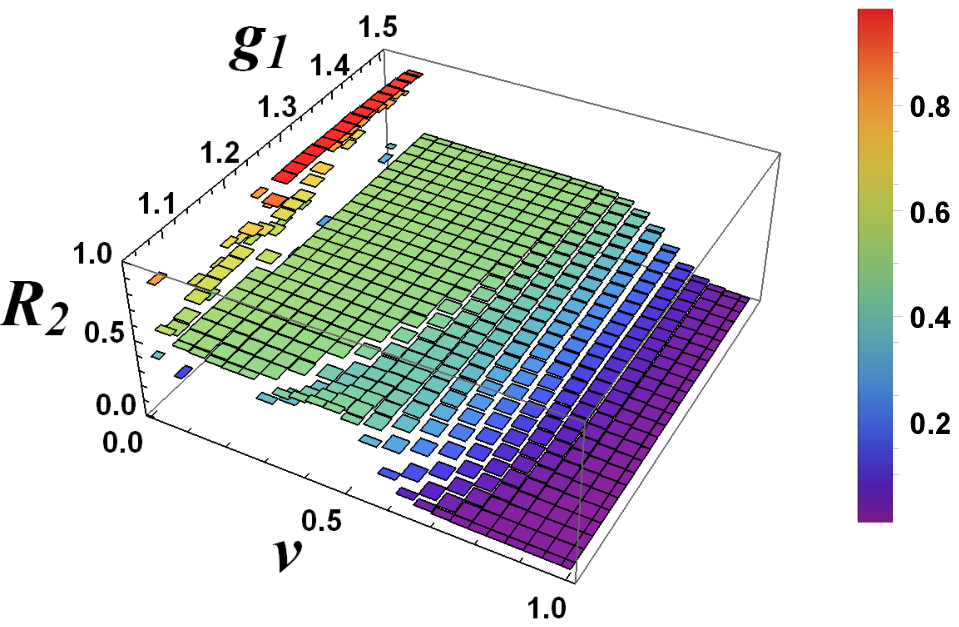}
	\includegraphics[scale=0.55]{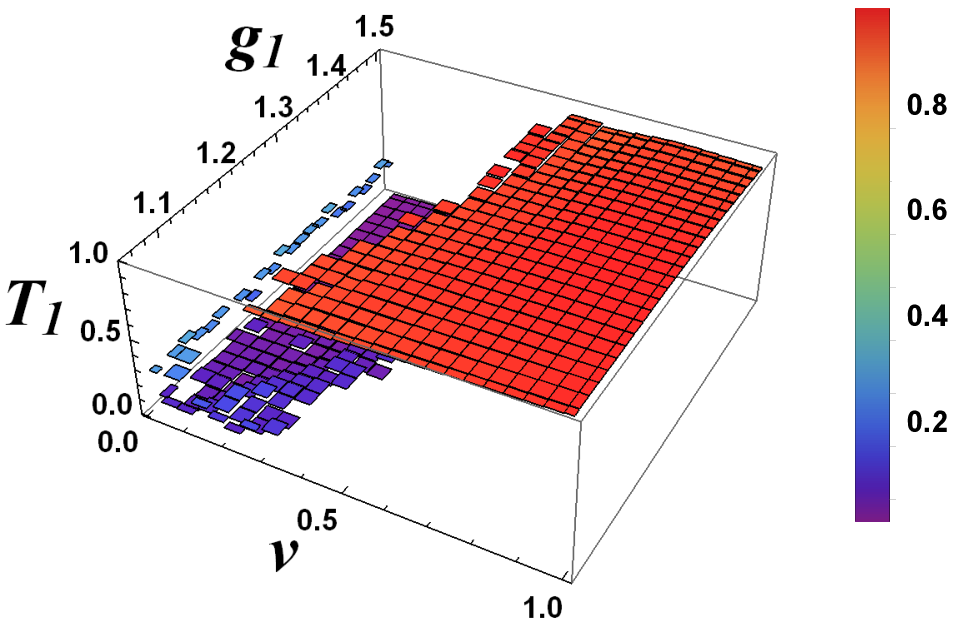}
	\includegraphics[scale=0.55]{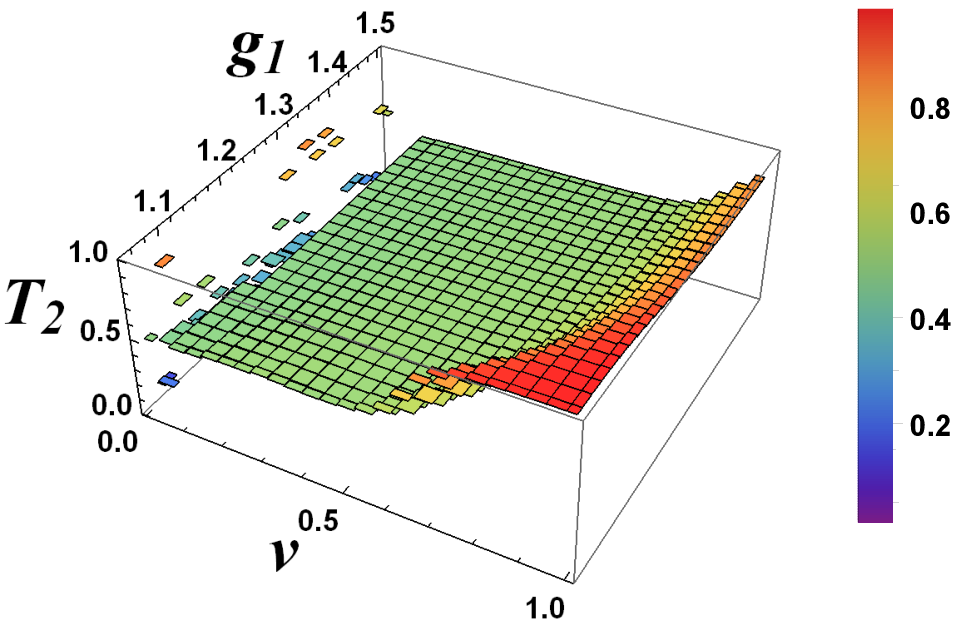}
	\caption{Reflection and transmission coefficients of the components $\psi_1$ and $\psi_2$ propagating through RM potential barriers from $x_0$ = 10 versus $v$ and $g_{1}$. The parameters used are $g_{12} = -0.5$ and $g_2$ = 1.}
	\label{figR2b}
\end{figure}

\begin{figure}[bt]\centering
	\includegraphics[scale=0.45]{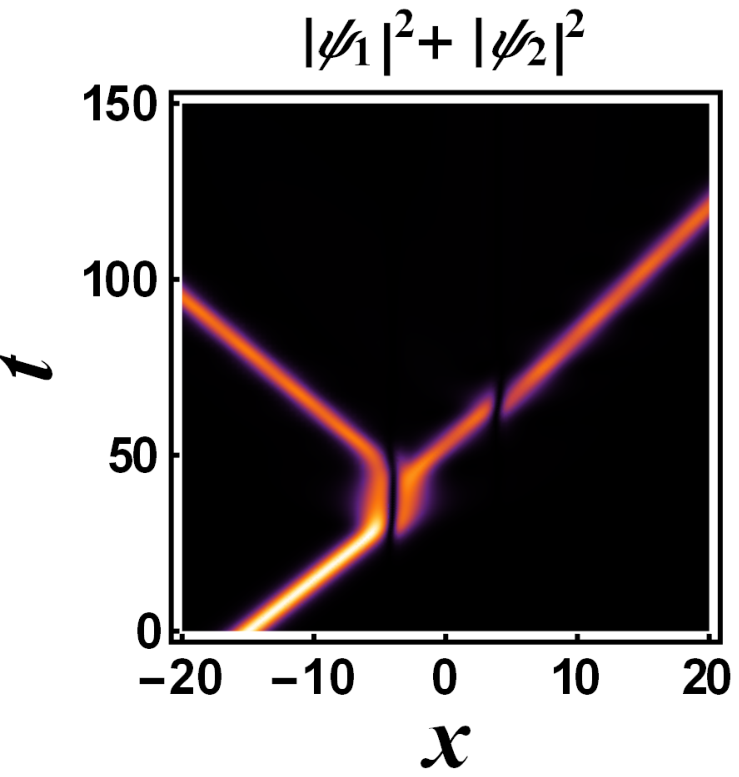}
	\includegraphics[scale=0.45]{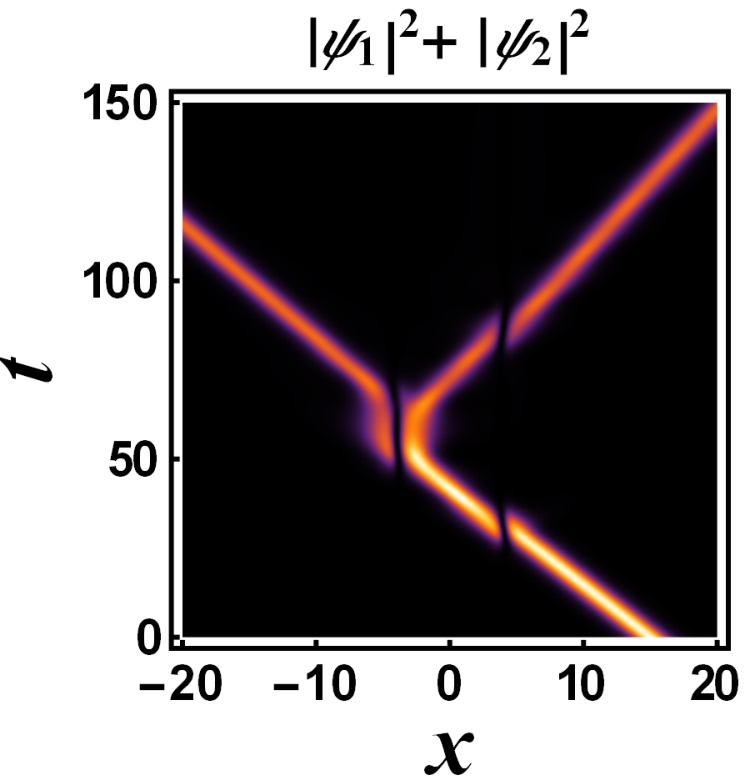}
	\caption{Propagation of components through asymmetric RM potential barriers at $v$= 0.34. The parameter of initial position $x_0$= $\pm$ 15 is used. Bidirectional segregation of composite BB soliton is observed. Other parameters used are $g_{12} = -0.5$, $g_1$ = 1.05, and $g_2$ = 1.}
	\label{figR3b}
\end{figure}

\section{Variational Approach versus Numerical Computation}\label{variational}

We use a comparative analysis between numerical computation and variational approach to validate our results for the unidirectional segregation of composite BB soliton. In this section, we will show how our numerical results for the computation of the unidirectional segregation are in excellent agreement with those generated by the variational approach. The starting point of our analysis is the Lagrangian density,

\begin{align}
\label{eq:secIV_lagrangian_density}
\mathcal{L}  &= \frac{i}{2}\Big(\psi_{1}^{*} \frac{\partial \psi_{1}}{\partial t}   - \psi_{1} \frac{\partial \psi_{1}^{*}}{\partial t} \Big) -\frac{1}{2} \Big \lvert \frac{\partial \psi_{1} }{\partial x}\Big \rvert^{2} +\frac{g_{1}}{2} \lvert  \psi_{1} \rvert^{4}  \nonumber \\ &
+ \frac{i}{2}\Big(\psi_{2}^{*} \frac{\partial \psi_{2}}{\partial t}  - \psi_{2} \frac{\partial \psi_{2}^{*}}{\partial t}\Big) -\frac{1}{2}  \Big \lvert \frac{\partial \psi_{2}}{\partial x} \Big \rvert^{2} +\frac{g_{2}}{2} \lvert  \psi_{2} \rvert^{4} \nonumber \\ & + g_{12} |\psi_{1}|^{2} |\psi_{2}|^{2} + V(x) [|\psi_{1}|^{2} + |\psi_{2}|^{2}].
\end{align}

Using the Euler-Lagrange equation with Eq.~\eqref{eq:secIV_lagrangian_density} we obtain the following coupled NLSE,

\begin{align}
\label{eq:secIV_eq1}
\nonumber
& i \frac{\partial \psi_{1}}{\partial t} + \frac{1}{2}  \frac{\partial^2 \psi_{1}}{\partial x^2} +\left[g_{1} |\psi_{1}|^2 + g_{12} |\psi_{2}|^2 + V\left(x\right) \right] \psi_{1} =0, \\
& i \frac{\partial \psi_{2}}{\partial t} + \frac{1}{2}  \frac{\partial^2 \psi_{2}}{\partial x^2} +\left[g_{2} |\psi_{2}|^2 + g_{12} |\psi_{1}|^2  + V\left(x\right) \right] \psi_{2} =0,
\end{align}

which is identical to Eq.~\eqref{Manakovsystem}. We adopt the following variational ansatz as the BB soliton solutions to Eq.~\eqref{eq:secIV_eq1}

\begin{align}
\label{eq:secIV_ansatz}
\psi_{1}(x,t) &= A \, \mathrm{sech} \Big(\frac{x+\xi_{1}}{a}\Big)\, \mathrm{e}^{i[\phi + v_{1} (x+\xi_{1}) + b (x+\xi_{1})^2]},  \nonumber \\
\psi_{2}(x,t) &= A \, \mathrm{sech} \Big(\frac{x+\xi_{2}}{a}\Big)\, \mathrm{e}^{i[\phi + v_{2} (x+\xi_{2}) + b (x+\xi_{2})^2]}.
\end{align}

The variational parameters $A(t)$, $\xi_{1,2}(t)$, $a(t)$, $\phi(t)$, $v_{1,2}(t)$, and $b(t)$ represent the amplitude, center-of-mass position, width, phase, velocity, and the chirp of the solitons, respectively. We use the normalization condition,

\begin{align}
\label{eq:secIV_norm}
\int_{-\infty}^{\infty} dx\; \left| \psi\right|^2 =2 A^2 a = N,
\end{align}

to reduce the number of variational parameters by one variable where we define the amplitude as a function of the width, $a$, and the normalization constant, $N$. The potential in Eq.~\eqref{eq:secIV_eq1} for both components takes the form,

\begin{align}
\label{eq:secIV_potential}
V(x) = l_{00}  +  l_{01}\,\Big[\Theta(x-q_{1}) - \Theta(x-q_{2})\Big]\nonumber \\
 +  l_{02}\, \Big[\Theta(x-q_{3}) - \Theta(x-q_{4})\Big],
\end{align}
where $\Theta(x)$ is the Heaviside unit step function, $l_{00}$, $l_{01}$, $l_{02}$, $q_{1}$, $q_{2}$, and $q_{3}$ are potential parameters to be suitably selected  for numerical computations. The choice of this particular potential simplifies the analytical calculations but will not limit the validity of our main results. Due to the fact that soliton scattering through a potential well causes its width to shrink while soliton scattering through a potential barrier will cause its width to expand substantially, then we selected the usage of potential well rather than barrier in our present variational approach.


\begin{figure}[!h]
	\centering
	\textbf{~~~~~~~~~~Double square well potential}\par\medskip
	\includegraphics[width=1\linewidth]{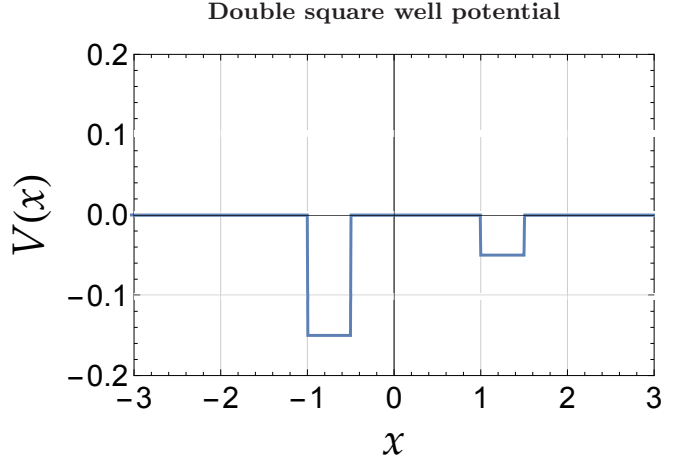}
	\caption{Asymmetric double well potential. We plot the potential, Eq.~\eqref{eq:secIV_potential}, with parameters: $l_{00}=0$, $l_{01}= -0.15$, $l_{02} = -0.05$, $q_1 = -1$, $q_{2} = -0.5$,  $q_{3} = 1$,  $q_{4} = 1.5$.}
	\label{secIV_potential}
\end{figure}

The double square well potential, Eq.~\eqref{eq:secIV_potential}, captures the essential features of the Gaussian and Rosen-Morse double potential barriers where it allows for a unidirectional flow too~\cite{recentpre}.
In Fig.~\ref{secIV_potential}, we plot the potential. In this section we are using the parameters: $g_1 = g_2 =1$, $g_{12} = 0.05$, $l_{00} = 0$, $l_{01} = -0.15$, $l_{02} = -0.05$, $q_1 = -1$, $q_{2} = -0.5$,  $q_{3} = 1$,  $q_{4} = 1.5$, for the variational and numerical calculations unless otherwise noted. In our analysis, we fix the potential position and change the launching point of the vector soliton such that when we set $\beta=-1$ ($\beta$= 1), the BB soliton is coming from the right (left) of the potential in Fig.~\ref{secIV_potential}.

Inserting our variational ansatz Eq.~\eqref{eq:secIV_ansatz}, into the Lagrangian density, Eq.~\eqref{eq:secIV_lagrangian_density}, and integrating with respect to $x$ from  $-\infty$ to $\infty$, results in the effective Lagrangian of the system as a function of the variational parameters,

\begin{align}
\label{eq:secIV_Lagrangian}
L & = -\frac{N}{3 a^2} + \frac{(g_{1}+g_{2})N^2}{6 a} -\frac{1}{3} N \pi^2 a^2 b^2 - \frac{1}{2} N (v^2_{1}+v^2_{2}) \nonumber \\ &
-\frac{1}{6} N \pi^2 a^2 b^{\prime} - N(v_{1} \xi^{\prime}_{1} +v_{2} \xi^{\prime}_{2} + 2 \phi^{\prime}) \nonumber \\ &
+\frac{1}{2} N  \Big \{ 2 l_{00} + l_{01} \Big[-\mathrm{tanh}\Big(\frac{q_{1}+\xi_{1}}{a}\Big) +\mathrm{tanh}\Big(\frac{q_{2}+\xi_{1}}{a}\Big)\Big] \nonumber \\ &
+ l_{02} \Big[-\mathrm{tanh}\Big(\frac{q_{3}+\xi_{1}}{a}\Big) +\mathrm{tanh}\Big(\frac{q_{4}+\xi_{1}}{a}\Big) \Big] \Big \} \nonumber \\ &
+\frac{1}{2} N  \Big \{ 2 l_{00} + l_{01} \Big[-\mathrm{tanh}\Big(\frac{q_{1}+\xi_{2}}{a}\Big) +\mathrm{tanh}\Big(\frac{q_{2}+\xi_{2}}{a}\Big)\Big] \nonumber \\ &
+ l_{02} \Big[-\mathrm{tanh}(\frac{q_{3}+\xi_{2}}{a}) +\mathrm{tanh}\Big(\frac{q_{4}+\xi_{2}}{a}\Big)\Big] \Big \} \nonumber \\ &
+\frac{g_{12} N^2}{a^2} \mathrm{csch}^2 \Big(\frac{\xi_{1}-\xi_{2}}{a}\Big) [a -(\xi_{1}-\xi_{2}) \mathrm{coth} \Big(\frac{\xi_{1}-\xi_{2}}{a}\Big)].
\end{align}

Applying the Euler-Lagrange equations for each variational parameter yields a system of ordinary differential equations that describe their time evolution (See Appendix~\ref{appendix}).

In Fig.~\ref{figva-num}, we plot the BB soliton's trajectory calculated from the variational approach (solid lines) and compare the result to the numerical calculation for soliton velocity $v=0.2$. We find that even when we allow for an internal oscillation between the two components in BB soliton, the agreement is good between the two methods.

\begin{figure}[bt]\centering
	\textbf{Unidirectional segregation by variational and numerical method}\par\medskip
	\includegraphics[scale=0.45]{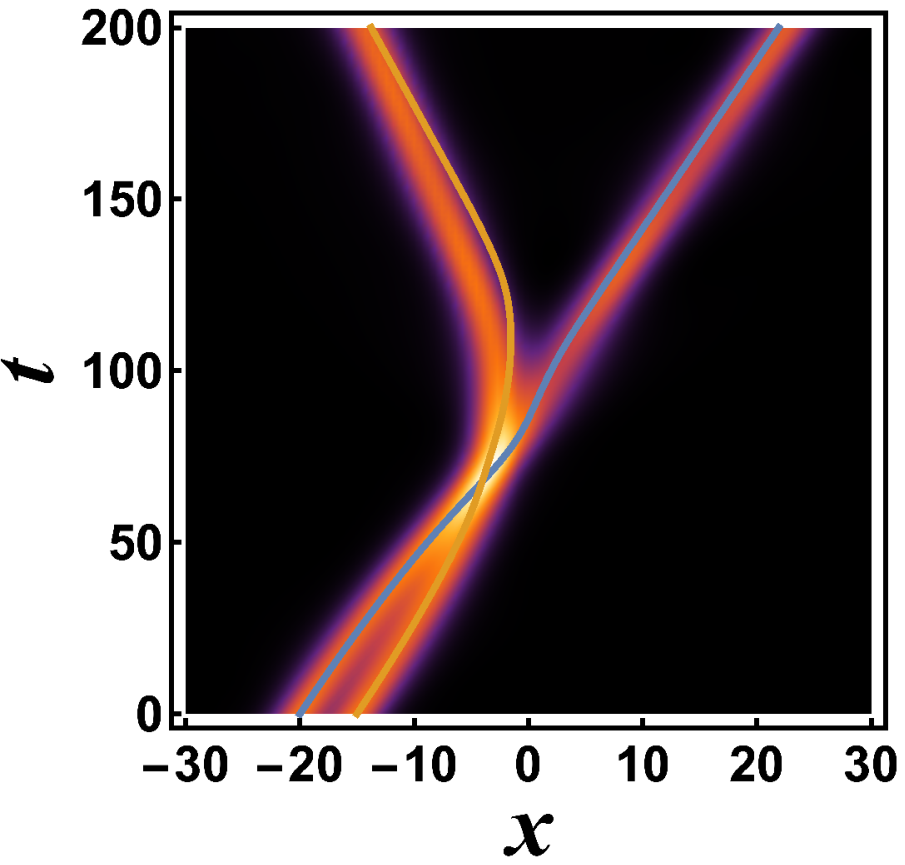}
	\includegraphics[scale=0.45]{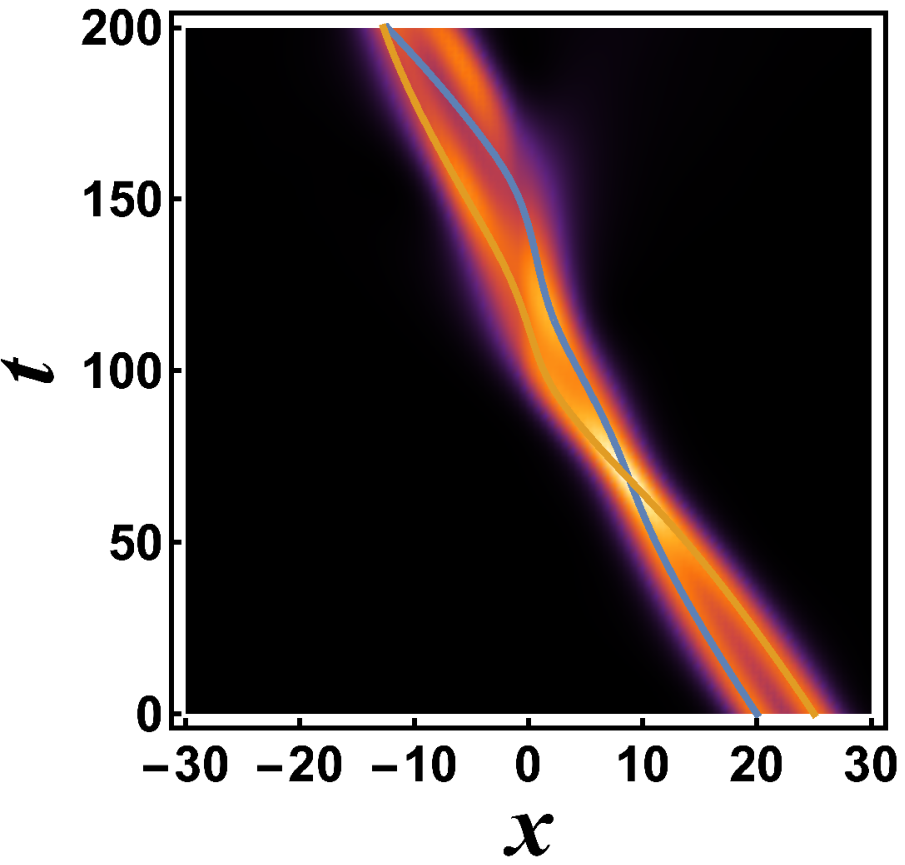}
	\caption{Propagation of components through asymmetric double square well potential, see Fig.~\ref{secIV_potential}, for $g_{12} = 0.05$ at $v$= 0.2. Solid lines are results from variational calculation. Intensity plots are results of numerical calculation. Unidirectional segregation of composite BB soliton is achieved with an excellent agreement between variational and numerical method.}
	\label{figva-num}
\end{figure}

The transport coefficients, given in Eq.~\eqref{eq:secII_R,T,L}, can be used to calculate also the reflectance, R, transmittance, T, and trapping, L, of the BB soliton obtained from the variational calculations in terms of velocity. In this case, the integration limits will be from the box edge to the potential position for R and T depending on whether the soliton comes from right or left. For the trapping coefficient, the integration covers the potential area only.

\begin{figure}[bt]\centering
\textbf{Variational calculation for right moving two component solitons}\par\medskip
	\includegraphics[scale=0.75]{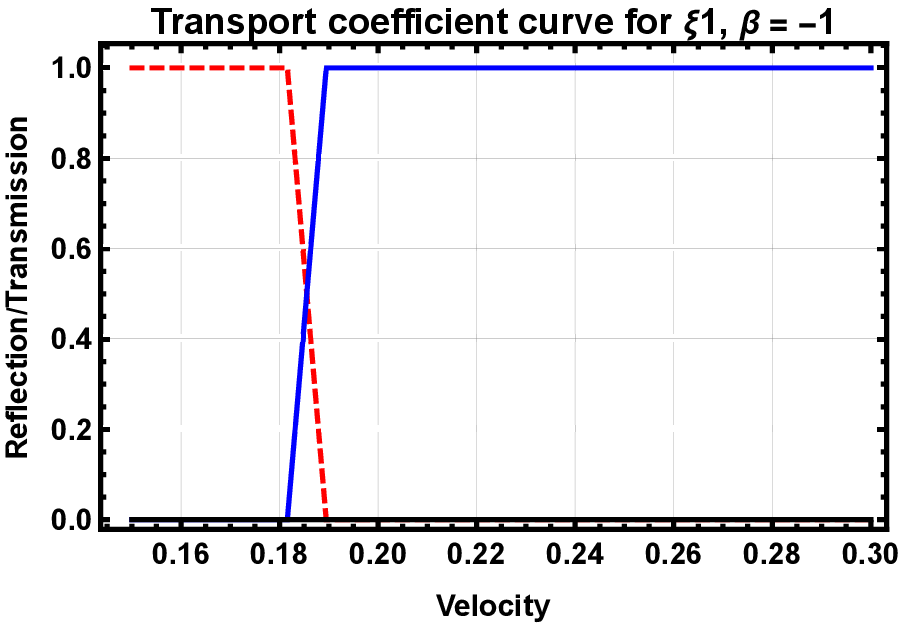}
	\includegraphics[scale=0.75]{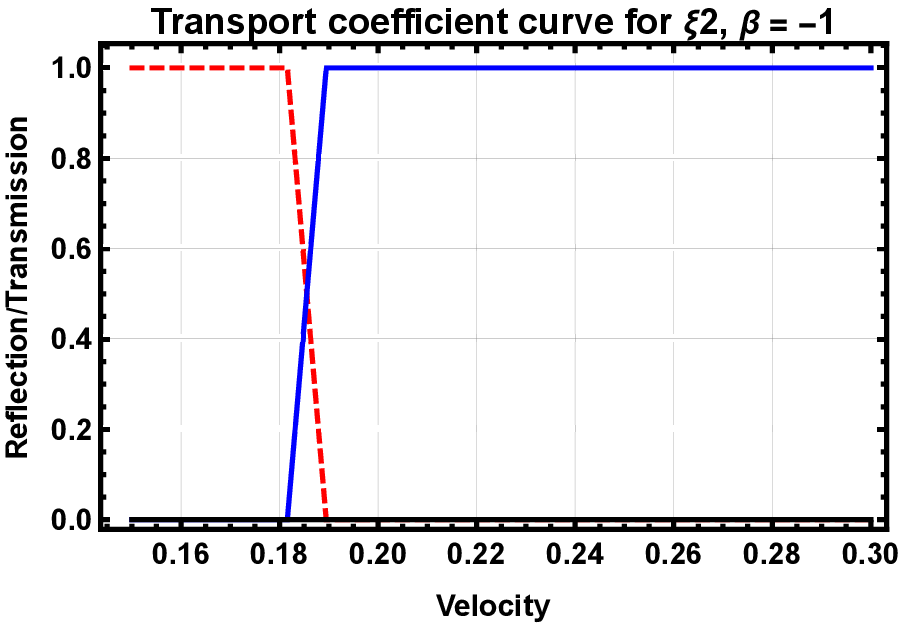}
	\caption{Analytically calculated transport coefficients in terms of velocity for the propagation of two right moving components through asymmetric double square potential well for $g_{12} = 0.05$. Upper panel shows first component while lower panel shows second component of the initial wave functions \eqref{eq:secIV_ansatz}. There is no segregation in this direction.}
	\label{TRL-VA}
\end{figure}

\begin{figure}[bt]\centering
\textbf{Variational calculation for left moving two component solitons}\par\medskip
	\includegraphics[scale=0.75]{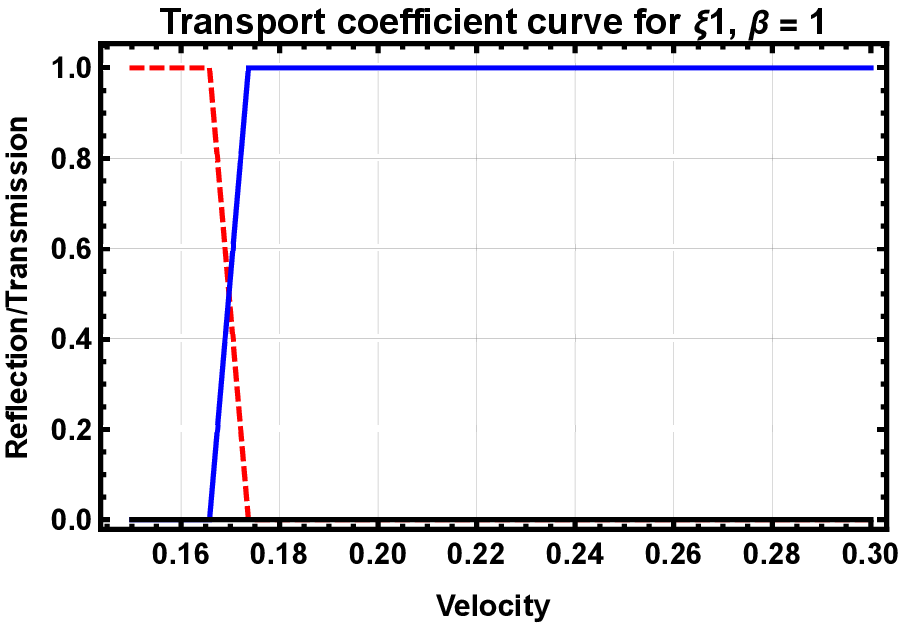}
	\includegraphics[scale=0.75]{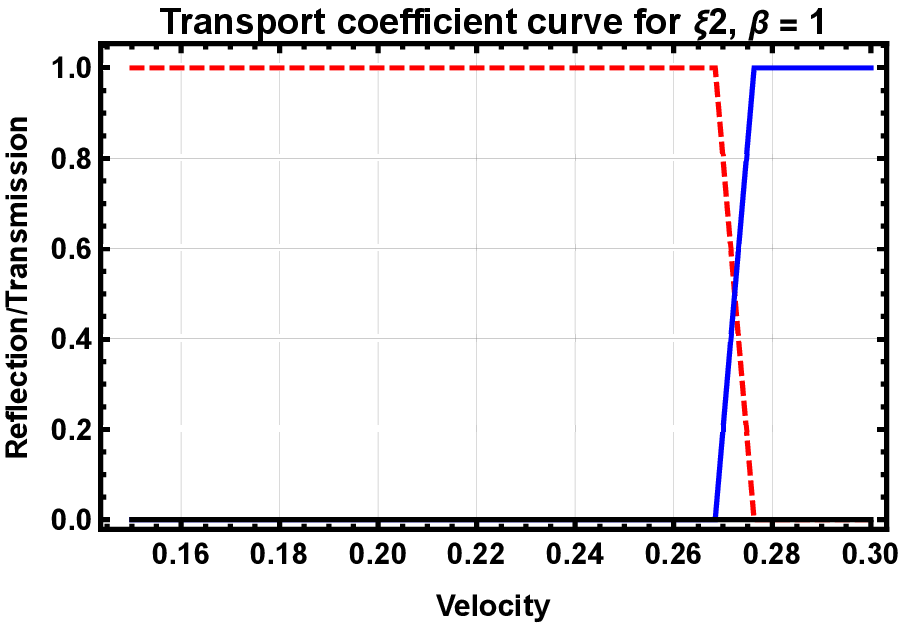}
	\caption{Analytically calculated transport coefficients in terms of velocity for the propagation of two left moving components through asymmetric double square potential well for $g_{12} = 0.05$. Upper panel shows first component while lower panel shows second component of the initial wave functions \eqref{eq:secIV_ansatz}. Segregation is observed in this direction.}
	\label{TRL-VA2}
\end{figure}

In Fig.~\ref{TRL-VA}, we see that for $\beta=-1$ (i.e., sending the BB soliton from the right), the transport coefficient curves are the same for the two components. That is, the two components move through the potential without segregation. But in Fig.~\ref{TRL-VA2}, for $\beta=1$, (i.e., sending the BB soliton from the left) we find that there is a window for the splitting of the two components when the velocity range is $0.17 < v < 0.27$. For example, a BB soliton with $v=0.2$ as in Fig.~\ref{figva-num}, should split in one direction and not the other as seen from Figs.~\ref{TRL-VA} and \ref{TRL-VA2}.

\begin{figure}[bt]\centering
\textbf{Numerical calculation for right moving two component solitons}\par\medskip
	\includegraphics[scale=0.75]{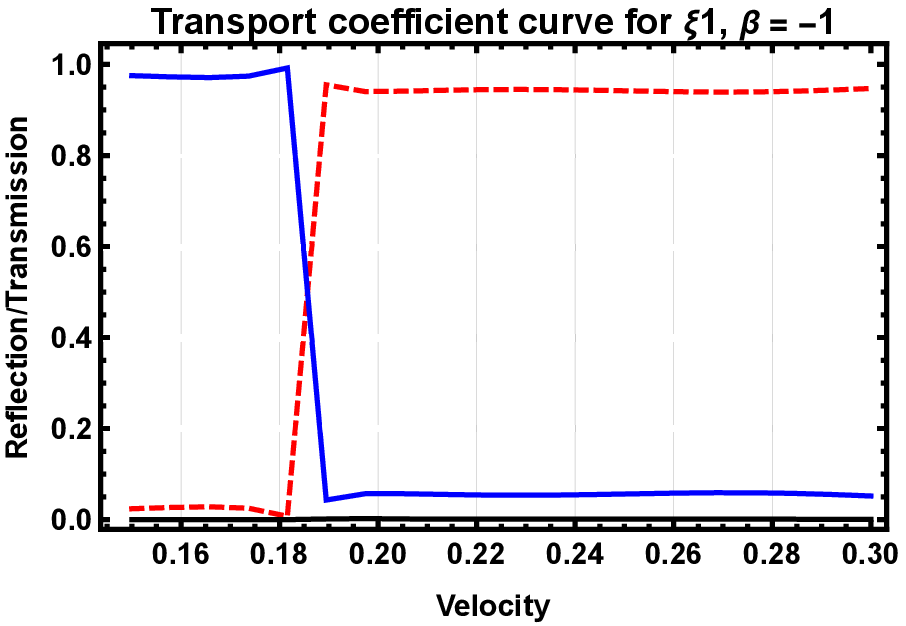}
	\includegraphics[scale=0.75]{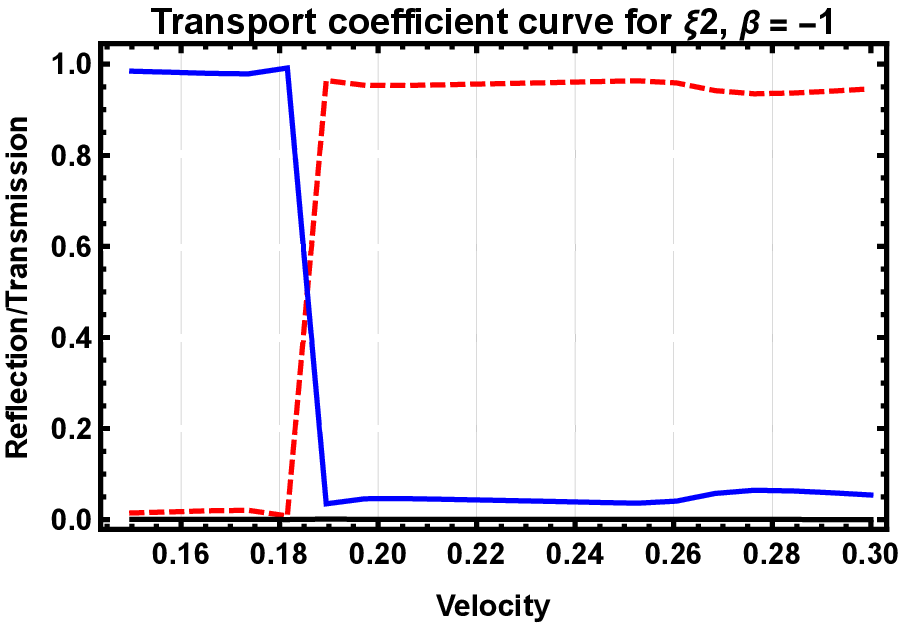}
	\caption{Numerically calculated transport coefficients in terms of velocity for the propagation of two right moving components through asymmetric double square potential well for $g_{12} = 0.05$. Upper panel shows first component while lower panel shows second component of the initial wave functions \eqref{eq:secIV_ansatz}. There is no segregation in this direction.}
	\label{TRL-num2}
\end{figure}

\begin{figure}[bt]\centering
\textbf{Numerical calculation for left moving two component solitons}\par\medskip
	\includegraphics[scale=0.75]{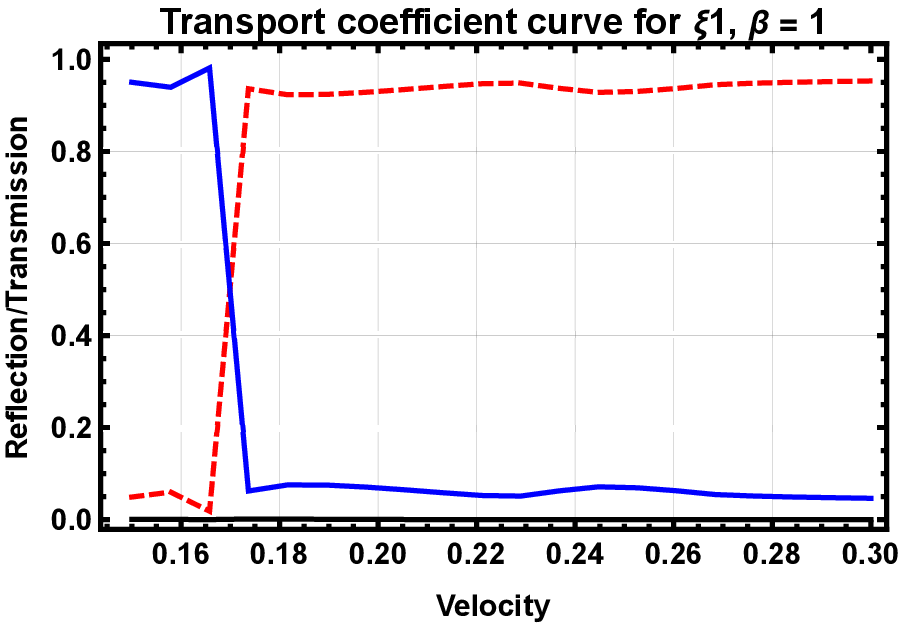}
	\includegraphics[scale=0.75]{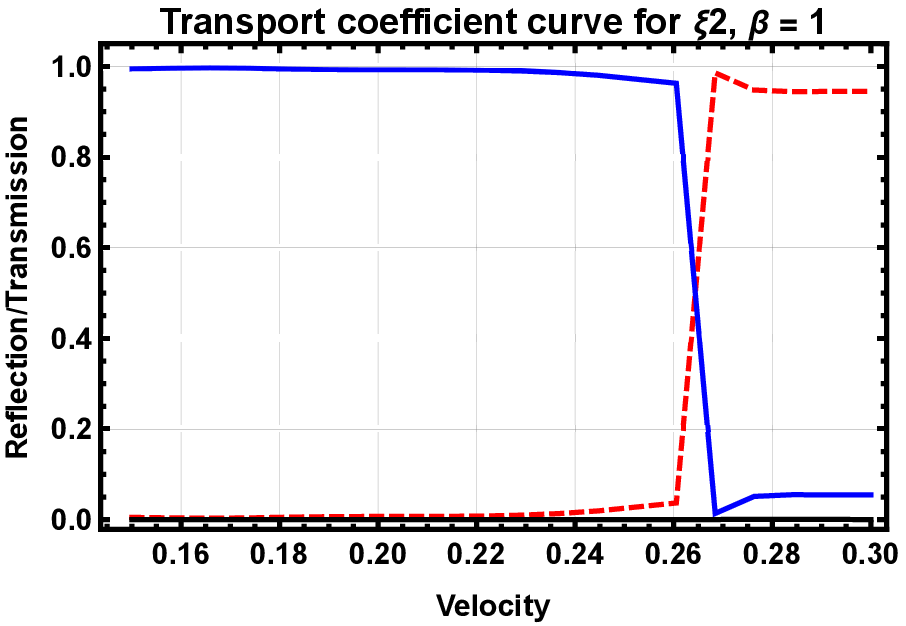}
	\caption{Numerically calculated transport coefficients in terms of velocity for the propagation of two left moving components through asymmetric double square potential well for $g_{12} = 0.05$. Upper panel shows first component while lower panel shows second component of the initial wave functions \eqref{eq:secIV_ansatz}. Segregation is observed in this direction.}
	\label{TRL-num}
\end{figure}

In Figs.~\ref{TRL-num2} and~\ref{TRL-num}, we plot the transport coefficients for the numerical simulation of a BB soliton interacting with the same potential. We find a good agreement between the results obtained from the numerical simulation compared to the variational analysis predictions.

\section{Results and Discussion}\label{results}

 Here, we summarize our main results: Firstly, we achieved unidirectional flow of composite BB soliton passing through asymmetric double potential barriers in the absence of nonlinear coupling ($g_{12} = 0$). We considered two types of potentials for our numerical simulations. We choose  asymmetric RM and Gaussian double barrier potentials. Our results for the unidirectional flow are achieved with potential barriers rather than potential wells, as in Ref.~\cite{usamaasad}. Secondly, we achieved also a unidirectional flow of composite BB soliton in the presence of attractive mean field inter-component coupling, i.e., $g_{12} > 0$ using the two selected potentials. Interestingly, we find a change in polarity in unidirectional flow for $g_{12} > 0.329$ for RM potential and $g_{12} > 0.316$  for Gaussian potential. Both components of the BB soliton remained  invariant throughout the propagation. Thirdly, we found segregation or splitting of composite BB soliton into its two components in the presence of repulsive mean field interaction coupling, $g_{12} < 0$, through both types of considered potentials. We also observed the shuttle motion between the barriers in our study. However, we restricted ourselves to the parameter regime for the unidirectional flow and segregation.\\
In addition, we achieved unidirectional segregation by varying nonlinear strength of one of the components ($g_1$) in the absence of nonlinear interaction coupling ($g_{12} = 0$). Further, we realised that an extremely small value of repulsive coupling ($g_{12} = -0.004$) can destroy the unidirectional segregation for such a case and results in bidirectional segregation. We also found unidirectional segregation of composite BB soliton using variational calculations and compared our results with numerical computations. We obtained an excellent agreement between analytical and numerical analysis. Our results are applicable to all-optical data processing and we strongly believe that this work is an important contribution to the effort made towards the realization of optical devices e.g., optical diode, interferometer. Another application would be the realization of quantum logic gates where two solitons are usually needed to code a qubit. The scattering dynamics of the two solitons through the potential may set up a protocol equivalent to a two-qubit logic gate such as CNOT gate.

\section*{Acknowledgment} The authors acknowledge the support of UAE University through grants UAEU-UPAR(4) 2016 and UAEU-UPAR(11) 2019. H.B. acknowledges the support of KFUPM under research group project DF191053.

\appendix
\numberwithin{equation}{section}
\numberwithin{figure}{section}
\numberwithin{table}{section}

 \section{\\ Propagation of bright-bright solitons through asymmetric Gaussian potential barriers}\label{appendix1}

\begin{figure}[bt]\centering
	\includegraphics[scale=0.3]{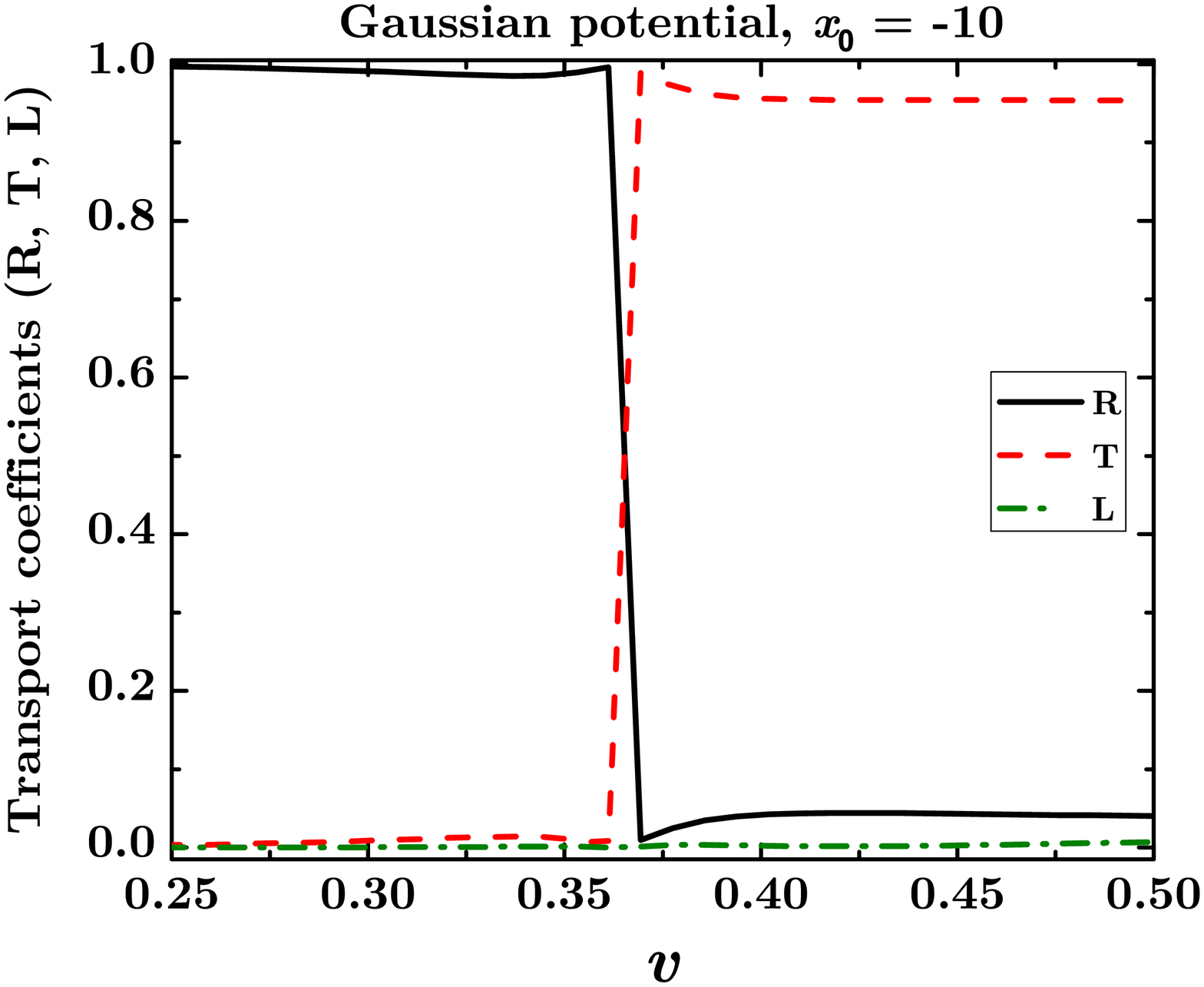}
	\includegraphics[scale=0.3]{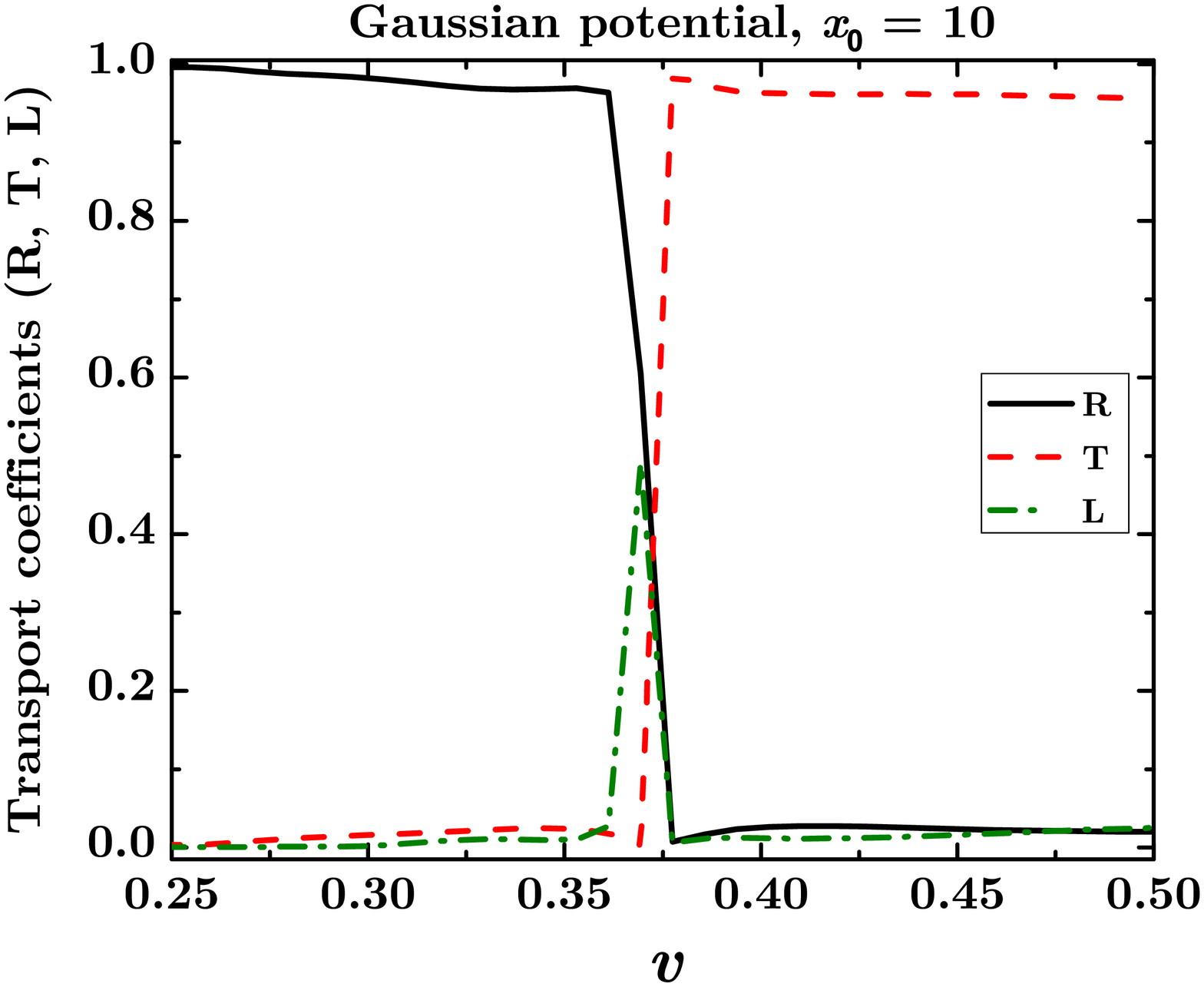}
	\caption{Transport coefficients in terms of velocity for the propagation of the $\psi_1$ component through asymmetric Gaussian barriers for $g_1 = g_2 = 1$, $g_{12} = 0$ from $x_0$ = -10 (upper panel) and $x_0$ = 10 (lower panel).}
	\label{fig:v}
\end{figure}

The transport coefficients for the Gaussian double potential barriers are shown in Fig.~\ref{fig:v}. For the right moving BB solitons scattered through Gaussian barriers, the critical velocity $v_c$ = 0.369 whereas that for the left moving BB solitons $v_c$ = 0.377. Furthermore, the velocity window for our new scheme with Gaussian barriers (0.365 $\leq v \leq$ 0.372) is very much comparable to the previous study of Gaussian potential wells \cite{usamaasad} ( 0.304 $\leq v \leq$ 0.310).

\subsection{Unidirectional flow for uncoupled components with  $g_{12} = 0$}

\begin{figure}[bt]\centering
\textbf{Gaussian potential}\par\medskip
	\includegraphics[scale=0.45]{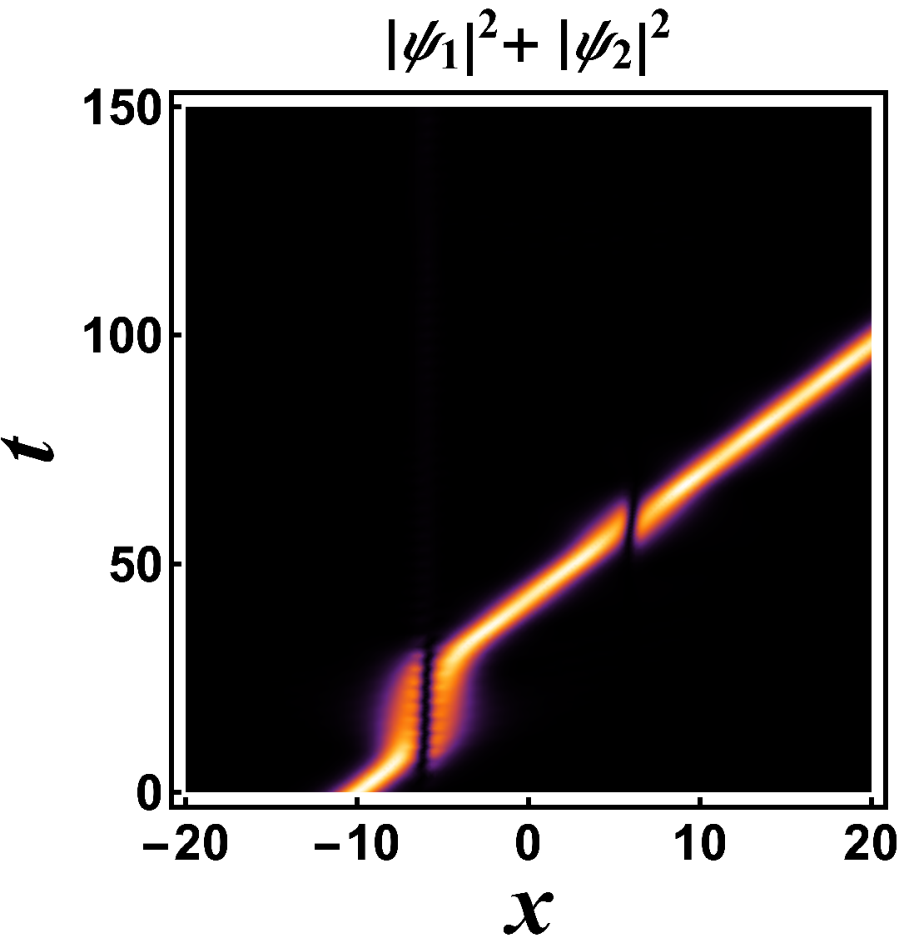}
	\includegraphics[scale=0.45]{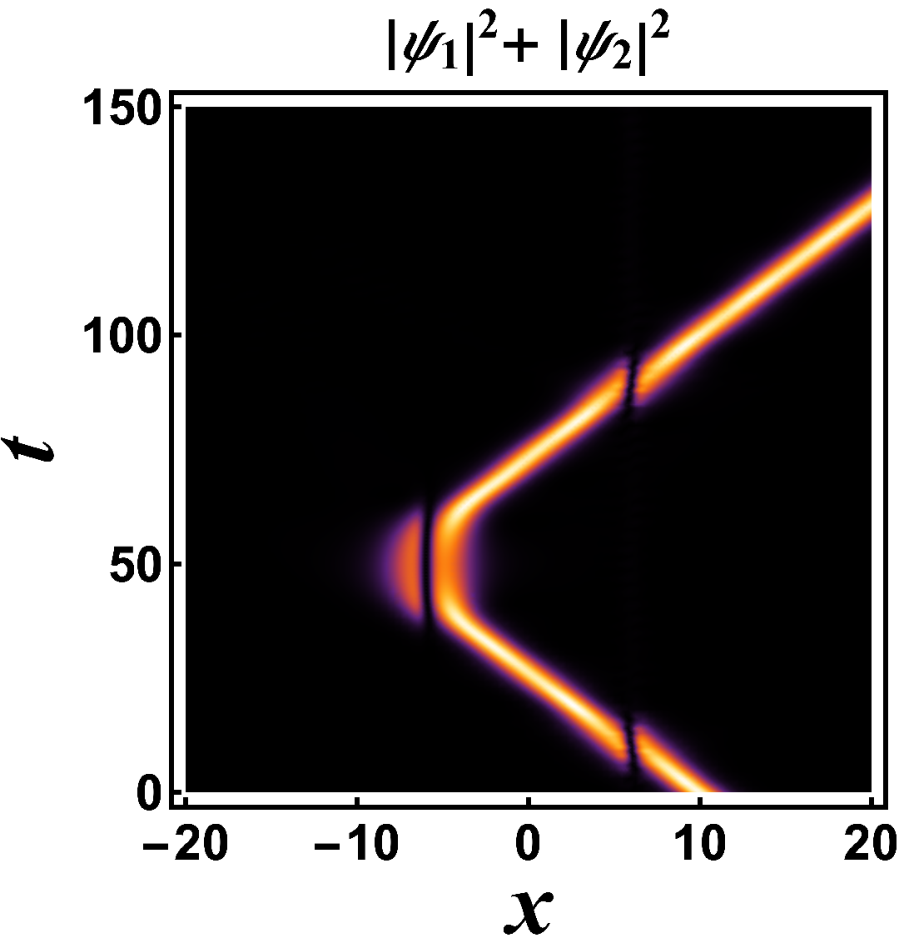}
	\caption{Propagation of composite BB soliton through asymmetric Gaussian potential barriers for $g_{12} = 0$ with $g_1 = g_2 = 1$ at $v$ = 0.368. Both components are identical. Left and right subfigures are results of initial propagation from $x_0$ = -10 and $x_0$ = 10, respectively.}
	\label{fig3g}
\end{figure}

 The unidirectional flow is also achieved for uncoupled components of BB solitons through Gaussian double potential barriers. Fig.~\ref{fig3g} shows the propagation of the BB soliton components $\psi_1$ and $\psi_2$ with a critical velocity $v_c$ = 0.368, incident from $x_0 = \mp10$ through asymmetric Gaussian double barrier potentials.

\subsection{Polarity reversal in unidirectional flow with $g_{12} > 0$}

\begin{figure}[bt]\centering
\textbf{Gaussian potential, $x_0$ = -10}\par\medskip
	\includegraphics[scale=0.55]{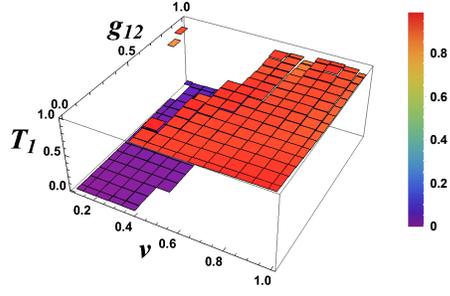}
	\includegraphics[scale=0.55]{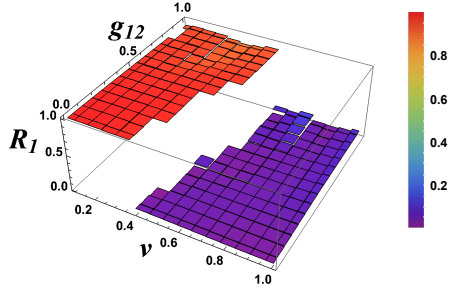}
\par\bigskip
\textbf{Gaussian potential, $x_0$ = 10}\par\medskip
\includegraphics[scale=0.55]{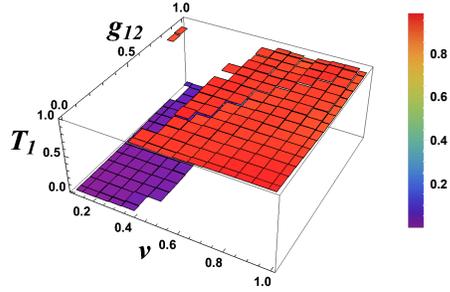}
	\includegraphics[scale=0.55]{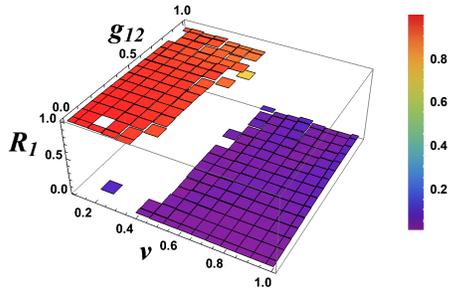}
	\caption{Transmission and reflection coefficients of the component $\psi_1$  propagating from $x_0$ = -10 (upper two) and $x_0$ = 10 (lower two) through Gaussian barriers versus $v$ and $g_{12}$. Other parameters are $g_1 = g_2 = 1$.}
	\label{fig5g}
\end{figure}

 The important characteristic of \textit{polarity reversal} is also observed through the propagation of BB solitons via the asymmetrical Gaussian double potential barriers. Fig.~\ref{fig5g} describes the transmission and reflection coefficients of the component $\psi_1$ passing through the RM potential barriers from $x_0$= $\pm$10, versus the velocity and positive mean field coupling.

\begin{figure}[bt]\centering
	\textbf{Gaussian potential}\par\medskip
	\includegraphics[scale=0.45]{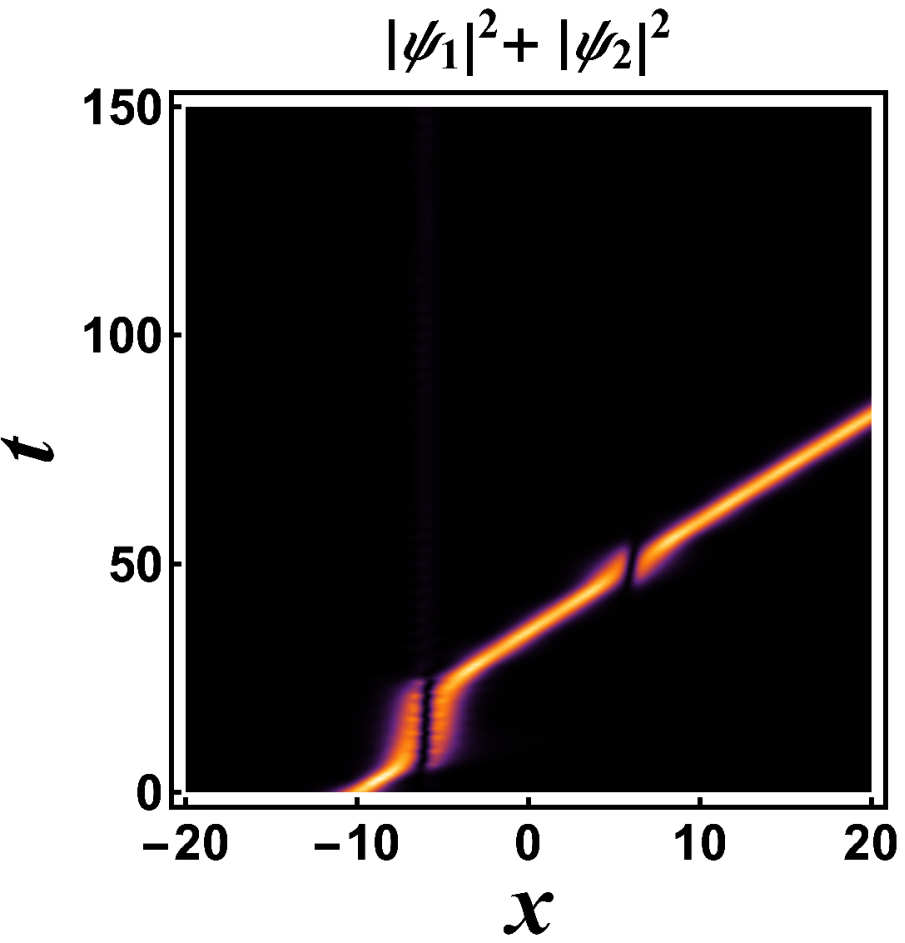}
	\includegraphics[scale=0.45]{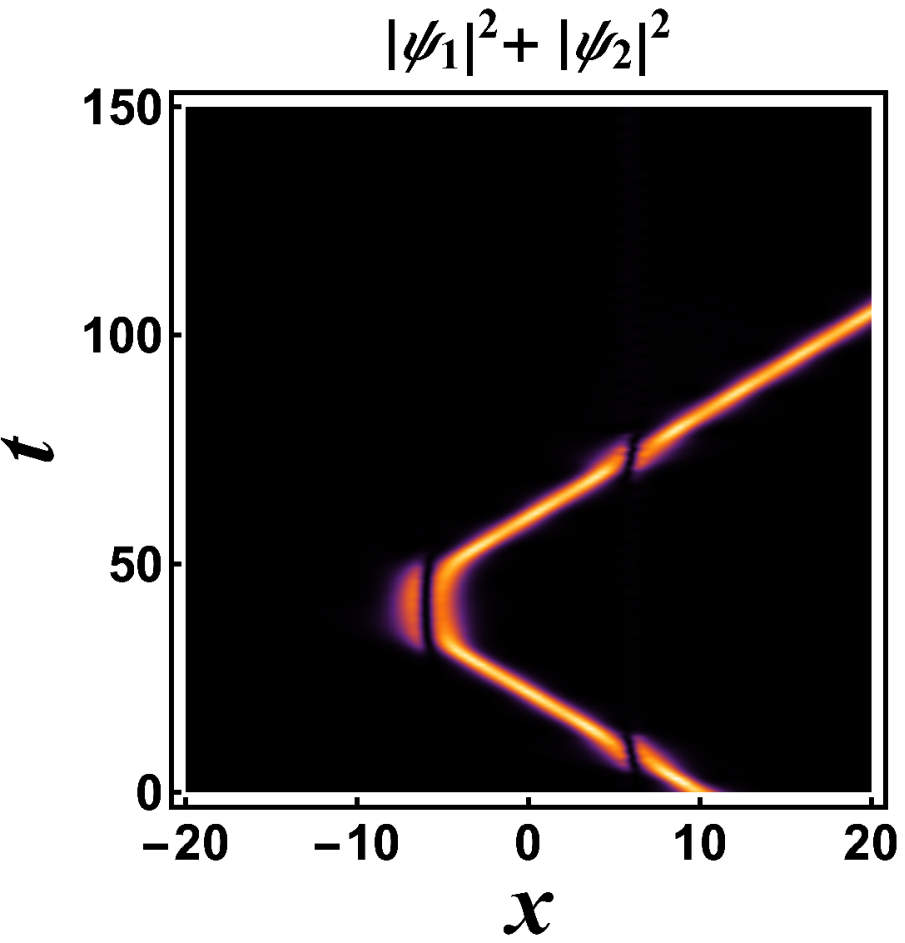}
	\caption{Propagation of composite BB soliton through asymmetric Gaussian potential barriers for $g_{12} = 0.3$ with $g_1 = g_2 = 1$ at $v$ = 0.452. Both components are identical. Left and right subfigures are results of initial propagation from $x_0$ = -10 and $x_0$ = 10, respectively.}
	\label{fig6g}
\end{figure}

From the spatiotemporal plots, it is observed that for lower coupling $g_{12} \le 0.312$, both components
exhibit the diode behavior with "right polarity" as shown in Fig.~\ref{fig6g} obtained for g12 = 0.3 which is similar to the one achieved in the case g12 = 0.

\begin{figure}[bt]\centering
	\textbf{Gaussian potential, $v$ = 0.454}\par\medskip
	\includegraphics[scale=0.45]{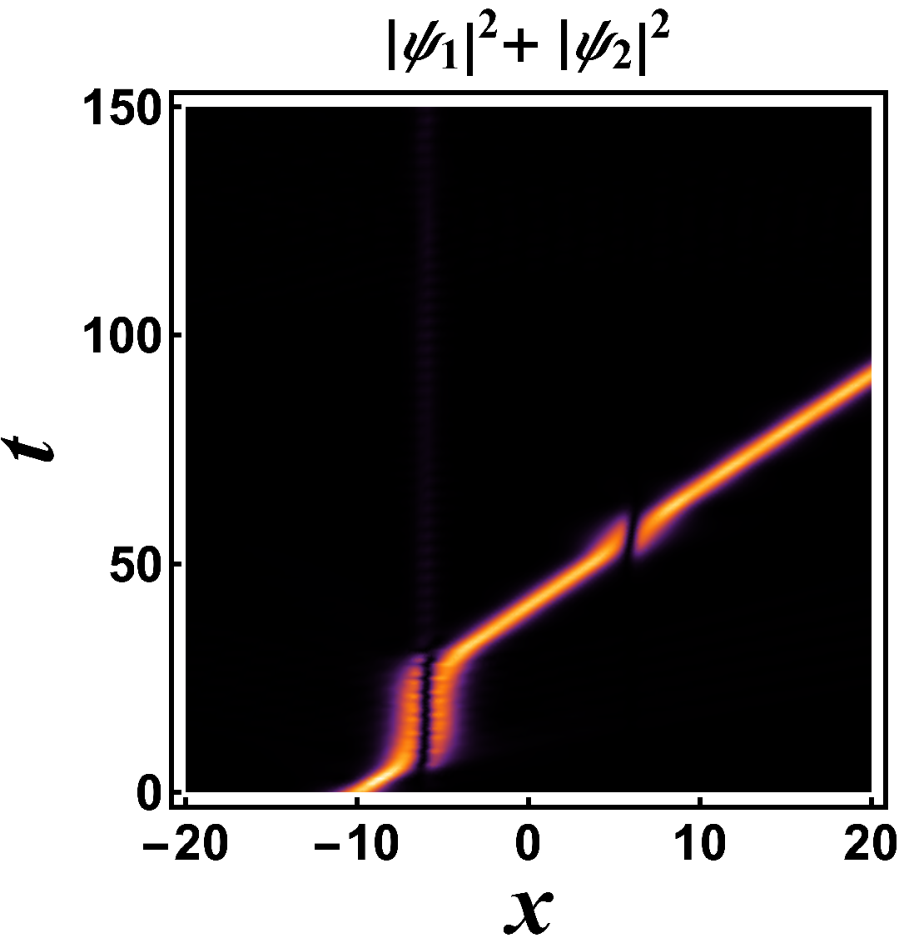}
	\includegraphics[scale=0.45]{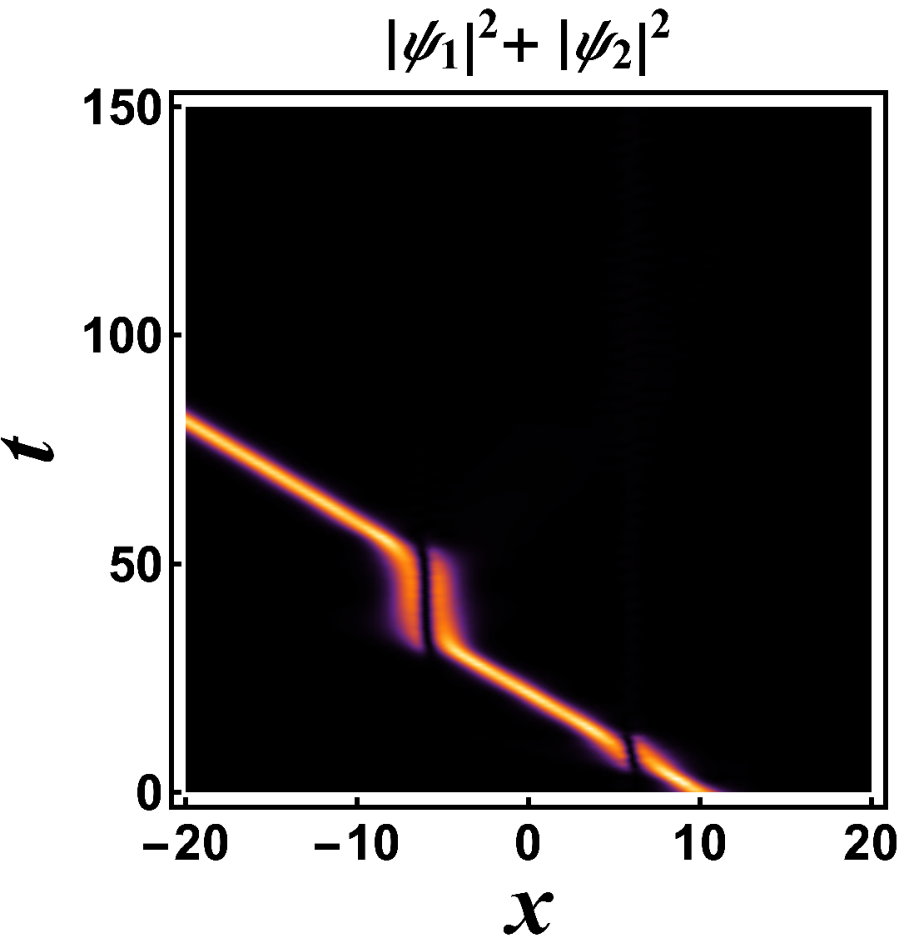}
	\par\bigskip
	\textbf{Gaussian potential, $v$ = 0.453}\par\medskip
	\includegraphics[scale=0.45]{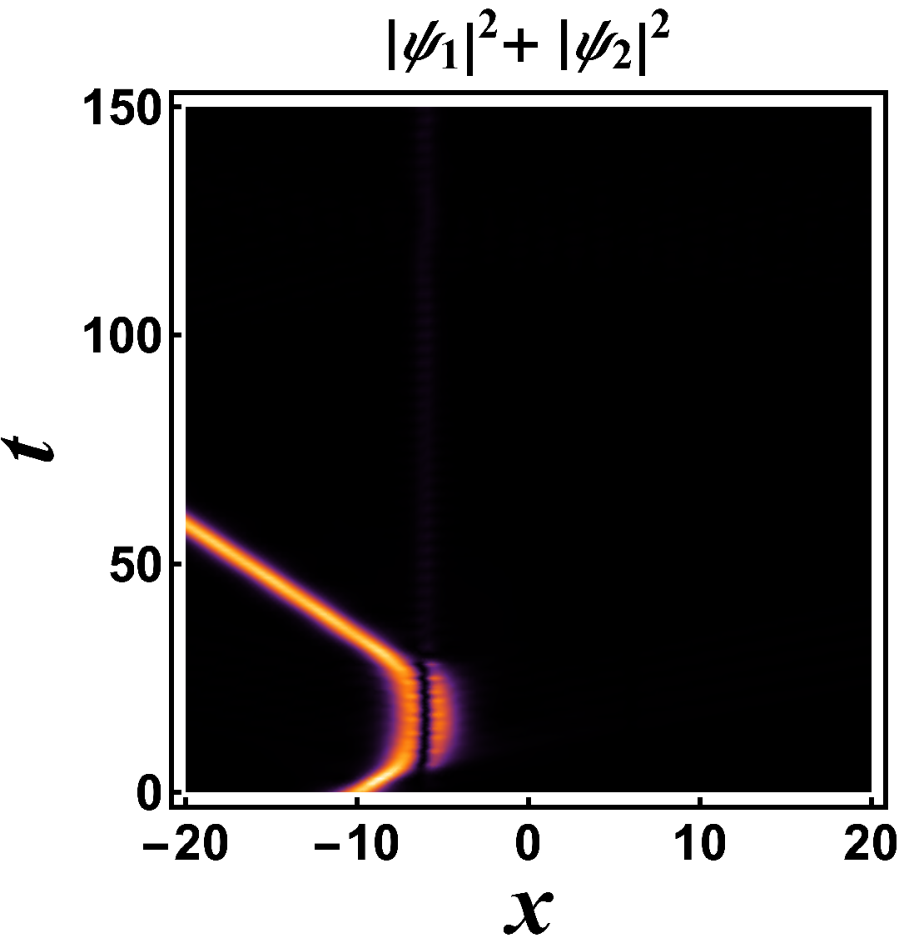}
	\includegraphics[scale=0.45]{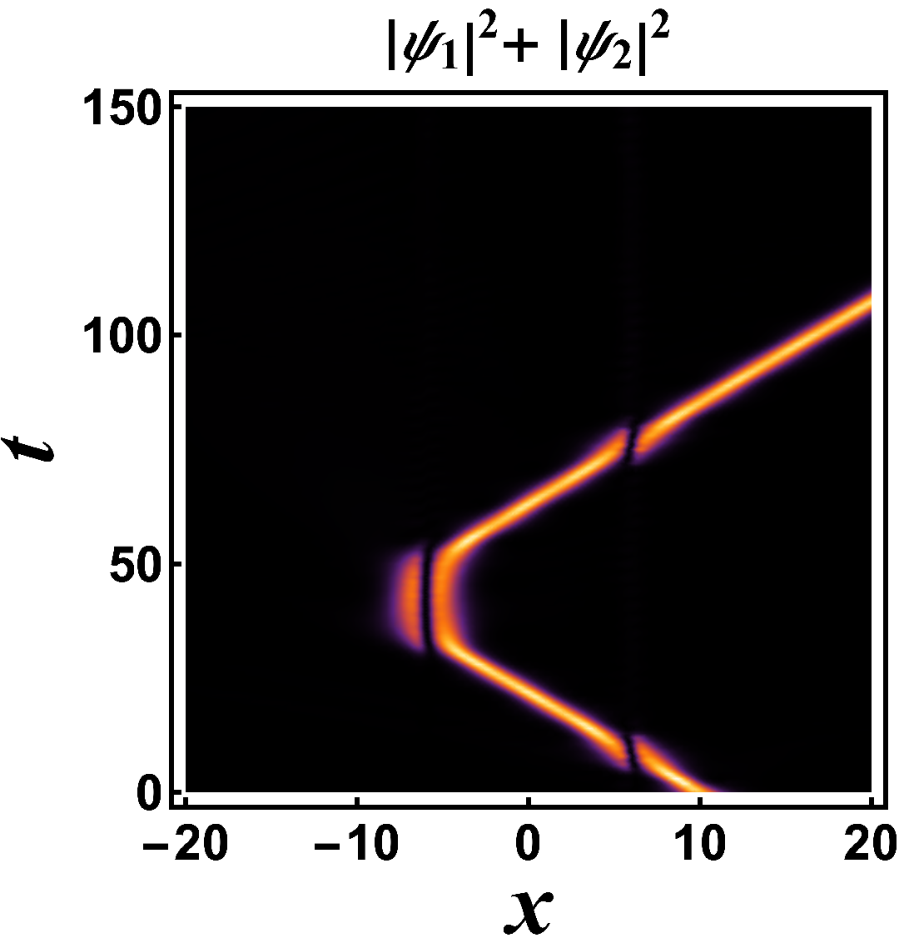}
	\caption{Propagation of composite BB soliton through asymmetric Gaussian potential barriers for $g_{12} = 0.315$ with $g_1 = g_2 = 1$. Upper panel shows full transmission at $v$ = 0.454 while lower panel shows full reflection at $v$ = 0.453, from both left and right directions. There is no unidirectional flow in the range of coupling strength 0.313 $\le$ $g_{12}$ $\le$ 0.316.}
	\label{fig7g}
\end{figure}

\begin{figure}[bt]\centering
	\textbf{Gaussian potential}\par\medskip
	\includegraphics[scale=0.45]{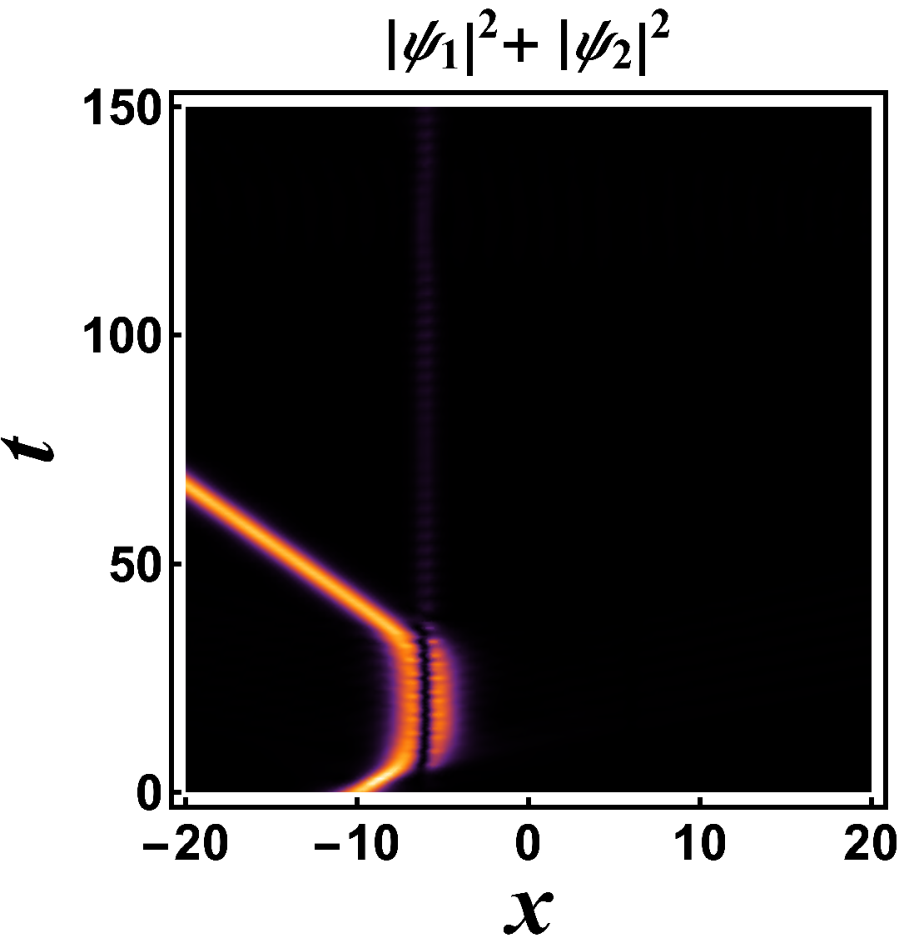}
	\includegraphics[scale=0.45]{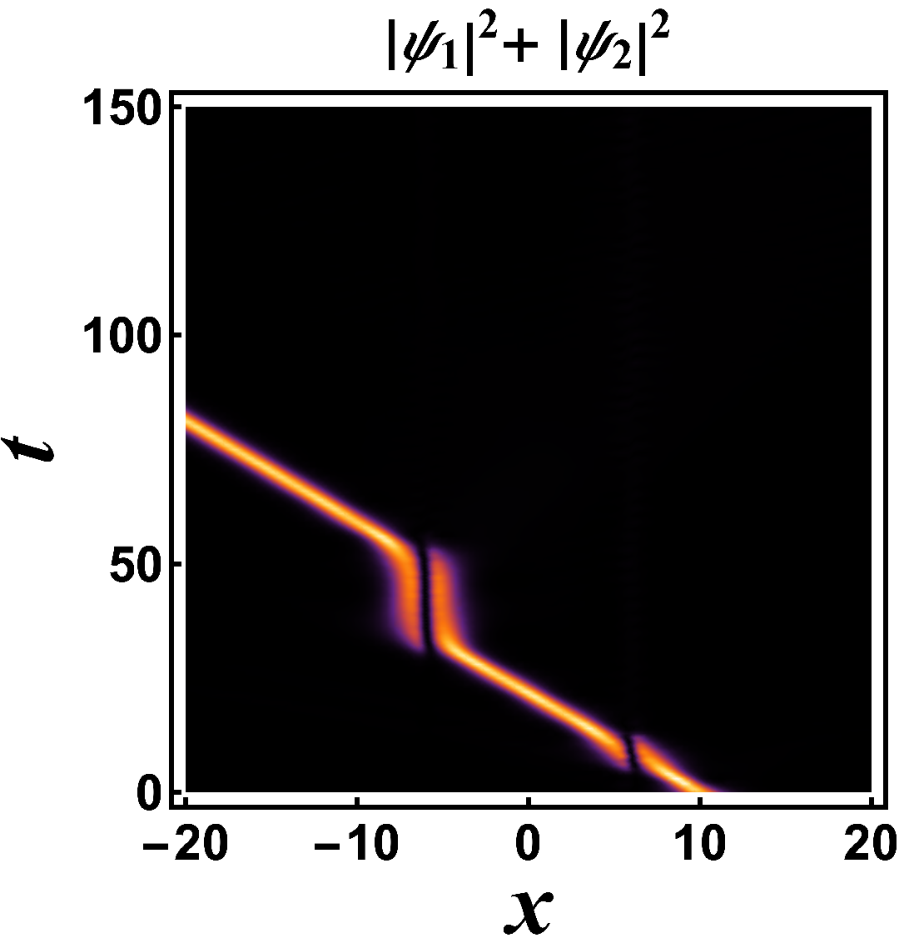}
	\caption{Propagation of composite BB soliton through asymmetric Gaussian potential barriers for $g_{12} = 0.317$ with $g_1 = g_2 = 1$ at $v$ = 0.454. Both components are identical. Left and right subfigures are results of initial propagation from $x_0$ = -10 and $x_0$ = 10, respectively. The polarity reversal phenomenon in unidirectional flow is achieved by comparing with Fig.~\ref{fig6g}}
	\label{fig8g}
\end{figure}

\begin{figure}[bt]\centering
	\textbf{~~~~~~~~Polarity Reversal through Gaussian potential}\par\medskip
	\includegraphics[scale=0.55]{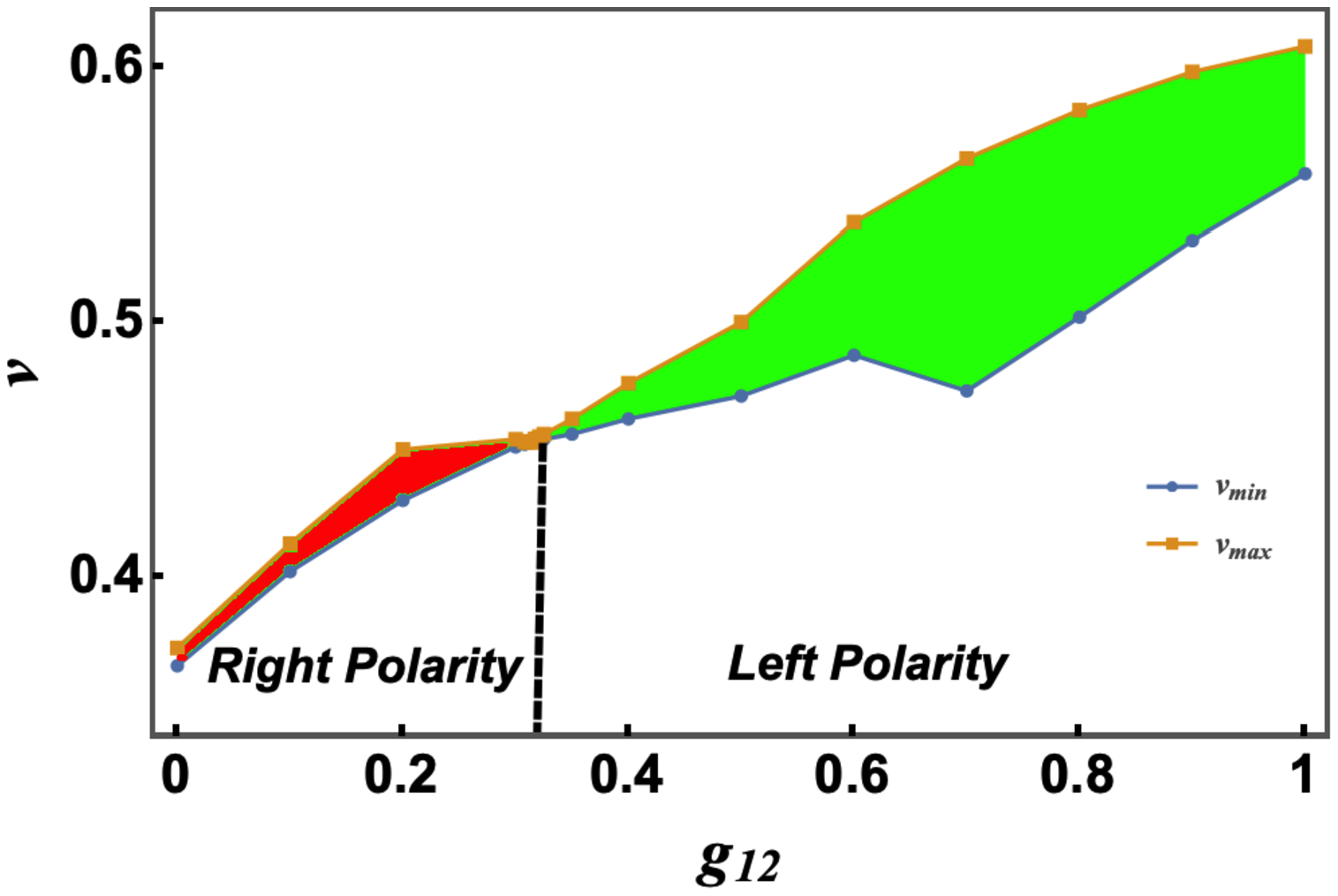}
	\caption{Borders of velocity window for the unidirectional flow ($v_{min}, v_{max}$) vs $g_{12}$ through the Gaussian potential barriers with $g_1 = g_2 = 1$. Full reflection for $v$ $\leq$ 0.453 and full transmission for $v$ $\geq$ 0.454 is obtained with the range of coupling strength 0.313 $\le$ $g_{12}$ $\le$ 0.316, hence no unidirectional flow is observed at this specific range of $g_{12}$. Away from this point of $g_{12} = 0.316$, we find polarity reversal in unidirectional flow. The shaded region shows the velocity window for the unidirectional flow. The red color shows the right polarity while the green color shows the left polarity of the unidirectional flow. The data used to generate this figure is listed in Table~\ref{gaussiantable}.}
	\label{fig:gaussian-reversal}
\end{figure}

For Gaussian potential barriers, we observe no unidirectional flow in the range of interaction coupling $0.313 \leq g_{12} \leq 0.316$ where the composite BB solitons propagating from both directions show full transmission for $v \geq 0.457$, while it exhibits maximum reflection for both right and left moving composite BB solitons for $v$ $<$ 0.457 as shown in Fig.~\ref{fig7g}. The velocity window for the diode functionality at different $g_{12}$ values are tabulated in Table \ref{gaussiantable}. Further, for $g12 \ge 0.317$, exactly the reverse phenomena is achieved through Gaussian potential barriers and the polarity of unidirectional flow is reversed from right to left polarity which can be seen by comparing Fig.~\ref{fig8g} with Fig.~\ref{fig6g}. This phenomena is purely due to the increase in $g_{12}$ above certain
critical value 0.316, which is demonstrated by Fig.~\ref{fig:gaussian-reversal}.

\subsection{Unidirectional segregation with $g_{12} < 0$}

\begin{figure}[bt]\centering
		\textbf{Gaussian potential, $x_0$ = -10}\par\medskip
		\includegraphics[scale=0.55]{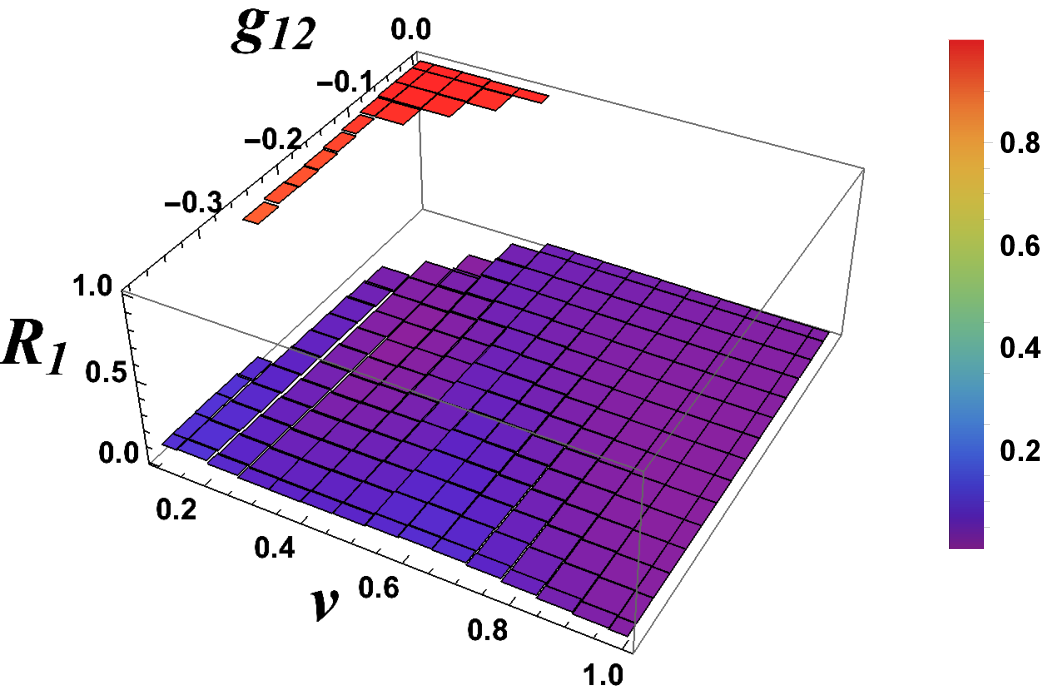}
		\includegraphics[scale=0.55]{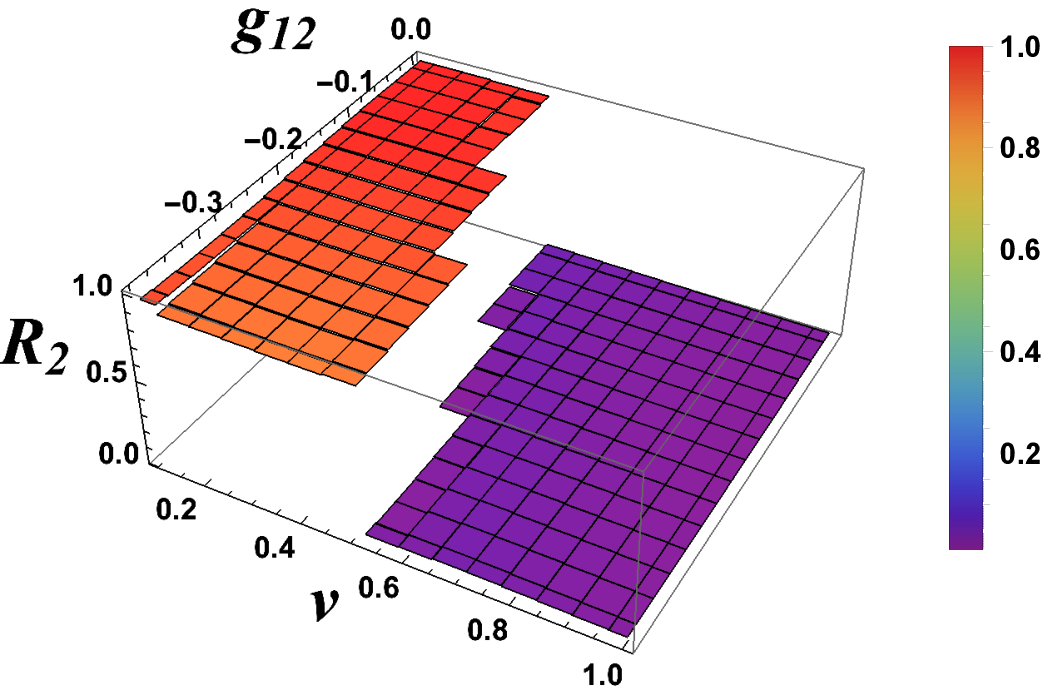}
		\par\bigskip
		\textbf{Gaussian potential, $x_0$ = 10}\par\medskip
		\includegraphics[scale=0.55]{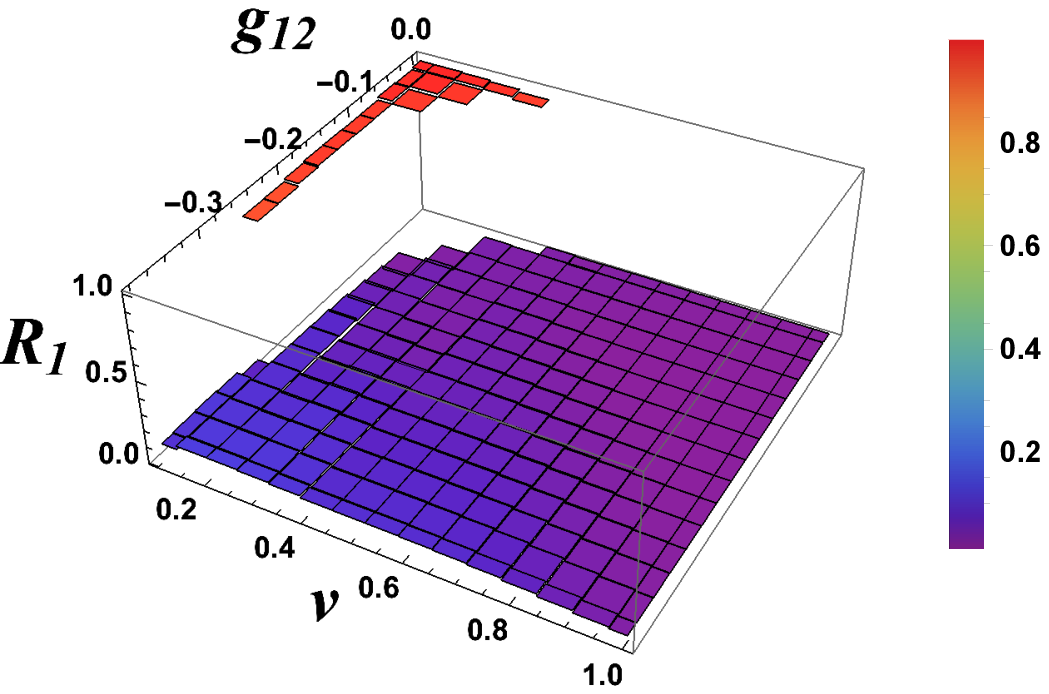}
		\includegraphics[scale=0.55]{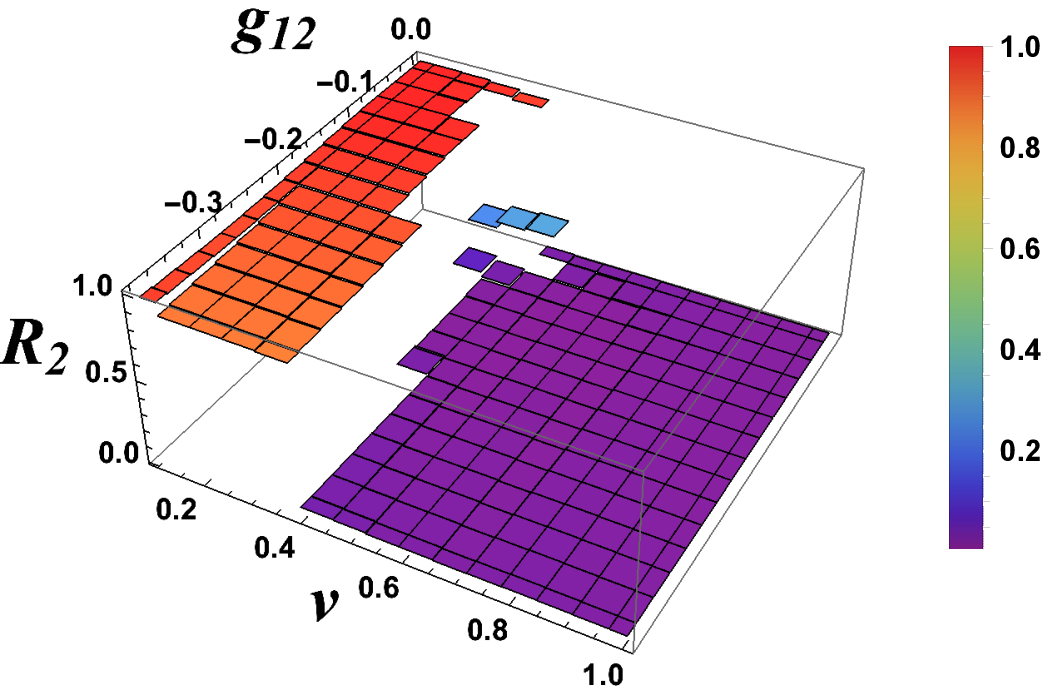}
		\caption{Reflection coefficients of the components $\psi_1$ and $\psi_2$ propagating through Gaussian potential barriers from $x_0$ = -10 (upper two) and $x_0$ = 10 (lower two) versus $v$ and $g_{12}$. Other parameters are $g_1 = g_2 = 1$.}
\label{fig9g}
\end{figure}

Almost the same behavior is observed in the variation of reflection coefficients for the BB soliton components propagating through Gaussian potential barriers from $x_0$ = $\pm$ 10 versus velocity and $g_{12}$ values which is shown in Fig.~\ref{fig9g}.

\begin{table*}[!h]
	\textbf{Gaussian Potential}
	\begin{center}
		\begin{tabular}{ |c|c|c| }
			\hline
			\multicolumn{3}{|c|}{\textbf{Unidirectional flow with composite BB solitons}} \\
			\hline
			\textbf{Interaction strength} & \multicolumn{2}{|c|}{\textbf{Velocity window}} \\
			\textbf{$g_{12}$} & \multicolumn{2}{|c|}{\textbf{$ v_{min} \le v \le v_{max}$}} \\
			\hline\hline
			0 & 0.365 $\le$ $v$ $\le$ 0.372&\cellcolor{red}\\\cline{1-2}
			0.1 & 0.402 $\le$ $v$ $\le$ 0.413&\cellcolor{red}\\\cline{1-2}
			0.2 & 0.43 $\le$ $v$ $\le$ 0.45 &\cellcolor{red}\\\cline{1-2}
			0.3 & 0.451 $\le$ $v$ $\le$ 0.454&\cellcolor{red}\\\cline{1-2}
			0.308 & 0.452 $\le$ $v$ $\le$ 0.453&\cellcolor{red}\\\cline{1-2}
			0.309-0.312 & 0.453&\cellcolor{red}\multirow{-6}{*}{\rotatebox{90}{\textit{\textbf{Right polarit}y}\,}}\\
			\hline\hline
			\textbf{0.313-0.316} &\multicolumn{2}{|c|}{ \textbf{no unidirectional flow}} \\
			\hline\hline	
			0.317-0.320 & 0.454&\cellcolor{green}\\\cline{1-2}
			0.321 & 0.454 $\le$ $v$ $\le$ 0.455&\cellcolor{green} \\\cline{1-2}
			0.325 &  0.454 $\le$ $v$ $\le$ 0.456&\cellcolor{green} \\\cline{1-2}
			0.35 &  0.456 $\le$ $v$ $\le$ 0.462&\cellcolor{green}\\\cline{1-2}
			0.4 & 0.462 $\le$ $v$ $\le$ 0.476&\cellcolor{green} \\\cline{1-2}
			0.5 & 0.471 $\le$ $v$ $\le$ 0.5&\cellcolor{green} \\\cline{1-2}
			0.6 & 0.487 $\le$ $v$ $\le$ 0.539&\cellcolor{green} \\\cline{1-2}
			0.7 & 0.473 $\le$ $v$ $\le$ 0.564&\cellcolor{green} \\\cline{1-2}
			0.8 & 0.502 $\le$ $v$ $\le$ 0.583&\cellcolor{green} \\\cline{1-2}
			0.9 & 0.532 $\le$ $v$ $\le$ 0.598&\cellcolor{green} \\\cline{1-2}
			1 & 0.558 $\le$ $v$ $\le$ 0.608&\multirow{-11}{*}{\rotatebox{90}{\textbf{\textit{Left polarity}\,\,}}}\cellcolor{green}\\
			\hline			
		\end{tabular}
		\caption {The velocity window for unidirectional flow of composite BB solitons with different coupling strengths. For the Gaussian potential barriers with the range of coupling strength 0.313 $\le$ $g_{12}$ $\le$ 0.316 with $g_1 = g_2 = 1$, we find full reflection for $v$ $\leq$ 0.453 and full transmission for $v$ $\geq$ 0.454, hence no unidirectional flow is observed at this specific range of $g_{12}$. Away from this point, we find polarity reversal in unidirectional flow.}
		\label{gaussiantable}
	\end{center}
\end{table*}

\onecolumngrid
\section{}\label{appendix}
Using the effective Lagrangian, Eq.~\eqref{eq:secIV_Lagrangian}, the Euler--Lagrange equations lead to the following equations of motion for the variational parameters,
\begin{align}
\label{eq:EOMs_c0}
\nonumber
& \frac{g_{12} N^2}{a^3} \mathrm{csch}^2 \left(\frac{\xi_{1}-\xi_{2}}{a}\right) \Big[a -4 \left(\xi_{1} - \xi_{2}\right) \mathrm{coth} \left(\frac{\xi_{1}-\xi_{2}}{a}\right)\Big]
+ \frac{g_{12} N^2}{a^4} \left(\xi_{1}-\xi_{2}\right)^2\, \mathrm{csch}^4 \left(\frac{\xi_{1}-\xi_{2}}{a}\right) \Big[2 +\mathrm{cosh} \left(2\frac{\xi_{1}-\xi_{2}}{a}\right) \Big]
\\ \nonumber &
+ \frac{N l_{01}}{2 a^2} \Big[-\left(q_{2}+\xi_{1}\right) \mathrm{sech}^2 \left(\frac{q_{2}+\xi_{1}}{a}\right) +\left(q_{1}+\xi_{1}\right) \mathrm{sech}^2\left(\frac{q_{1}+\xi_{1}}{a}\right)
-\left(q_{2}+\xi_{2}\right) \mathrm{sech}^2\left(\frac{q_{2}+\xi_{2}}{a}\right)
+\left(q_{1}+\xi_{2}\right) \mathrm{sech}^2\left(\frac{q_{1}+\xi_{2}}{a}\right)\Big] \\ \nonumber &
+\frac{N l_{02}}{2 a^2}   \Big[-\left(q_{4}+\xi_{1}\right) \mathrm{sech}^2\left(\frac{q_{4}+\xi_{1}}{a}\right)
+\left(q_{3}+\xi_{1}\right) \mathrm{sech}^2\left(\frac{q_{3}+\xi_{1}}{a}\right)
-\left(q_{4}+\xi_{2}\right) \mathrm{sech}^2\left(\frac{q_{4}+\xi_{2}}{a}\right)
+\left(q_{3}+\xi_{2}\right) \mathrm{sech}^2\left(\frac{q_{3}+\xi_{2}}{a}\right)\Big] \\  &
+ \frac{2 N}{3 a^3} - \frac{\left(g_{1}+g_{2}\right)N^2}{6 a^2} -\frac{1}{3} N \pi^2 a \left(2 b^2 + b^{\prime}\right) = 0,
\end{align}

\begin{align}
\label{eq:EOMs_c1}
& \frac{3 g_{12} N^2}{a^2} \mathrm{coth} \left(\frac{\xi_{1}-\xi_{2}}{a}\right) \mathrm{csch}^2 \left(\frac{\xi_{1}-\xi_{2}}{a}\right)
+ \frac{N l_{01}}{2 a}\, \Big[-\mathrm{sech}^2 \left(\frac{q_{1}+\xi_{1}}{a}\right) +\mathrm{sech}^2\left(\frac{q_{2}+\xi_{1}}{a}\right)\Big] \\ \nonumber &
+\frac{N l_{02}}{2 a} \Big[-\mathrm{sech}^2 \left(\frac{q_{3}+\xi_{1}}{a}\right) +\mathrm{sech}^2\left(\frac{q_{4}+\xi_{1}}{a}\right)\Big]
- \frac{g_{12} N^2}{a^3} \left(\xi_{1} - \xi_{2}\right)  \mathrm{csch}^4 \left(\frac{\xi_{1}-\xi_{2}}{a}\right) \Big[2+ \mathrm{cosh}\left(2\frac{\xi_{1}-\xi_{2}}{a}\right)\Big] + N v^{\prime}_{1} = 0,
\end{align}

\begin{align}
\label{eq:EOMs_c2}
&-\frac{3 g_{12} N^2}{a^2}\,\mathrm{coth}\left(\frac{\xi_{1}-\xi_{2}}{a}\right)\,\mathrm{csch}^2 \left(\frac{\xi_{1}-\xi_{2}}{a}\right) + \frac{N l_{01}}{2 a}\Big[-\mathrm{sech}^2 \left(\frac{q_{1}+\xi_{2}}{a}\right) +\mathrm{sech}^2\left(\frac{q_{2}+\xi_{2}}{a}\right)\Big] \\ \nonumber &
+\frac{N l_{02}}{2 a} \Big[-\mathrm{sech}^2 \left(\frac{q_{3}+\xi_{2}}{a}\right) +\mathrm{sech}^2 \left(\frac{q_{4}+\xi_{2}}{a}\right)\Big]
+ \frac{g_{12} N^2}{a^3} \left(\xi_{1} - \xi_{2}\right)\, \mathrm{csch}^4 \left(\frac{\xi_{1}-\xi_{2}}{a}\right) \Big[2+ \mathrm{cosh}\left(2\frac{\xi_{1}-\xi_{2}}{a}\right)\Big] + N v^{\prime}_{2} = 0,
\end{align}

\begin{eqnarray}
\label{eq:EOMs_xi1}
- N v_{1} - N \xi^{\prime}_{1} = 0,
\end{eqnarray}
\begin{eqnarray}
\label{eq:EOMs_xi2}
- N v_{2} - N \xi^{\prime}_{2} = 0,
\end{eqnarray}
\begin{eqnarray}
\label{eq:EOMs_a}
-\frac{2}{3} N \pi^2 a^2 b + \frac{1}{3} N \pi^2 a a^{\prime} = 0.
\end{eqnarray}

\twocolumngrid

\end{document}